\pdfoutput=1
\documentclass[PAPER, atlasdraft=false, texlive=2023, UKenglish, subfigure, backref=false, cernpreprint , texmf, orcidlogo]{atlasdoc}
 
\usepackage{atlaspackage}
\usepackage{xcolor}
\usepackage{arydshln}
\usepackage{multirow}
\usepackage{graphicx}
\usepackage{graphbox}
\usepackage{seqsplit}
\usepackage{amsfonts}
\usepackage{ANA-SUSY-2018-42-PAPER-defs}
\usepackage{atlasbiblatex}
 
\usepackage[jetetmiss,BSM]{atlasphysics}
 
\makeatletter\let\expandableinput\@@input\makeatother
\graphicspath{{logos/}{figures/}}
 
\usepackage{ANA-SUSY-2018-42-PAPER-defs}

\AtlasTitle{Search for heavy, long-lived, charged particles with large ionisation energy loss in $pp$ collisions at $\sqrt{s} = 13~\text{TeV}$ using the ATLAS experiment and the full Run~2 dataset}
 
\AtlasJournal{JHEP}
\AtlasJournalRef{JHEP 06 (2023) 158}
\AtlasDOI{10.1007/JHEP06(2023)158}
\AtlasAbstract{
This paper presents a search for hypothetical massive, charged, long-lived particles with the \mbox{ATLAS} detector at the LHC using an integrated luminosity of $139~\mathrm{fb^{-1}}$ of proton--proton collisions at $\sqrt{s}=13~\text{TeV}$. These particles are expected to move significantly slower than the speed of light and should be identifiable by their high transverse momenta and anomalously large specific ionisation losses, ${\mathrm{d}}E/\mathrm{d}x$. Trajectories reconstructed solely by the inner tracking system and a ${\mathrm{d}}E/\mathrm{d}x$ measurement in the pixel detector layers provide sensitivity to particles with lifetimes down to ${\cal O}(1)~\text{ns}$ with a mass, measured using the Bethe--Bloch relation, ranging from $100~\text{GeV}$ to $3~\text{TeV}$. Interpretations for pair-production of $R$-hadrons, charginos and staus in scenarios of supersymmetry compatible with these particles being long-lived are presented, with mass limits extending considerably beyond those from previous searches in broad ranges of lifetime.}

\author{The ATLAS Collaboration}
 
\AtlasRefCode{SUSY-2018-42}
\PreprintIdNumber{CERN-EP-2022-029}

\hypersetup{pdftitle={ATLAS document},pdfauthor={The ATLAS Collaboration}}
 
\begin{document}
 
\maketitle

\clearpage

\section{Introduction}
\label{sec:intro}

A large number of physics models that extend the Standard Model (SM) predict the existence of new, massive, long-lived particles (LLPs)
which have a decay length long enough to be observed using particle detectors. These particles appear in proposed solutions to the gauge hierarchy problem, including supersymmetric (SUSY) models~\cite{Golfand:1971iw,Volkov:1973ix,Wess:1974tw,Salam:1974ig,Wess:1974jb,Ferrara:1974pu,Fayet:1976et} that either conserve or violate $R$-parity.\footnote{$R$-parity is a quantum number defined as $(-1)^{3(B-L)+2S}$ where $S$ is the particle spin and $L$ and $B$ are, respectively, its lepton and baryon number.} Within SUSY models, supersymmetric particles (sparticles) may have lifetimes that depend on the mass-hierarchy parameters or on the size of any $R$-parity-violating coupling.
 
SUSY models are theoretical extensions of the SM which relate fermions and bosons. They postulate that each SM particle has a SUSY partner (a sparticle) that is a boson if the corresponding particle is a fermion, and vice-versa. The sparticles are named so as to recall their SM partner: the gluino, stau, wino and higgsino are the SUSY partners of the gluon, $\tau$-lepton, $W$ boson and Higgs boson. The chargino is a linear combination of the charged wino and the charged higgsinos, while the neutralino is a linear combination of the bino, the neutral wino and the neutral higgsino, where the bino is the SUSY partner of the SM U(1) weak hypercharge gauge field.
 
\begin{figure}[tbp]
\begin{center}
\subfigure[]{
\label{fig:diagram_gluino}
\includegraphics[align=c,height=40mm]{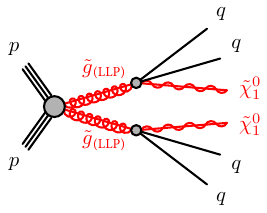}
}
\subfigure[]{
\label{fig:diagram_chargino}
\includegraphics[align=c,height=40mm]{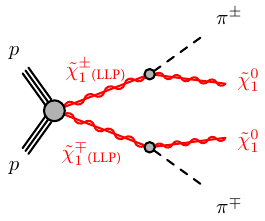}
}
\subfigure[]{
\label{fig:diagram_slepton}
\includegraphics[align=c,height=40mm]{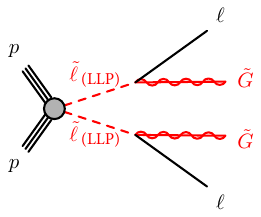}
}
\caption{Production diagrams for (a) pair-produced gluinos which form $R$-hadrons, (b) pair-produced charginos, and (c) pair-produced sleptons.}
\label{fig:diagrams}
\end{center}
\end{figure}

This study is sensitive to SUSY along with many other models of new physics beyond the Standard Model (BSM physics models), particularly those that predict the production of massive particles with lifetimes exceeding ${\cal O}(1)~\text{ns}$ at LHC energies. In this paper, the results of the search are interpreted in the context of pair production, from proton--proton ($pp$) collisions at $\sqrt{s}=13~\TeV$, of several different long-lived sparticles of charge $|q|=1$. Within the SUSY framework, several different processes yielding LLPs are present, and the production diagrams of the processes covered by this search are shown in Figure~\ref{fig:diagrams}. Gluinos (\gluino) can be long-lived if $R$-parity is conserved and the squark mass scale is very high, as proposed by mini-split SUSY scenarios~\cite{Giudice:2004tc,ArkaniHamed:2004fb}. Although the gluino itself is neutral, a long-lived gluino is predicted to hadronise, involving SM quarks or gluons with its QCD interaction and forming a colour-neutral state referred to as an $R$-hadron~\cite{FAIRBAIRN20071}, which can be electrically charged. Charginos~(\chinoonepm) can be long-lived when they and the counterpart neutralino (\ninoone) are almost mass-degenerate, for instance in anomaly-mediated SUSY-breaking (AMSB) models predicting a `pure wino' scenario~\cite{Giudice:1998xp,Randall:1998uk}. Staus~(\stau) can also be long-lived when a quasi-massless gravitino ($\gravino$) is assumed as the lightest neutral sparticle and the coupling between the stau and gravitino is very weak~\cite{Alwall:2008ve,Alwall:2008ag,Alves:2011wf}.
 
Extensive searches for charged or neutral LLPs have been carried out in the ATLAS and CMS experiments. Limits on gluino $R$-hadrons have been set by various complementary search techniques, including an explicit requirement of decay-vertex or displaced jet reconstruction~\cite{SUSY-2016-08, CMS-EXO-19-013, CMS-EXO-19-021}, and time-of-flight (ToF) and its optional combination with pixel \dEdx~\cite{SUSY-2016-32,SUSY-2016-02,CMS-EXO-15-010}. Very long-lived $R$-hadrons, eventually trapped within the calorimeters, were searched for by looking for decays during empty beam bunch-crossings~\cite{SUSY-2018-05,CMS-EXO-16-004}. The search presented here is agnostic with regard to decay activity, and is therefore also sensitive when the mass difference between the gluino and the neutralino is small, a situation in which displaced decay detection becomes more challenging.
 
Limits on long-lived charginos, for pure-wino~(AMSB, \cite{Giudice:1998xp,Randall:1998uk}) as well as pure-higgsino~\cite{FUKUDA2018306} scenarios, have also been set using a disappearing-track signature~\cite{SUSY-2018-19,SUSY-2016-06,CMS-SUS-19-005, CMS-EXO-19-010,CMS-EXO-16-044} and using the ToF technique~\cite{SUSY-2016-32,SUSY-2016-02}. Masses up to $660~(210)~\GeV$ are excluded in scenarios where the chargino is a pure wino (higgsino) for theoretically preferred lifetimes of each, and the largest excluded mass is $860~(720)~\GeV$ at lifetimes of 1--1.3~ns~\cite{SUSY-2018-14}. For a stable chargino, masses up to $1.09~\TeV$ are excluded using the ToF technique~\cite{SUSY-2016-32}. Long-lived sleptons in the context of gauge-mediated SUSY-breaking (GMSB)~\cite{Alwall:2008ve,Alwall:2008ag,Alves:2011wf} were searched for by a dedicated displaced-leptons search for stau (\stau), smuon (\smu) and selectron (\sel) cases separately, and masses up to $340~\GeV$, $680~\GeV$ and $720~\GeV$, respectively, are excluded for lifetimes of 0.1~ns~\cite{SUSY-2018-14}. For a stable \stau, masses up to $430~\GeV$ are excluded using the ToF technique~\cite{SUSY-2016-32, CMS-EXO-15-010}. The search presented in this paper is more sensitive than previous ATLAS searches for most of the lifetime range of $\tau \gtrsim 1~\text{ns}$ and is complementary to the other search strategies mentioned above.
 
A charged particle produced by a collision propagates outwards in space until its decay, and specific ionisation losses (\dEdx) along its path produce hit records on the detector layers. Appropriate fitting of these hits reconstructs the trajectory of the particle, referred to as a track. Charged LLPs with masses greater than ${\cal O}(100)~\GeV$ produced at the LHC are expected to be significantly slower than the speed of light, and therefore should have \dEdx significantly higher than any SM particle of the same momentum, following the Bethe--Bloch relation. The pixel detector~\cite{ATLAS-TDR-11,ATLAS-TDR-19,PIX-2018-001} at the core of the ATLAS detector\footnote{ATLAS uses a right-handed coordinate system with its origin at the nominal interaction point in the centre of the detector and the $z$-axis coinciding with the axis of the beam pipe. The $x$-axis points from the interaction point to the centre of the LHC ring, and the $y$-axis points upward. Cylindrical coordinates ($r$, $\phi$) are used in the transverse plane, $\phi$ being the azimuthal angle around the $z$-axis. The pseudorapidity is defined in terms of the polar angle $\theta$ as $\eta$ = $-$ln tan($\theta$/2).}~\cite{PERF-2007-01} is able to measure \dEdx within a radius of about 13~cm from the $pp$ collisions, and therefore can be used to identify LLPs with lifetime in excess of ${\cal O}(1)~\text{ns}$. To a good approximation, this identification method does not depend on the way the LLP interacts in the ATLAS calorimeters~\cite{ATLAS-TDR-02,ATLAS-TDR-03} or on the LLP decay mode. Consequently, this search can capture a broad range of possible signals in BSM models predicting charged LLPs, without heavily depending on the specific features of these models.
 
This analysis uses the full Run~2 dataset, corresponding to an integrated luminosity of $139~\mathrm{fb^{-1}}$, and is an update of previous ATLAS searches performed in both Run~1 and Run~2~\cite{SUSY-2011-03, SUSY-2014-09, SUSY-2016-03, SUSY-2016-31}. Similar searches were performed by the CMS experiment~\cite{CMS-EXO-15-010,CMS-EXO-12-026}. In the last ATLAS search targeting the same signature, with 36~fb$^{-1}$ of Run~2 data, no significant deviation from SM background expectations was observed. Results were interpreted assuming the pair production of $R$-hadrons. $R$-hadrons with lifetimes above 1~ns were excluded at the 95\% confidence level (CL) of the $\mathrm{CL}_\mathrm{s}$ prescription~\cite{Read:2002hq}, with lower limits on the gluino mass ranging between $1.29~\TeV$ and $2.06~\TeV$. In the case of $R$-hadrons stable enough to leave the detector (detector-stable), the lower limit on the gluino mass at the 95\% CL is $1.89~\TeV$.
In the previous ATLAS search~\cite{SUSY-2016-31}, a mild excess of data events over the background prediction was observed with a local significance of $2.4\sigma$ in the mass range between $500~\GeV$ and  $800~\GeV$ for the stable-particle selection. This has motivated an effort to improve the analysis sensitivity over a wide mass range from around several hundred $\GeV$ to a few \TeV, as explained further in Section~\ref{sec:overview}. Accordingly, this search aims to cover a wide range of production cross-sections, including both electroweak and strong production of charged LLPs.

This paper is organised as follows. Section~\ref{sec:detector} describes the ATLAS experiment and, in more detail, its pixel detector.
Section~\ref{sec:overview} presents the search strategy and emphasises the model-independence of the method used. Section~\ref{sec:data_sim} describes the dataset used in the analysis and the simulation models used in interpreting the results. Section~\ref{sec:correct_calib} explains the various corrections and calibrations applied to the pixel \dEdx measurement in order to make optimal use of this variable. It also explains and characterises the mass reconstruction obtained through the Bethe--Bloch relation.
Section~\ref{sec:selection} describes the selection criteria applied to the events and candidate LLP tracks in this search, and the studies to optimise these criteria. Section~\ref{sec:background} focuses on the data-driven background estimation, while Section~\ref{sec:syst} examines systematic uncertainties in the background estimation and signal efficiency. Finally, Section~\ref{sec:results} discusses the combined signal significance and its interpretation, and Section~\ref{sec:conclusion} summarises the most important results of this search.


\FloatBarrier
\section{ATLAS detector and \dEdx measurement}
\label{sec:detector}
 
The ATLAS detector is a general-purpose detector with a forward--backward-symmetric cylindrical layout covering nearly $4\pi$ in solid angle~\cite{PERF-2007-01}.
It consists of an inner detector (ID) tracking system, to measure the trajectories of charged particles, surrounded by a 2~T solenoid, followed by calorimeters to measure the energy of particles that interact electromagnetically or hadronically, and
a muon spectrometer (MS) inside toroidal magnets to provide additional tracking for muons. The detector is hermetic within its $\eta$ acceptance and can therefore measure the missing transverse momentum ($\vec{p}_\mathrm{T}^{~\mathrm{miss}}$, with magnitude $\MET$) associated with each event.
A two-level trigger system is used to select events~\cite{TRIG-2016-01}. The first-level trigger selects events from a bunch crossing (the LHC bunch crossings have a separation of 25~ns), is implemented in hardware and uses a subset of the detector information. This is followed by a software-based high-level trigger, which runs calibration and prompt reconstruction algorithms, reducing the event recording rate to about 1~kHz. An extensive software suite~\cite{ATL-SOFT-PUB-2021-001} is used in the reconstruction and analysis of real and simulated data, in detector operations, and in the trigger and data acquisition systems of the experiment.
 
The ID is made of three detector systems organised in concentric regions covering $|\eta|<2.5$. The outermost system (TRT)~\cite{IDET-2015-01} is made of densely packed 4-mm-diameter cylindrical drift tubes covering $|\eta|<2$.  The TRT covers the radial region from 60~cm to 100~cm and provides many (${>}30$) measurements with 0.13~mm ($r$--$\phi$) accuracy, thus contributing to momentum measurement accuracy and robustness. The radial region from 30~cm to 60~cm is equipped with silicon microstrip detectors (SCT), providing at least four layers of double-sided strip modules with a 40~mrad stereo angle along trajectories within $|\eta|<2.5$~\cite{IDET-2013-01}. The innermost region is covered by a silicon pixel detector~\cite{ATLAS-TDR-11,ATLAS-TDR-19,PIX-2018-001}, which, being crucial for this measurement, is described below in some detail.
 
The pixel detector provides, on average, four precision measurements for each track in the region $|\eta| <2.5$ at radial distances of 3.4~cm to 13~cm from the LHC beam line.
Compared to the other layers, the innermost pixel layer (IBL)~\cite{ATLAS-TDR-19,PIX-2018-001} has smaller-area pixels, reduced thickness, faster electronics, and provides charge measurements with lower resolution and dynamic range. At normal incidence and without any radiation damage, the average charge released by a minimum-ionising particle (MIP) in a pixel sensor is approximately $20\,000~e^{-}$ ($16\,000~e^{-}$ for the IBL) and the charge threshold (i.e.\ the minimum value for a charge to be recorded) is typically set to $3500\pm40~e^{-}$ ($2500\pm40~e^{-}$ for the IBL). If the charge released in a pixel exceeds the IBL dynamic range (which is set at approximately $30\,000~e^{-}$) an overflow bit ($\IBLOF$) is set. The overflow mechanism is not present in the outer pixel layers, and hits exceeding the dynamic range (${\sim}200\,000~e^{-}$) are lost.  The presence of an IBL overflow bit ($\IBLOF=1$) indicates that a high specific-ionisation charge was deposited locally in the IBL and this is much more likely to happen for a heavy, charged-LLP track than for a SM particle track. Tracks with an IBL overflow bit undergo a dedicated analysis treatment, as described in Section~\ref{sec:correct_calib}.
 
The time-over-threshold (\ToT), i.e.\ the time interval with the signal above a preset threshold, is digitised and recorded to 8~bits (4~bits for the IBL).
The \ToT is approximately proportional to the ionisation charge~\cite{BaBar_ToT} and allows the specific ionisation of a track to be calculated, as explained below.
The charge released by a track crossing the pixel detector is rarely contained within just one pixel; neighbouring pixels registering hits are joined together using a connected component analysis~\cite{PERF-2012-05} to form clusters. The charge of a cluster is calculated by summing the charges of all pixels belonging to the cluster after calibration corrections.
The \dEdx measurement assigned to each track is then calculated by averaging the ionisation measurements (charge collected in the cluster per unit track length in the sensor) of its individual clusters. The specific ionisation follows a Landau distribution, and to reduce the effect of the tails of this distribution a truncated average (\dEdxtrunc) is evaluated after removing the highest \dEdx cluster, or the two highest \dEdx clusters in the rare case of more than four pixel clusters on a track. Clusters including pixels at the sensor edges are dropped, as part of the charge may escape detection. The $\IBLOF=1$ clusters are never used to calculate the \dEdxtrunc, as their \dEdx is only known to be above a given value.
A track is considered for this analysis if the \dEdxtrunc is calculated using at least two clusters after removal of those meeting the criteria defined above. The average number of clusters used for \dEdxtrunc calculation is $\approx 2.7$ per track.
 
Finally, the \dEdxtrunc is corrected for variations of the pixel detector conditions during the data-taking period (e.g.\ charge losses due to radiation damage) and for the residual $\eta$-dependence, as described in detail in Section~\ref{sec:correct_calib}. The output is referred to as `corrected \dEdx' (\dEdxcorr) and is the variable used in the signal selection for the search. This variable, like the \textit{restricted energy loss}~\cite{10.1093/ptep/ptaa104}, has no logarithmic rise at high values of $\beta\gamma$ and no sensitivity to radiative effects.
 
The $\beta\gamma$ of a particle can be calculated from the \dEdxcorr of its track using the Bethe--Bloch relation. A meaningful $\beta\gamma$ value can only be estimated in the range $0.3\lesssim\beta\gamma\lesssim 0.9$ using the pixel detector. The lower limit is a consequence of the \ToT dynamic range, while the upper limit is due to the proximity of the MIP regime which begins at $\beta\gamma\approx$~3 and where  \dEdxcorr  becomes quasi-independent of $\beta\gamma$.


\section{Analysis overview}
\label{sec:overview}
 
\begin{figure}[tbp]
\begin{center}
\subfigure[]{
\label{fig:mcTruth_betaGamma}
\includegraphics[width=0.47\textwidth]{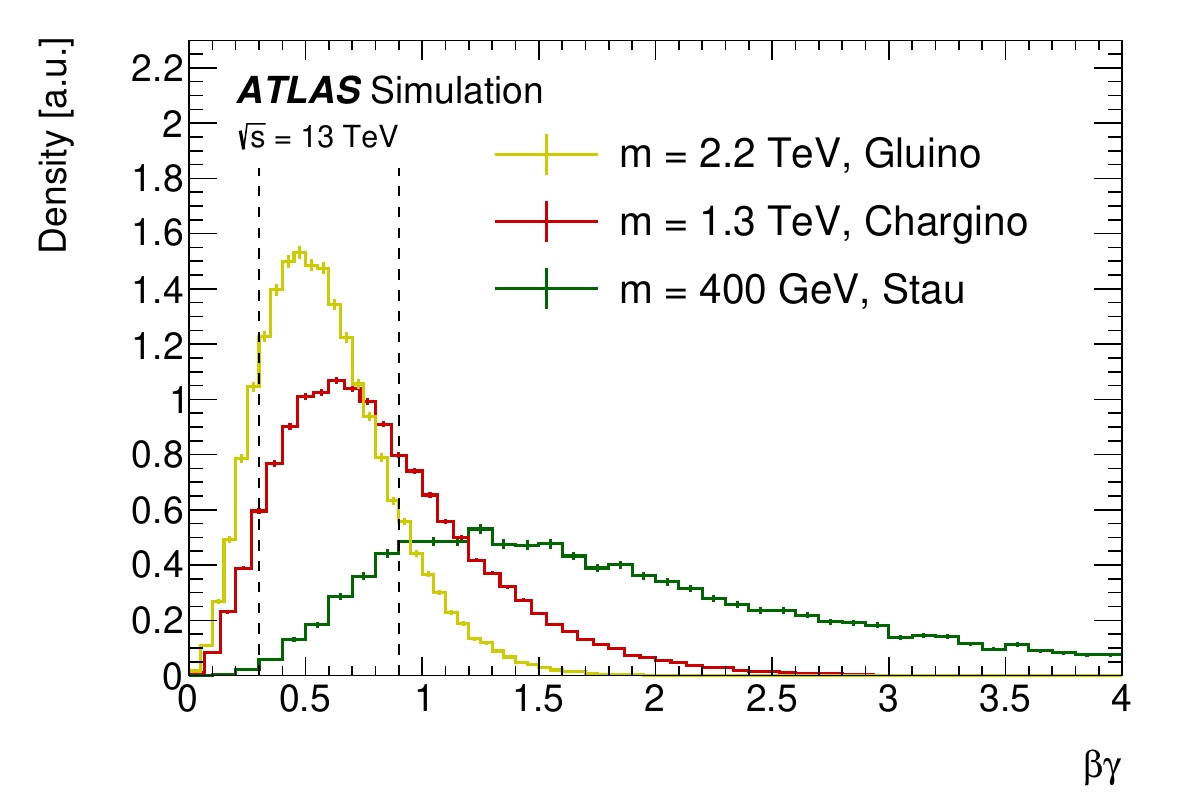}
}
\subfigure[]{
\label{fig:mcTruth_eta}
\includegraphics[width=0.47\textwidth]{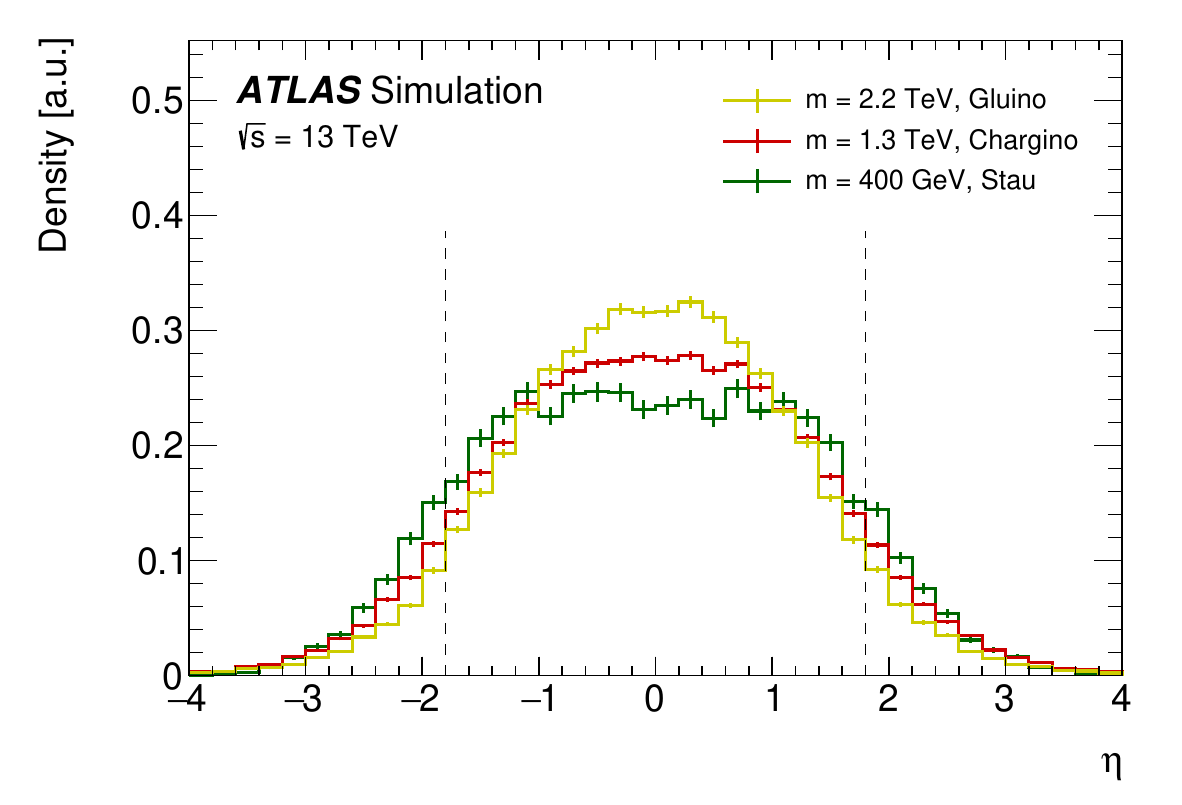}
}
\subfigure[]{
\label{fig:mcTruth_betaGamma_Gluino}
\includegraphics[width=0.47\textwidth]{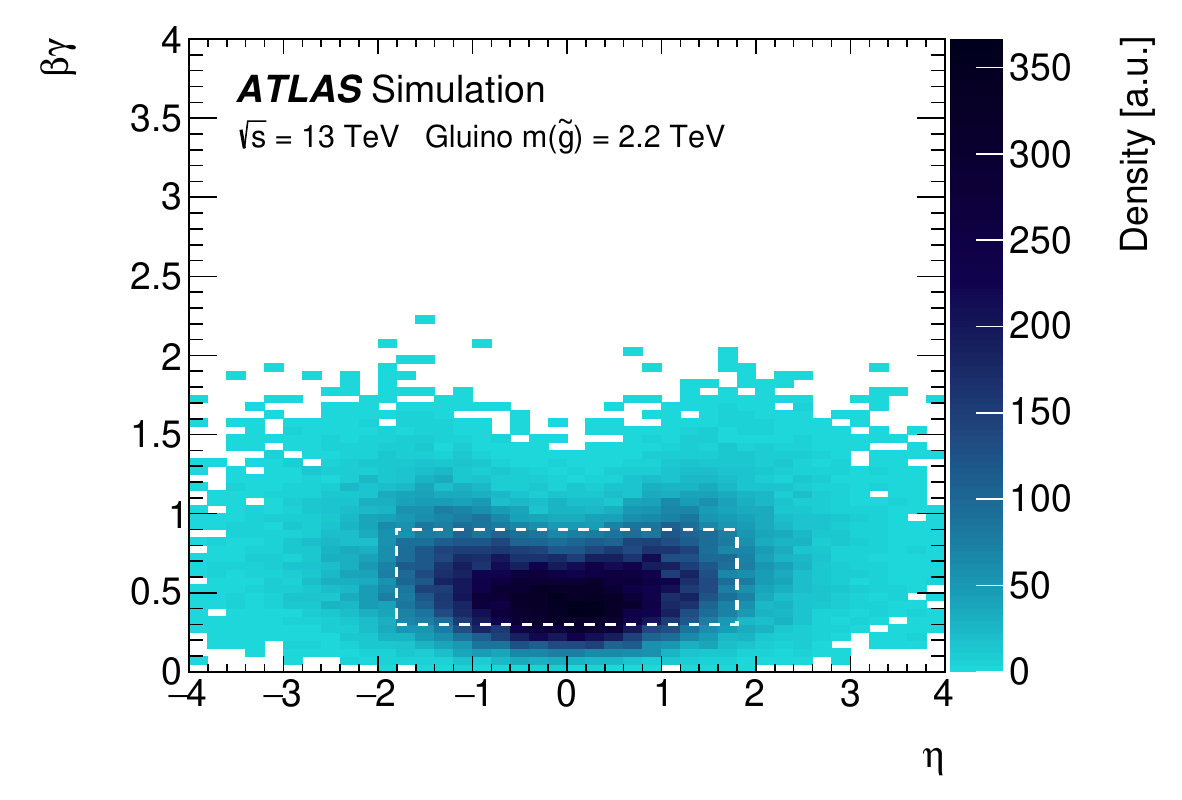}
}
\subfigure[]{
\label{fig:mcTruth_betaGamma_Chino}
\includegraphics[width=0.47\textwidth]{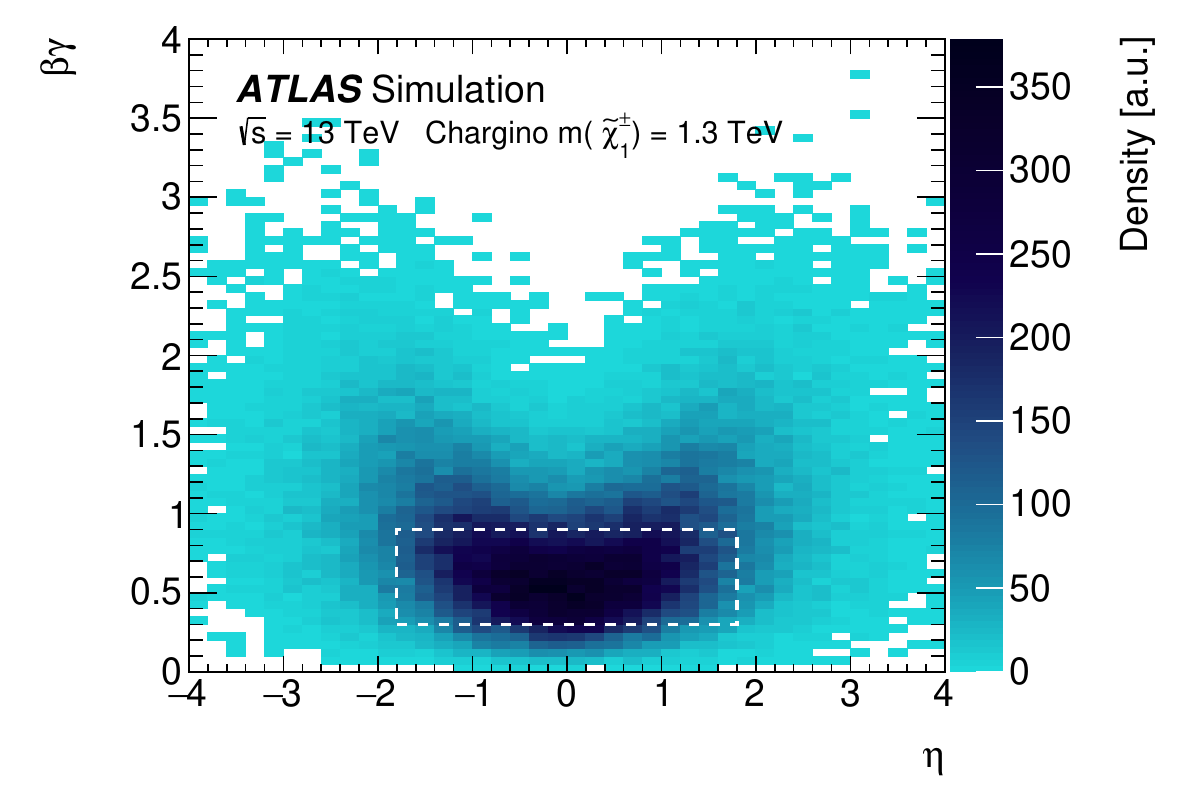}
}
\caption{\subref{fig:mcTruth_betaGamma} $\beta\gamma$ distribution and \subref{fig:mcTruth_eta} $\eta$ distribution of various benchmark signal particles at the generator level, with the density displayed in arbitrary units. Preselections are not applied. The ranges $0.3<\beta\gamma<0.9$ of \subref{fig:mcTruth_betaGamma} and $|\eta| < 1.8$ of \subref{fig:mcTruth_eta} indicated by vertical dashed lines approximately correspond to the acceptance of this search from the $\dEdx$ and $\eta$ requirements respectively. Panels~\subref{fig:mcTruth_betaGamma_Gluino} and \subref{fig:mcTruth_betaGamma_Chino} illustrate the distribution in the $\eta\mbox{--}\beta\gamma$ plane for the gluino and chargino samples respectively. The area within the dashed lines corresponds to the approximate acceptance of the search. The correlation between $\beta\gamma$ and $|\eta|$ loosely reflects the relation $\beta\gamma=p/m=\pt\cosh(\eta)/m$.}
\label{fig:mcTruth}
\end{center}
\end{figure}

The search strategy consists of looking for isolated tracks with high transverse momentum, $\pt$, and large specific ionisation. The reconstructed mass is then calculated for each track using a parameterisation of the Bethe--Bloch relation and depends on its momentum and \dEdxcorr. It is noteworthy that the effective $\beta\gamma$ range of $0.3 \lesssim \beta\gamma \lesssim 0.9$ matches that of pair-produced heavy charged LLPs over wide mass ranges, as illustrated in Figure~\ref{fig:mcTruth_betaGamma}. The $\beta\gamma$ distribution reflects the opening of relativistic phase space near the mass threshold, and the distribution peaks at lower values for larger LLP masses. This is a common feature of pair-production kinematics from an off-shell propagator, including the pair production of SM particles such as top quarks, and hence is not strongly dependent on the BSM dynamics of the LLP. Only tracks produced centrally ($|\eta| < 1.8$) are selected.  This removes background tracks at high $|\eta|$ while retaining the acceptance for LLPs as shown in Figure~\ref{fig:mcTruth}, and also limits the pixel layers used in the \dEdx calculation to those within the barrel region, thereby simplifying the measurement.
 
Events are selected using the lowest-threshold unprescaled calorimetric \MET trigger, which is based on the magnitude of the negative of the vectorial energy sum measured in the calorimeters corrected to take into account the multiple $pp$ interactions in each event~\cite{TRIG-2019-01}. In events with metastable LLPs (i.e. LLPs which decay inside the ATLAS detector), the measured $\MET$ originates mainly from neutralinos or gravitinos which carry away unmeasured momentum. In events where the LLPs are detector-stable and decay outside \mbox{ATLAS}, the LLPs leave only modest energy depositions in the calorimeters, even in the $R$-hadron case, and only a fraction of them are reconstructed as a muon owing to their late arrival time in the muon spectrometer. Therefore, most of the momentum of each LLP is not accounted for in the measurement of \MET, and only QCD initial-state radiation (ISR) provides a visible contribution that results in a measured imbalance of transverse momentum. Due to the presence of neutralinos or gravitinos, the $\MET$ trigger efficiency is higher for metastable LLPs than for detector-stable LLPs. On the other hand, the track reconstruction efficiency is higher for  detector-stable LLPs and penalises particles with lifetimes shorter than 10~ns, which may not have crossed enough detector layers. Further selections are applied to triggered events and candidate tracks as detailed in Section~\ref{sec:selection}.
 
The signal can form a peak in the reconstructed mass distribution and thus be observed as an excess of events over the expected background.  The search takes into account the mass resolution for the signal processes. Full-length ID tracks have a relative $1/\pt$ resolution of approximately 4\% (40\%) at $\pt=100~\GeV~(1~\TeV)$ for $|\eta|<0.5$. This resolution is approximately proportional to $\pt$ in this high-momentum range and is dominated by the intrinsic position resolution and residual misalignment of the tracker~\cite{ATLAS-CONF-2014-047}. Low-lifetime LLPs have shorter tracks, resulting in larger momentum uncertainties and larger mass uncertainties. Reflecting this momentum resolution and the \dEdx dispersion, the reconstructed mass distribution has a considerable width. For a predefined set of target mass-value hypotheses, the corresponding set of mass windows is defined so that each window captures approximately 70\% of the expected signal at the given target mass, but differs slightly between lifetimes less than or equal to 1~ns and greater than 1~ns. Here, the choice of mass windows for lifetimes shorter than 1~ns accounts for the poorer mass resolution due to the shorter LLP tracks.
The mass windows are common to sleptons, charginos and $R$-hadrons of the same mass and are defined as explained in Section~\ref{sec:masswindows}.

The background is mostly due to SM processes generating high-\pt tracks with a large \dEdx that is randomly produced according to the Landau distribution of MIPs. The background yield and its distribution in the reconstructed mass spectrum is estimated in a fully data-driven approach, as described in Section~\ref{sec:background}. Data control samples are used to parameterise the momentum and $\dEdx$ distributions and their interdependence, and then to generate pseudo-data which predict the background distribution. Potential signal contamination is minimised in these background samples by inverting some of the selection criteria.
 
\FloatBarrier


\section{Data and simulation samples}
\label{sec:data_sim}
 
The analysis is conducted with 139 $\text{fb}^{-1}$ of $pp$ collision data which satisfy the ATLAS data quality requirements~\cite{DAPR-2018-01}. The dataset was taken during Run~2 of the LHC from 2015 to 2018, at a centre-of-mass energy of 13 \TeV. The average number of collisions per bunch-crossing (\pileup, $\langle\mu\rangle$) is approximately 34. A dedicated 21~$\text{pb}^{-1}$ low-\pileup dataset with $\langle\mu\rangle\sim 0.4$ taken in 2017 is used for the \dEdx-to-$\beta\gamma$ calibration. In this dataset, tracks are reconstructed if they have $\pT > 100~\MeV$, while the minimum $\pT$ requirement in the standard dataset is $500~\MeV$.
 
To optimise the analysis selection, Monte Carlo (MC) samples were produced to simulate events containing long-lived gluinos, charginos, and staus, corresponding to the production diagrams shown in Figure~\ref{fig:diagrams}. These three signal models are complementary in this study. The gluino samples have large production cross-sections and are suited to probing the high-mass frontier beyond 2~\TeV. The slepton samples have production cross-sections that are several orders of magnitude smaller than for gluinos of the same mass, and are suited to probing the mass range from a few to several hundred~\GeV. The chargino sample cross-sections have intermediate values and are useful in probing the mass range from $500~\GeV$ to $1.3~\TeV$.

Gluino pair production was simulated for gluino masses ranging from 400~\GeV\ to 3~\TeV\ and lifetimes ranging from 1~ns to stable within a simplified model inspired by a split-SUSY scenario~\cite{Giudice:2004tc,ArkaniHamed:2004fb}. The events were generated by \MGNLO[2.6.2] with up to two additional partons at leading order, and interfaced to \PYTHIA[8.240]~\cite{Sjostrand:2014zea} using the A14 set of tuned parameters (`tune')~\cite{ATL-PHYS-PUB-2014-021} and the \NNPDF[2.3lo] parton distribution function (PDF) set for parton showering and hadronisation, with decays of bottom and charm hadrons performed by \EVTGEN[1.6.0]~\cite{Lange:2001uf}. The CKKW-L merging scheme~\cite{Catani:2001cc,Hoeche:2009rj} was applied to combine the matrix element with the parton shower. The long-lived gluino, which carries colour charge, hadronises to form a colourless composite particle called an \Rhadron.  The details of the \Rhadron simulation are given in Ref.~\cite{ATL-PHYS-PUB-2019-019}. The nominal cross-section values were calculated at next-to-leading order (NLO) with resummation of next-to-leading logarithms (NLL) and their uncertainty were taken from an envelope of predictions using different PDF sets and factorisation and renormalisation scales~\cite{Borschensky:2014cia}. Each gluino decays into a neutralino and two quarks via a virtual squark at a very high mass scale. To probe decays with different kinematics, two sets of samples were produced: one with a fixed neutralino mass of $m(\ninoone) = 100~\GeV$, and the other one having a compressed spectrum with a fixed mass-splitting of $\Delta m(\gluino,\ninoone) = 30~\GeV$. These two series of mass parameters are complementary and illustrate the breadth of the search, reflecting the fact that it does not require explicit decay properties of the charged LLP.  
 
Samples with a combination of chargino--neutralino ($\chinoonepm\ninoone$) and chargino--chargino ($\chinoonep\chinoonem$) events were generated with nearly degenerate chargino and neutralino masses, motivated by the `pure wino' AMSB scenario~\cite{Giudice:1998xp,Randall:1998uk}. Each long-lived chargino decays into a neutralino and a pion, where the mass-splitting between the chargino and neutralino is set to approximately 160~\MeV, with $\tan\beta=5$ and a positive sign of the higgsino mass parameter. Although the AMSB model has a specific preference for the chargino's lifetime $(\mathcal{O}(0.2)~\mathrm{ns})$ and mass relation via the loop dynamics, this theoretical constraint was artificially loosened for experimental benchmarking, and lifetimes ranging from 1~ns to stable and chargino masses ranging from $400~\GeV$ to $1.6~\TeV$ were examined. Samples were produced using \MGNLO[2.6.2] with up to two additional partons at leading order in the matrix element, and interfaced to \PYTHIA[8.230] using the A14 tune and the \NNPDF[2.3lo] PDF set for parton showering and hadronisation, with decays of bottom and charm hadrons performed by \EVTGEN~1.6.0. The CKKW-L merging scheme~\cite{Catani:2001cc,Hoeche:2009rj} was applied to combine the matrix element with the parton shower. A 100\% branching ratio for $\chinoonepm \rightarrow \pi^{\pm}\ninoone$ is assumed. The cross-sections for the electroweak and strong production models are calculated at NLO in the strong coupling constant $\alphas$ using Prospino2~\cite{Beenakker:1996ch}.

Events with pair-produced staus, each of which decays into a $\tau$-lepton and a gravitino, were produced in the GMSB scenario~\cite{Alwall:2008ve,Alwall:2008ag,Alves:2011wf}. The stau masses range from 100~\GeV\ to 1~\TeV\ and the gravitino is massless. Samples with stau lifetimes ranging from 1~ns to stable were produced. Events were simulated with up to two additional partons at leading order using \MGNLO[2.6.1] with the \NNPDF[2.3lo] PDF set, and interfaced to \PYTHIA[8.230] using the A14 tune. The mixed states $\stau_{1,2}$ of the left- and right-handed staus ($\stau_\mathrm{L,R}$) were generated with a mixing angle $\sin\theta_{\stau} = 0.95$. Signal cross-sections were calculated at next-to-leading order in $\alphas$, with soft-gluon emission effects added at next-to-leading-logarithm accuracy~\cite{Beenakker:1999xh,Debove:2010kf,Fuks:2012qx,Fuks:2013vua,Fiaschi:2018hgm}.
 
Inelastic $pp$ interactions were generated using \Pythia[8.186] and \EVTGEN[1.6.0] with the \NNPDF[2.3lo] PDF set and the A3 tune~\cite{ATL-PHYS-PUB-2016-017}. The inelastic collisions were overlaid onto the hard-scattering process to simulate the effect of multiple $pp$ interactions. MC samples were reweighted to match the distribution of the average number of interactions per bunch crossing observed in data.
 
The MC events were passed through a full detector simulation~\cite{SOFT-2010-01} based on \GEANT~\cite{Agostinelli:2002hh}. The propagation and decays of charginos and staus were simulated within \GEANT, taking into account ionisation loss and interactions with detector material. The propagation of $R$-hadrons and their interactions were handled by \GEANT\ until their decay, at which point the decay chains and subsequent hadronisation were simulated by \PYTHIA[8], and then information about the outgoing particles was transferred back to \GEANT.
 
The \dEdx response of the pixel detector is also simulated in the \GEANT framework. It is based on a realistic charge-deposition model~\cite{WANG20181}, but due to the sensitivity of the \dEdx measurement to detector conditions, including radiation damage, the simulated track \dEdx and especially the probability that a track has a hit in the IBL overflow do not follow the data accurately enough for this analysis. Hence, the \dEdx response for simulated events was customised in this analysis by replacing it with values from a data-driven template which was derived from a study described in the following section. 


\section{\dEdx corrections and mass calibration}
\label{sec:correct_calib}

\begin{figure}[tbp]
\begin{center}
\subfigure[]{
\label{fig:IBL0}
\includegraphics[width=0.47\textwidth]{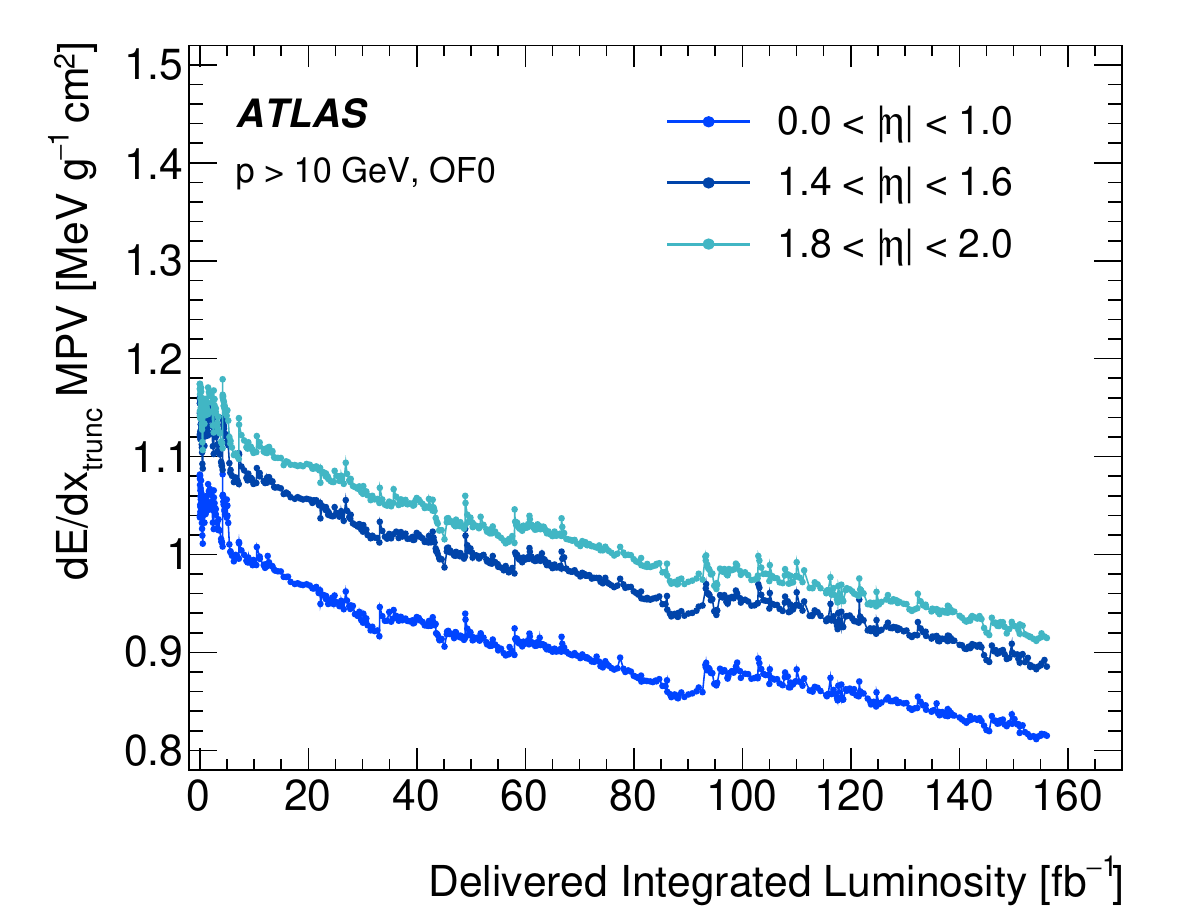}
}
\subfigure[]{
\label{fig:IBL1}
\includegraphics[width=0.47\textwidth]{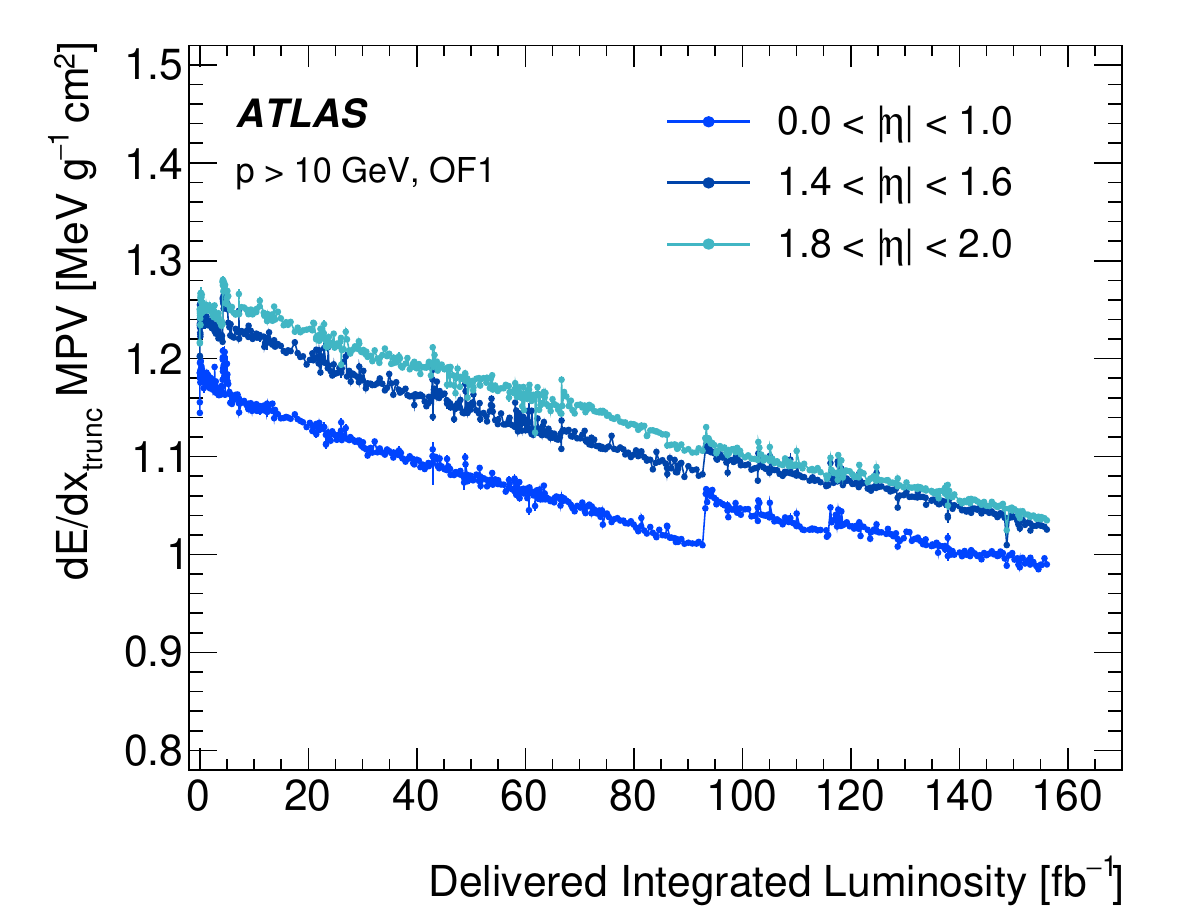}
}
\caption{The drift of MIP-MPV $\dEdxtrunc$ as a function of the delivered integrated luminosity in Run~2, divided into different pseudorapidity ranges, for MIP tracks of $p>10~\GeV$ \subref{fig:IBL0} without an IBL overflow (OF0) or \subref{fig:IBL1} with an IBL overflow (OF1). The lower $\dEdxtrunc$ in \subref{fig:IBL0} is due to the contribution of the IBL clusters which are restricted by the overflow  to be under $\approx$~2 MIPs.}
\label{fig:run2_dedx_vs_lumi}
\end{center}
\end{figure}

The most probable value (MPV) of the track $\dEdxtrunc$ as a function of delivered integrated luminosity is shown in Figure~\ref{fig:run2_dedx_vs_lumi} for $p>10~\GeV$ tracks, with and without a hit in the IBL overflow. As expected, the charge collection efficiency decreases with increasing integrated luminosity because of the damage induced in the silicon by the particle flux, and a decrease in the measured $\dEdxtrunc$ is therefore visible across the dataset. Large occasional jumps in the trend (e.g.\ at around $92~\mathrm{fb^{-1}}$) indicate changes to the pixel detector's charge calibration scheme, while smaller fluctuations correspond to regular updates of the charge and threshold calibrations during machine development periods or technical shutdowns. The effect of the radiation damage can be seen to depend on $|\eta|$, reflecting differences in the fluence over $|\eta|$. In addition, the fluence is higher in pixel sensors at smaller radii. However, the analysis does not compensate for the effect in each pixel sensor, since the $\dEdxtrunc$ of a track is calculated before such compensation can be made, and each pixel cluster's information is practically inaccessible in the data flow of the analysis. Corrections are therefore applied only to $\dEdxtrunc$ in this analysis.
 
To minimise these dependences, run-dependent corrections are calculated and applied. The run-dependent corrections are calculated separately for tracks in bins of $|\eta|$ and $\IBLOF$. In each run with sufficient data, an estimate of the $\dEdxtrunc$ peak for each ($|\eta|, \IBLOF$) bin is used to normalise the most probable $\dEdxtrunc$ to that of a reference run.\footnote{the reference run has been chosen in the middle of total integrated luminosity and close to the low luminosity run used for the calibration of $\dEdxtrunc$ into $\beta\gamma$} If a run does not have sufficient data, the correction from the closest run with sufficient data is applied. After the run-dependent corrections are applied, $|\eta|$-dependent corrections are applied separately in bins of $\IBLOF$ so that the MPV for MIPs is equalised over $|\eta|$. The resulting corrected \dEdx measurement is referred to as \dEdxcorr and is normalised for MIP-like tracks to peak at ${\sim}1~\dEdxunit$, the value for unirradiated silicon of the thickness used in the \mbox{ATLAS} pixel detector. Hereafter, the symbol `$\dEdx$' stands for \dEdxcorr.
 
\begin{figure}[tbp]
\begin{center}
\subfigure[]{
\label{fig:calib_lowpSlice}
\includegraphics[width=0.7\textwidth]{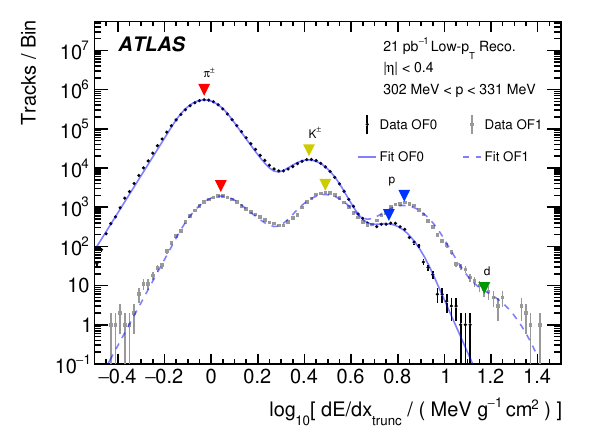}
}
\subfigure[]{
\label{fig:calib_Bethe}
\includegraphics[align=c,height=72mm]{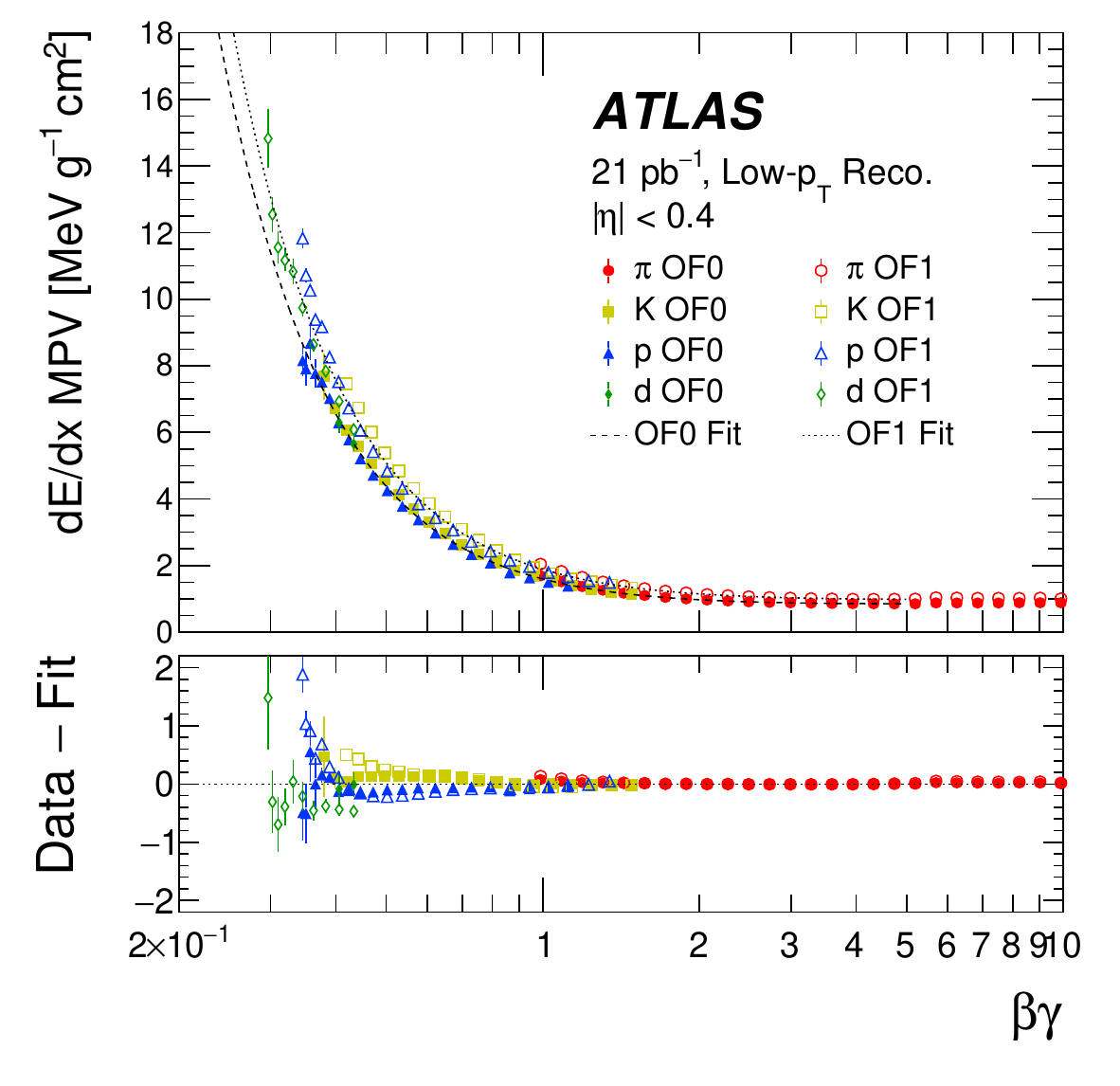}
}
\subfigure[]{
\label{fig:calib_IBLOF}
\includegraphics[align=c,height=72mm]{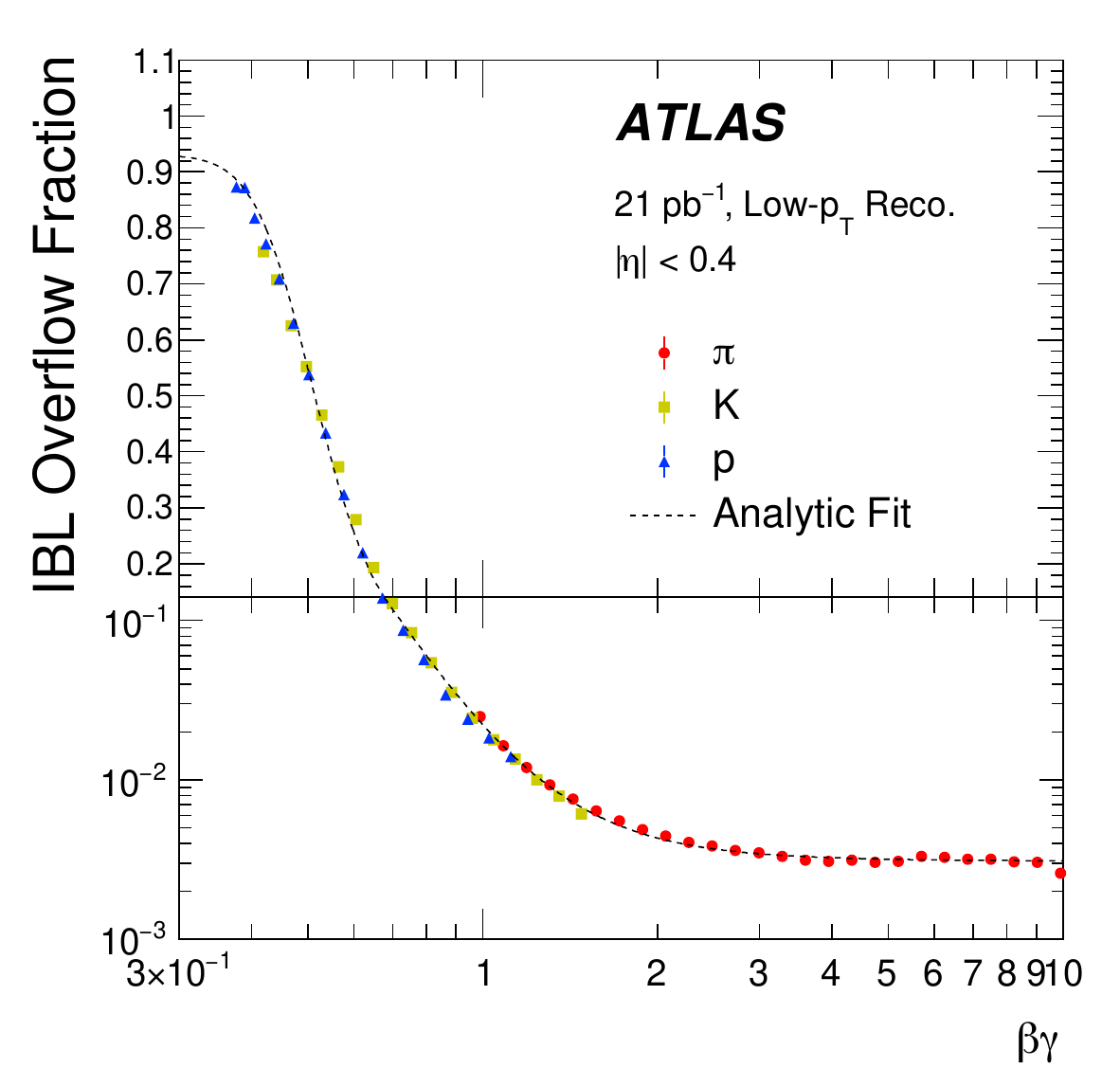}
}
\caption{\subref{fig:calib_lowpSlice} An example of a $\log_{10}\left[\dEdxtrunc/(\dEdxunit)\right]$ distribution of low-\pileup minimum-bias charged-particle tracks in a momentum slice of $302~\MeV < p < 331~\MeV$ in the pseudorapidity range $|\eta|<0.4$. Tracks are classified by $\IBLOF$. Triangle markers indicate identified MPVs corresponding to pions (red), kaons (yellow), protons (blue) and deuterons (green). The result of fitting a superposition of the $\dEdxtrunc$ distributions of identified particles is also overlaid for each data series. \subref{fig:calib_Bethe} MPV values of $\dEdxtrunc$ as a function of particle $\beta\gamma$ in $|\eta|<0.4$, classified by $\IBLOF$. The data values are fitted by a function respecting the Bethe--Bloch formula. \subref{fig:calib_IBLOF} The fraction of $\IBLOF=1$ tracks as a function of particle $\beta\gamma$ in $|\eta|<0.4$. The data values are fitted with an analytic function ${\displaystyle y_{0}+\frac{ 1-y_{0}-y_{1} }{ 1+a\left[\log_{10}(\beta\gamma)-x_{0}\right]^{n}} }$. In order to show all important features, a linear scale is used in the upper panel for the fraction above 0.14, while a logarithmic scale is used for the lower panel.}
\label{fig:lowPtCalibrationPlots}
\end{center}
\end{figure}

The \dEdx measurement provides a measurement of the particle $\beta\gamma$ through a calibrated relation between \dEdx and $\beta\gamma$. The calibration is performed in the range $0.3 < \beta\gamma < 5$, using a special low-\pileup dataset in which it is feasible to reconstruct tracks with \pT as low as 100~\MeV\ (see Section~\ref{sec:data_sim}). In a narrow low momentum slice, the contributions to the \dEdx spectrum from individual particle species can be resolved, as shown in Figure~\ref{fig:calib_lowpSlice}. The \dEdx distribution of the tracks in the special dataset, sliced into bins of momentum, $|\eta|$, and $\IBLOF$, is fitted with a superposition of Crystal Ball functions~\cite{Oreglia:1980cs,Skwarnicki:1986xj} modified to have double-sided Gaussian cores to extract the most probable value of \dEdx for pions, kaons, and protons. These values trace out the $\dEdx\mbox{--}\beta\gamma$ relationship, as shown in Figure~\ref{fig:calib_Bethe}. In the ID, charged particles are assumed to be charged pions by default, and the reconstructed momentum is unbiased for charged pions. For other particle species, however, the reconstructed momentum at the low-$\pt$ end below $300~\MeV$ is biased. This momentum bias was derived using the same low-\pt track reconstruction for a minimum-bias Monte Carlo simulation sample and removed by applying a correction. This correction is already included in Figure~\ref{fig:calib_Bethe}. The fitted peaks as a function of $\beta\gamma$ are then fitted with the following empirical function inspired by the original Bethe--Bloch formula
\begin{eqnarray}
\label{eq:mass_calib}
{\mathrm{MPV}}_{\dEdx}(\beta\gamma) = \frac{1+(\beta\gamma)^{2}}{(\beta\gamma)^{2}}\left(c_{0}+c_{1}\log_{10}(\beta\gamma)+c_{2}\left[\log
_{10}(\beta\gamma)\right]^{2}\right)~
\end{eqnarray}
where $c_{0}$, $c_{1}$, and $c_{2}$ are free parameters of the fit. Inversion of the above function provides an estimate of the charged-particle $\beta\gamma$ from the measured $\dEdx$.
Combined with the momentum measurement, the mass of the particle associated with the track can be calculated as $m_{\dEdx} \equiv p_\mathrm{reco}/\beta\gamma(\dEdxcorr)$. This reconstructed mass is hereafter simply denoted by `$m$'.
 
The fraction of $\IBLOF=1$ tracks as a function of $\beta\gamma$, shown in Figure~\ref{fig:calib_IBLOF} for $|\eta|<0.4$, is also monitored in the same low-\pileup dataset. Here, the power of $\IBLOF$ as a key discriminant of the search is clearly illustrated: the fraction of MIP tracks with an IBL overflow is highly suppressed by more than two orders of magnitude, but a substantial fraction of tracks with smaller $\beta\gamma$ have an IBL overflow. The $\dEdx\mbox{--}\beta\gamma$ calibration and the fraction of $\IBLOF=1$ tracks versus track $\beta\gamma$ form the basis of the data-driven template used to replace the simulated \dEdx and $\IBLOF$ values in Monte Carlo samples: when the track associated with the signal particle is reconstructed, first the label $\IBLOF$ is reassigned according to a binomial probability depending on the particle $(\beta\gamma,|\eta|)$, and then the $\dEdxcorr$ value is determined from the probability density distribution template corresponding to the $\IBLOF$ label of the $(\beta\gamma,|\eta|)$ slice.
 
\begin{figure}[tbp]
\begin{center}
\includegraphics[width=0.5\textwidth]{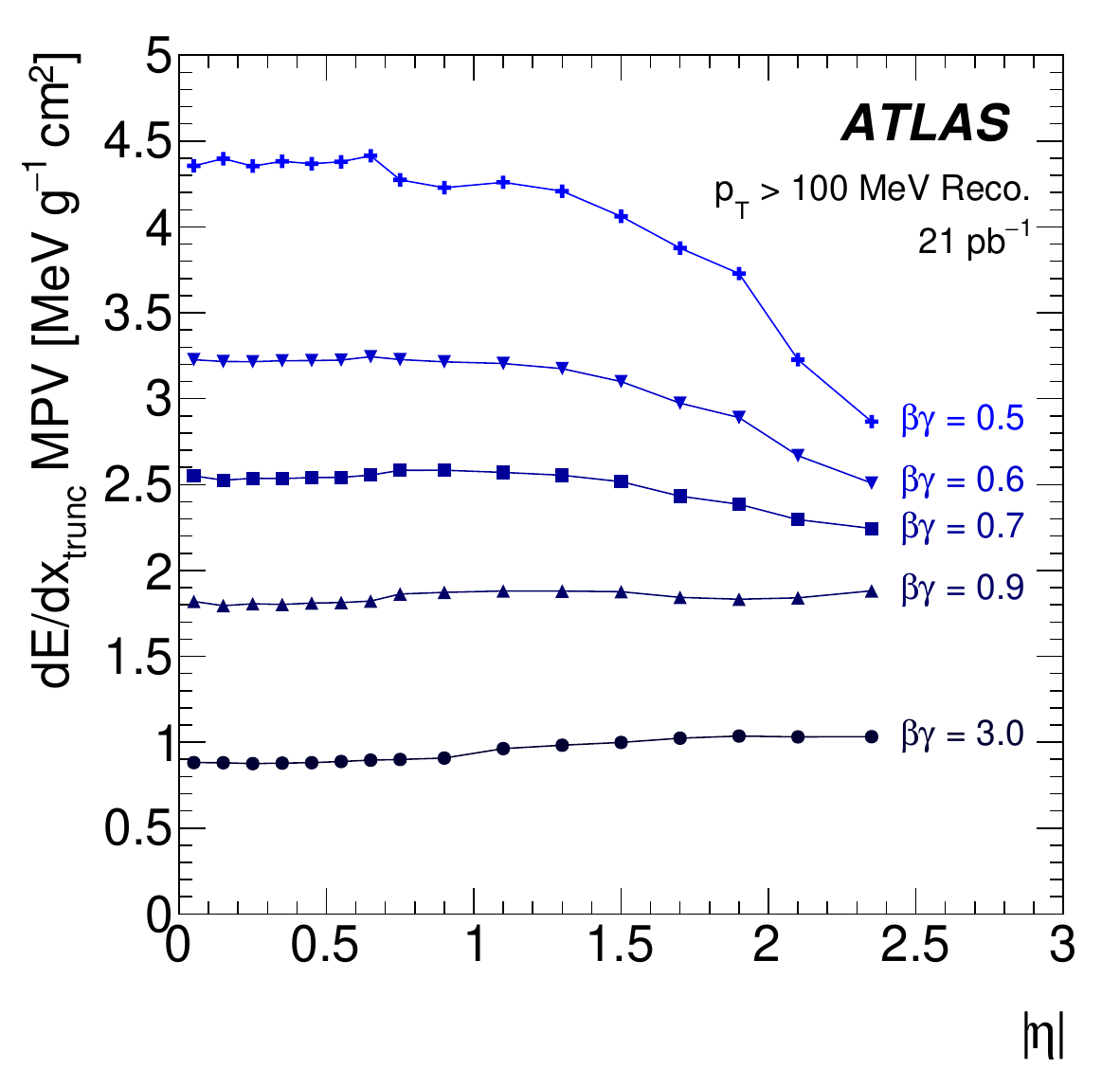}
\caption{$|\eta|$-dependence of the MPV of $\dEdxtrunc$ for various charged-particle $\beta\gamma$ values before MPV equalisation over $|\eta|$ for the $\IBLOF=0$ case. A value of $\beta\gamma=3$ is approximately the MIP case. In the signal region of this search, $\dEdxtrunc$ approximately corresponds to $\beta\gamma<0.9$. A low-\pileup dataset recorded in 2017 ($21~\mathrm{pb^{-1}}$) with $\pt>100~\MeV$ track reconstruction is used to probe the dependence on $\beta\gamma$.}
\label{fig:calib_etaDep}
\end{center}
\end{figure}

It is worth mentioning that the $|\eta|$-dependence of $\dEdxtrunc$ is not uniform over $\beta\gamma$, as illustrated in Figure~\ref{fig:calib_etaDep}. Even after the equalisation of the MPV over $|\eta|$ described above, the $\dEdxtrunc$ response requires different mass calibration functions depending  on  $|\eta|$. This is because not only the MPV but also the shape of the Landau distribution depends on $|\eta|$ (the Landau tails are reduced increasing the traversed silicon thickness).   The ionisation loss response is therefore treated by slicing in $|\eta|$ throughout this analysis.
 
\FloatBarrier


\section{Selection of events, tracks and mass windows}
\label{sec:selection}
 
In this section, criteria applied to events and tracks are described in detail. These selections except the mass windows described in Section~\ref{sec:masswindows} are summarised in Table~\ref{tbl:signalSelections}.

\begin{table}[htbp]
\centering
\caption{Summary of signal selection.}
\vspace{0.2cm}
\footnotesize
\renewcommand{\arraystretch}{1.5}
\begin{tabular}{l|l||ll}
\hline
\hline
\textbf{Category} &\textbf{Item} & \multicolumn{2}{c}{\textbf{Description}}\\
\hline
\hline
Event topology & Trigger & \multicolumn{2}{l}{Unprescaled lowest-threshold \MET trigger}\\
& \MET & \multicolumn{2}{l}{$\MET>170~\GeV$}\\
& Primary vertex & \multicolumn{2}{l}{The hard-scatter vertex must have at least two tracks}\\
\hline
\hline
\multicolumn{4}{l}{Events are required to have at least one track fulfilling \emph{all} criteria listed below; tracks sorted in \pt descending order}\\
\hline
Track kinematics & Momentum & \multicolumn{2}{l}{$\pt>120~\GeV$} \\
& Pseudorapidity & \multicolumn{2}{l}{$|\eta|<1.8$} \\
& $W^{\pm}\rightarrow\ell^{\pm}\nu$ veto & \multicolumn{2}{l}{$m_{\mathrm{T}}(\text{track},\vec{p}_{\mathrm{T}}^{~\text{miss}})>130~\GeV$}\\
&\\
Track quality & Impact parameters & \multicolumn{2}{l}{Track matched to the hard-scatter vertex; $|d_{0}|<2~\text{mm}$ and $|\Delta z_{0}\sin\theta|<3~\text{mm}$} \\
& Rel.\ momentum resolution & \multicolumn{2}{l}{$\sigma_{p}< \mathrm{max}\biggl(10\%, -1\%+90\%\times{\displaystyle\frac{|p|}{\TeV}}\biggr)$ and $\sigma_{p}<200\%$}\\
& Cluster requirement (1) & \multicolumn{2}{l}{At least two clusters used for the \dEdxtrunc calculation}\\
& Cluster requirement (2) & \multicolumn{2}{l}{Must have a cluster in the IBL (if this is expected), or}\\
& & \multicolumn{2}{l}{a cluster in the next-to-innermost pixel layer }\\
& & \multicolumn{2}{l}{(if this is expected while a cluster is not expected in IBL)}\\
& Cluster requirement (3) & \multicolumn{2}{l}{No shared pixel clusters and no split pixel clusters}\\
& Cluster requirement (4) & \multicolumn{2}{l}{$\mbox{Number of SCT clusters} > 5$}\\
&\\
Vetoes & Isolation & \multicolumn{2}{l}{$\biggl(\sum_\mathrm{trk} \pt\biggr) < 5~\GeV$  (cone size $\Delta R=0.3$)}\\
& Electron veto & \multicolumn{2}{l}{EM fraction~$<0.95$}\\
& Hadron and $\tau$-lepton veto & \multicolumn{2}{l}{$E_{\mathrm{jet}}/p_{\mathrm{track}}<1$}\\
\cdashline{2-4}
& Muon requirement & \multicolumn{2}{l}{\texttt{SR-Mu}: MS track matched to ID track;~~~\texttt{SR-Trk}: otherwise} \\
\hline
Pixel \dEdx
& \multirow{2}{*}{\texttt{Inclusive}} & \texttt{Low}:  &$\dEdx \in [1.8,2.4]~\dEdxunit$\\
&                                     & \texttt{High}: &$\dEdx > 2.4~\dEdxunit$\\
\cline{2-4}
&\multirow{3}{*}{Binned}  & \texttt{IBL0\_Low}:  &$\dEdx \in [1.8,2.4]~\dEdxunit$ and $\IBLOF=0$\\
&                                     & \texttt{IBL0\_High}: &$\dEdx > 2.4~\dEdxunit$ and $\IBLOF=0$\\
&                                     & \texttt{IBL1}:       &$\dEdx > 1.8~\dEdxunit$ and $\IBLOF=1$\\
\hline
\hline
\end{tabular}
\label{tbl:signalSelections}
\end{table}

\subsection{Event selection}
Events in the signal region are first selected with a trigger based on \MET, which is calculated using energy measurements in the calorimeter with corrections for multiple $pp$ interactions in each event~\cite{TRIG-2016-01}. The high-level \MET trigger threshold varies from 70~\GeV\ to 110~\GeV\ during the data-taking period. In the offline reconstruction of the recorded events, \MET is built from calibrated muons and electrons which pass baseline selections, and from calibrated jets reconstructed with the anti-$k_t$ jet clustering algorithm~\cite{antikt,Fastjet} with radius parameter $R=0.4$, using clusters of energy depositions in the calorimeter as inputs. A term that includes soft tracks not associated with any other objects in the event~\cite{PERF-2014-04} but consistent with the hard-scatter primary vertex (PV) (i.e.\ the vertex with the largest $\sum \pt^{2}$ for the associated tracks) is added to the \MET calculation. Events are required to have  $\MET~>~170$~\GeV\ to enhance the signal sensitivity and to ensure that the selected events are near the efficiency plateau of the trigger. To ensure a reliable  calculation of \MET, events are rejected if they contain a jet with $E_{\text{T}} > 20$~\GeV\ that is consistent with detector noise or beam-induced background, as determined from shower shape information. Unlike in standard ATLAS selections for jet-cleaning~\cite{ATLAS-CONF-2015-029}, a requirement on the relationship between track and calorimeter measurements of $\pt$ and a requirement on the fraction of jet energy deposited in the electromagnetic calorimeter are not applied, since they are found to be inefficient for signal events in which an LLP decays before or inside the calorimeters.

Events that pass the trigger and \MET selections are required to have a PV built from at least two reconstructed tracks each with $\pt$ above 500~\MeV, and must contain at least one candidate track that passes the track-level selections detailed below.
 
\subsection{Track selection}
 
All the track parameters are derived using only ID information, including the TRT hits. This choice is justified by the desire to remain agnostic about the decay products of the LLPs. When combining an ID track and muon spectrometer track segments, it is assumed that the combined track has $\beta$ = 1. Therefore, the addition of track information from the muon spectrometer would not improve the mass resolution of selected low-$\beta\gamma$ LLPs and would induce $\eta$-dependent effects related to the particle's time of flight.
 
In order to enrich the selected sample in potential signal events, candidate tracks are required to have $\pt > 120~\GeV$ and $|\eta| < 1.8$. To reject non-prompt background tracks and those inconsistent with the PV, the transverse impact parameter\footnote{The transverse impact parameter ($d_{0}$) is defined as the distance of closest approach in the transverse plane between a track and the beam line. The longitudinal impact parameter ($z_{0}$) corresponds to the $z$-coordinate distance between the point along the track at which the transverse impact parameter is defined and the primary vertex.} of candidate tracks, $|d_0|$, must be less than 2~mm, and the absolute value of the product of the longitudinal impact parameter relative to the $z$-position of the PV, $\Delta z_0$, and $\sin\theta$ of the track, must satisfy $|\Delta z_{0}\sin\theta|<3~\mathrm{mm}$. Reconstructed tracks must have at least six clusters across the SCT detectors,\footnote{In the SCT, clusters on each side of a double-sided strip module are individually counted as clusters; typically, two clusters are produced along a track in a module.} and to be considered a candidate, the track must have an associated cluster in the innermost active pixel detector module within the first two layers. Tracks are rejected if any pixel cluster is shared by two or more tracks (shared cluster), or if the shape of a pixel cluster is judged to be likely to arise from energy deposited by multiple particles (split cluster) by a neural-network algorithm applied to the pixel pattern of the cluster~\cite{PERF-2015-08,PERF-2012-05}. To enhance the selection of isolated tracks, the scalar sum of the $\pt$ of other tracks, with $\pt > 1~\GeV$ and consistent with the PV, in a cone of size $\Delta R = 0.3$ around the candidate track must be less than 5~\GeV.
 
To reject tracks from leptonic $W$ decays, the transverse mass, $m_\text{T}$,\footnote{$m_\text{T} \equiv \sqrt{2 \pt^{~\text{track}} \MET \left( 1-\cos\Delta\phi \right)}$, where $\Delta\phi$ is the azimuthal separation between the track and $\vec{p}_{\text{T}}^{~\text{miss}}$.} associated with the candidate track must be greater than 130~\GeV. Tracks from electrons are removed by considering any jets with $\pt > 20~\GeV$ that are within $\Delta R(\text{track},\text{jet}) = 0.05$ of the candidate track,\footnote{The angular separation of two objects $(i,j)$ is defined as $\Delta R(i,j)\equiv \sqrt{(\Delta\eta_{ij})^{2}+(\Delta\phi_{ij})^{2}}$.} and rejecting the track if any such jet deposits at least $95\%$ of its energy in the electromagnetic calorimeter. SM hadrons are removed by excluding tracks for which a nearby jet satisfying $\Delta R(\text{track},\text{jet}) < 0.05$ has a calibrated energy larger than the track momentum.
 
At least two pixel clusters, after discarding the cluster (or the two clusters) with the highest ionisation, must be included in the calculation of \dEdx to ensure it is robust.
 
The relative uncertainty in the momentum measurement depends linearly on the momentum, and the uncertainty upper bound should lie between 10\% and 200\%, and must satisfy
\begin{eqnarray*}
\sigma_{p} < \mathrm{max}\left(10\%, \left(-1\% + 90\%\times\frac{|p|}{\TeV}\right)\right)~\mbox{and}~\sigma_{p}<200\%
\end{eqnarray*}
where $\sigma_{p} \equiv \left|\delta(p^{-1})/(p^{-1})\right|$ is the relative uncertainty of the inverse of the charge-signed track momentum. This selection was chosen so as to maximise the statistical significance of the signal over the full lifetime and mass range of the LLPs under study.

The $\dEdx$ of the candidate track must be larger than $1.8~\dEdxunit$, which corresponds to a selection of approximately $1\%$ of the whole track set in data. This threshold value was used in all previous ATLAS searches based on $\dEdx$ measurement in the pixel detector~\cite{SUSY-2011-03, SUSY-2014-09, SUSY-2016-03, SUSY-2016-31} and is related to the data-driven background generation method. The background extrapolation from a `below-threshold' sample to the `above-threshold' behaviour is driven by the sample size. The larger the available sample, the higher the threshold setting can be. The larger integrated luminosity available for this measurement allows a higher threshold setting. The threshold setting was optimised by maximising the statistical significance of the signal, and a specific-ionisation threshold of  $2.4~\dEdxunit$ was found to be the optimal choice (approximately $0.15\%$ of the whole track set survives this cut). The interval $\dEdx \in [1.8,2.4]~\dEdxunit$ below this threshold contains a sizeable fraction of the signal, approximately $25\%\mbox{--}40\%$ for particles with a hypothetical mass of $2.2\mbox{--}0.4~\TeV$ respectively, but with less favourable statistical significance.
 
Two mutually exclusive intervals are considered: `\texttt{Low}' for $\dEdx \in [1.8,2.4]~\dEdxunit$ and `\texttt{High}' for $\dEdx>2.4~\dEdxunit$. The latter has the higher sensitivity, but the former also contributes to the sensitivity, especially for smaller LLP masses.
 
\begin{figure}[t]
\begin{center}
\subfigure[$\gluino, m(\gluino)=2.2~\TeV, m(\ninoone)=100~\GeV,\tau(\gluino)=10~\mathrm{ns}$]{
\label{fig:mcReco_dEdxP_Gluino}
\includegraphics[width=0.47\textwidth]{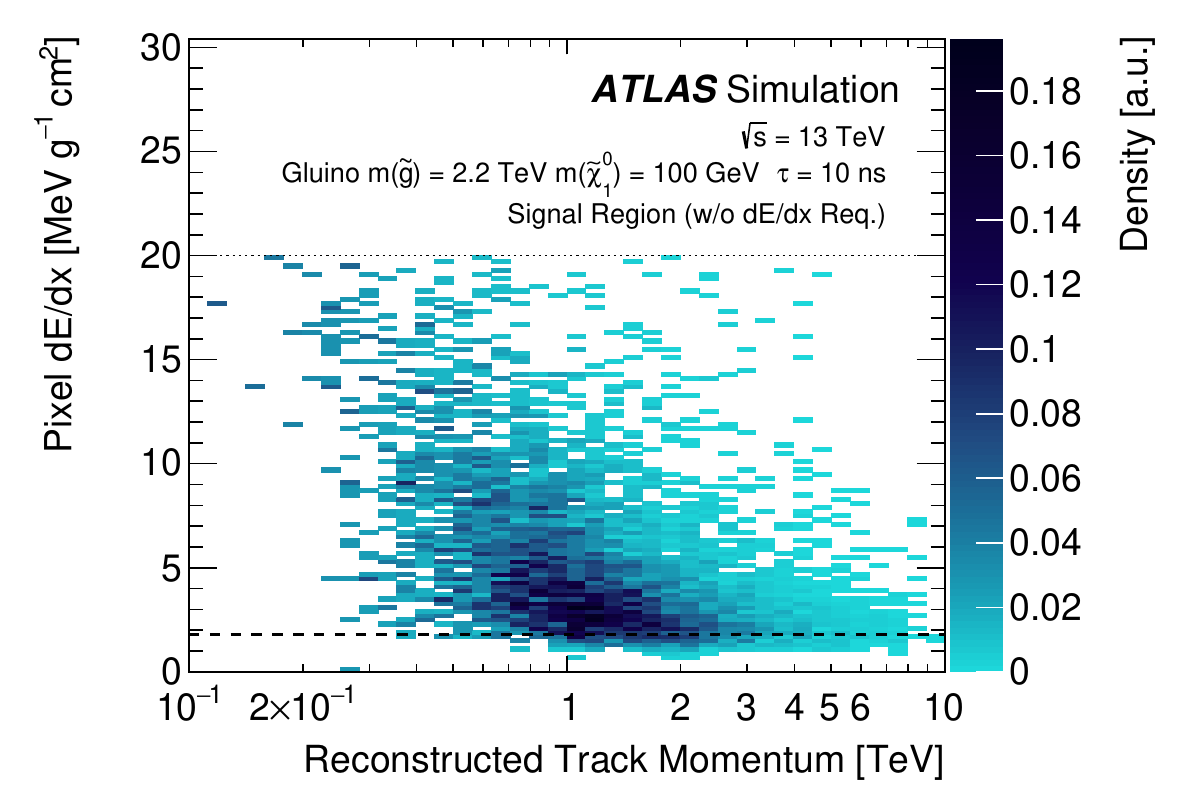}
}
\subfigure[$\chinoonepm, m(\chinoonepm)=1.3~\TeV, \tau(\chinoonepm)=10~\mathrm{ns}$]{
\label{fig:mcReco_dEdxP_Chargino}
\includegraphics[width=0.47\textwidth]{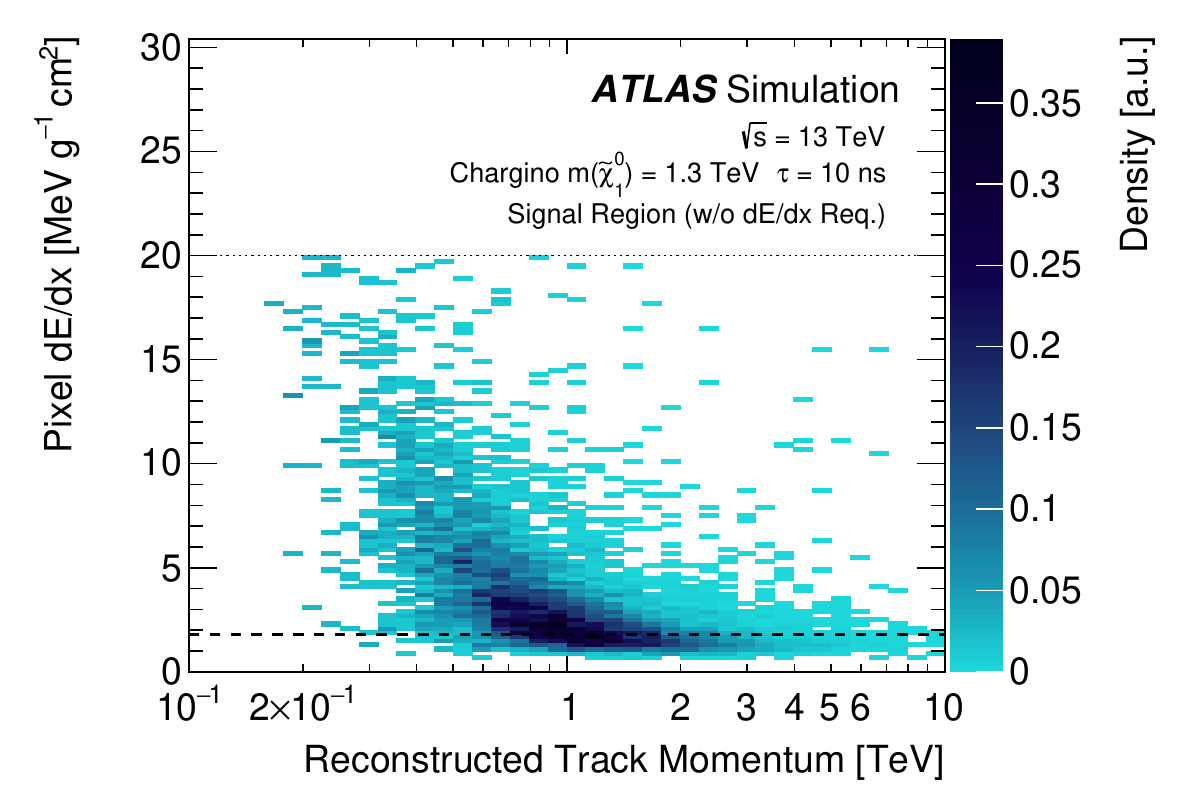}
}
\subfigure[$\stau, m(\stau)=400~\GeV, \tau(\stau)=10~\mathrm{ns}$]{
\label{fig:mcReco_dEdxP_Stau}
\includegraphics[width=0.47\textwidth]{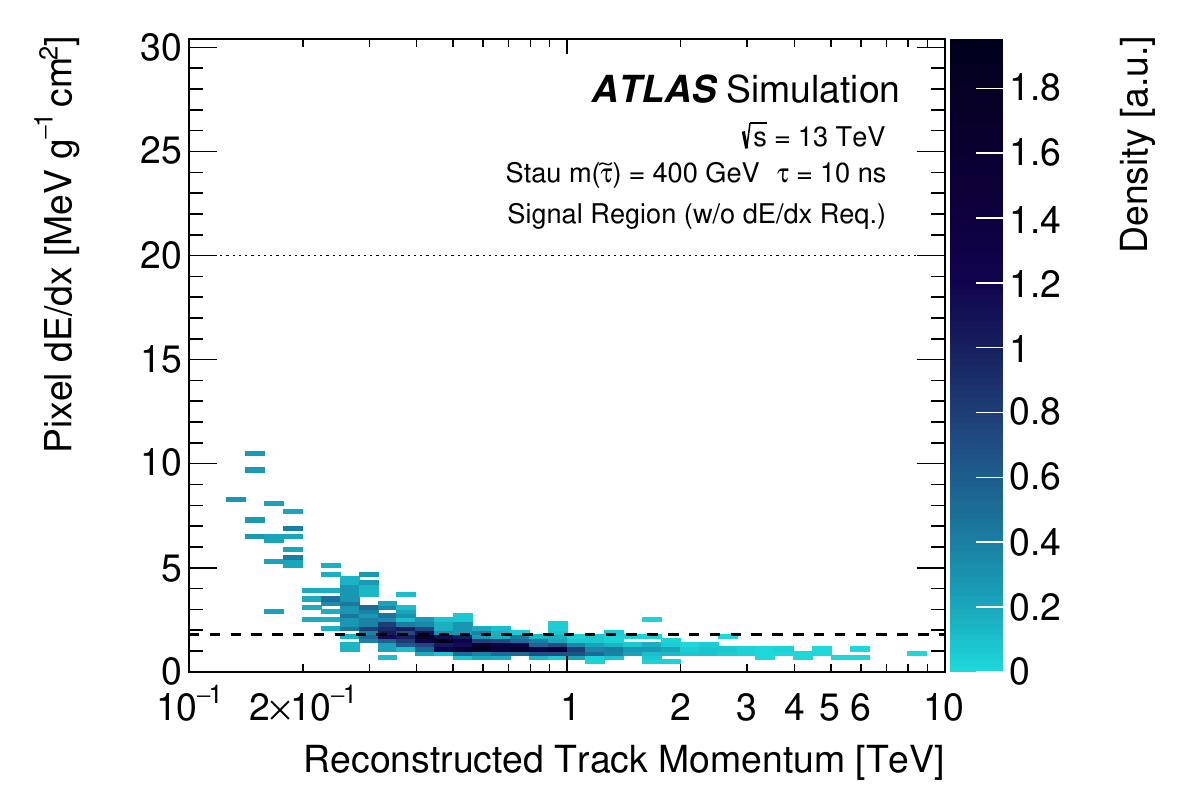}
}
\caption{Simulated density distribution of representative signal tracks in the $p$--$\dEdx$ plane after event selection except the $\dEdx$ requirement. The region above the dashed horizontal line at $\dEdx = 1.8~\dEdxunit$ corresponds to the signal region. No events are expected above $\dEdx = 20~\dEdxunit$ because of the limited dynamic range of the pixel electronics. The data-driven \dEdx template technique described in Section~\ref{sec:correct_calib} is used in the simulation of the \dEdx variable for these plots.}
\label{fig:mcReco}
\end{center}
\end{figure}

Figure~\ref{fig:mcReco} illustrates the distribution of representative signal samples in the reconstructed $p$--$\dEdx$ plane after the selection. While the Bethe--Bloch relation is clearly retained, a substantial amount of smearing is visible for higher LLP masses, reflecting the limited ID momentum resolution.

\subsection{Event subsamples}

\begin{figure}[t]
\begin{center}
\subfigure[$\tau=3~\mathrm{ns}$]{
\label{gluino_bar_3}
\includegraphics[width=0.31\textwidth]{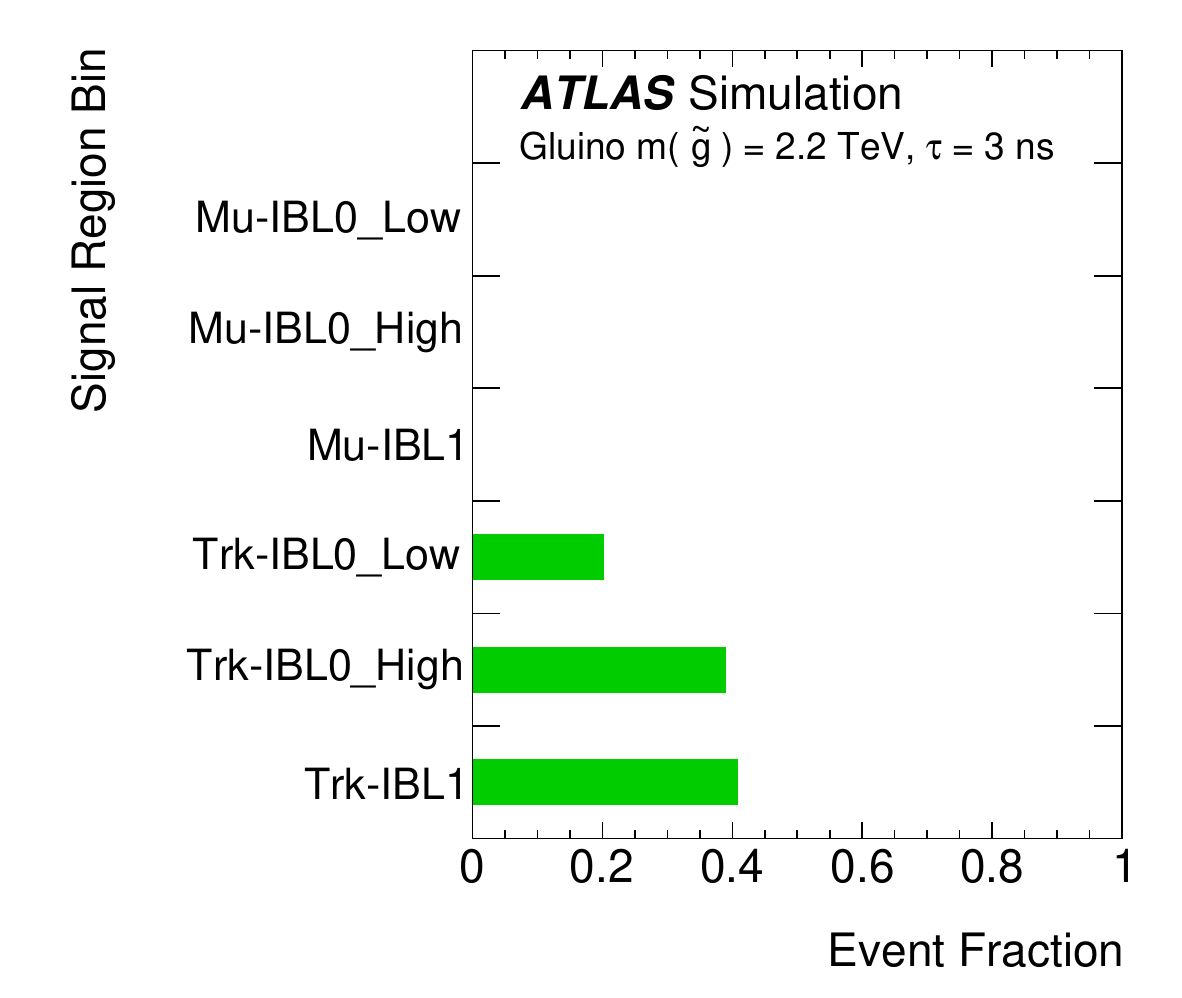}
}
\subfigure[$\tau=30~\mathrm{ns}$]{
\label{fig:gluino_bar_30}
\includegraphics[width=0.31\textwidth]{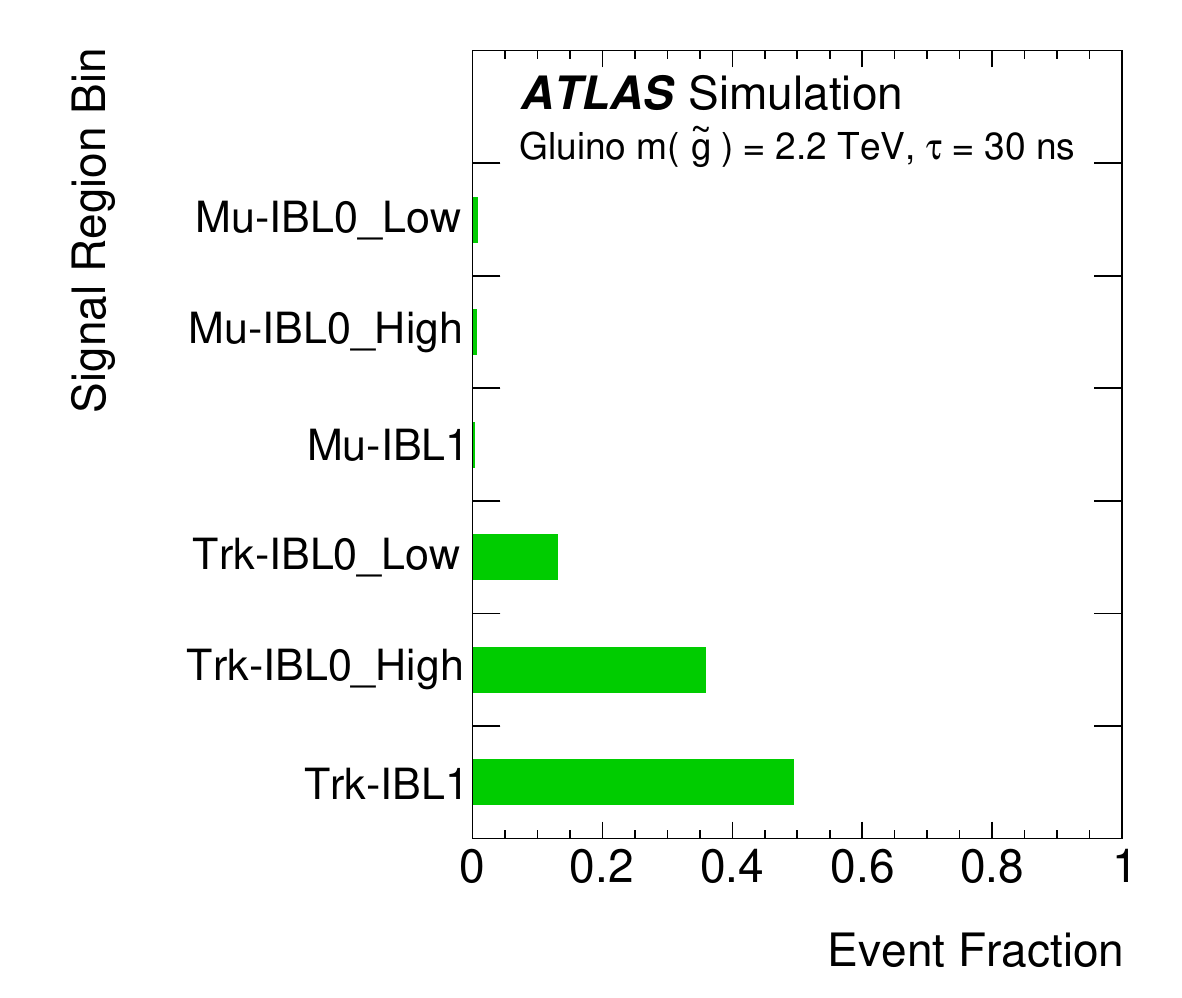}
}
\subfigure[stable]{
\label{fig:gluino_bar_3000}
\includegraphics[width=0.31\textwidth]{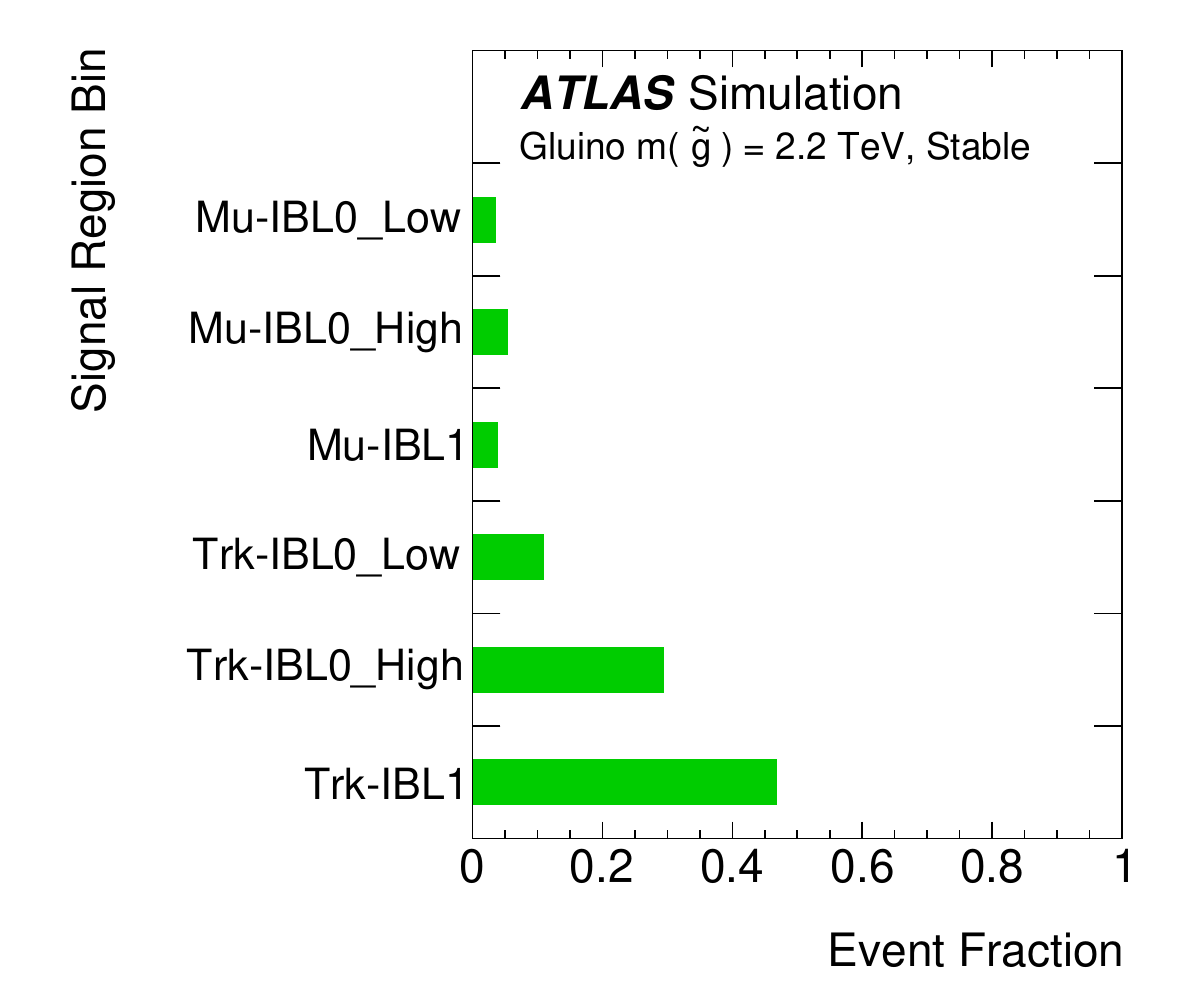}
}
\caption{Relative fraction of events in the signal-region bins for gluino scenarios with $m(\gluino) = 2.2~\TeV, m(\ninoone) = 100~\GeV$ and different lifetime values. The fractions of events in muon-identified bins increase with the lifetime. The total probability to pass the event selection is 9.5\% for $\tau=3~\mathrm{ns}$, 20.7\% for $\tau=30~\mathrm{ns}$ and 9.3\% if the gluino is stable.}
\label{fig:gluino_SRfraction}
\end{center}
 
\begin{center}
\subfigure[$\tau=4~\mathrm{ns}$]{
\label{chargino_bar_4}
\includegraphics[width=0.31\textwidth]{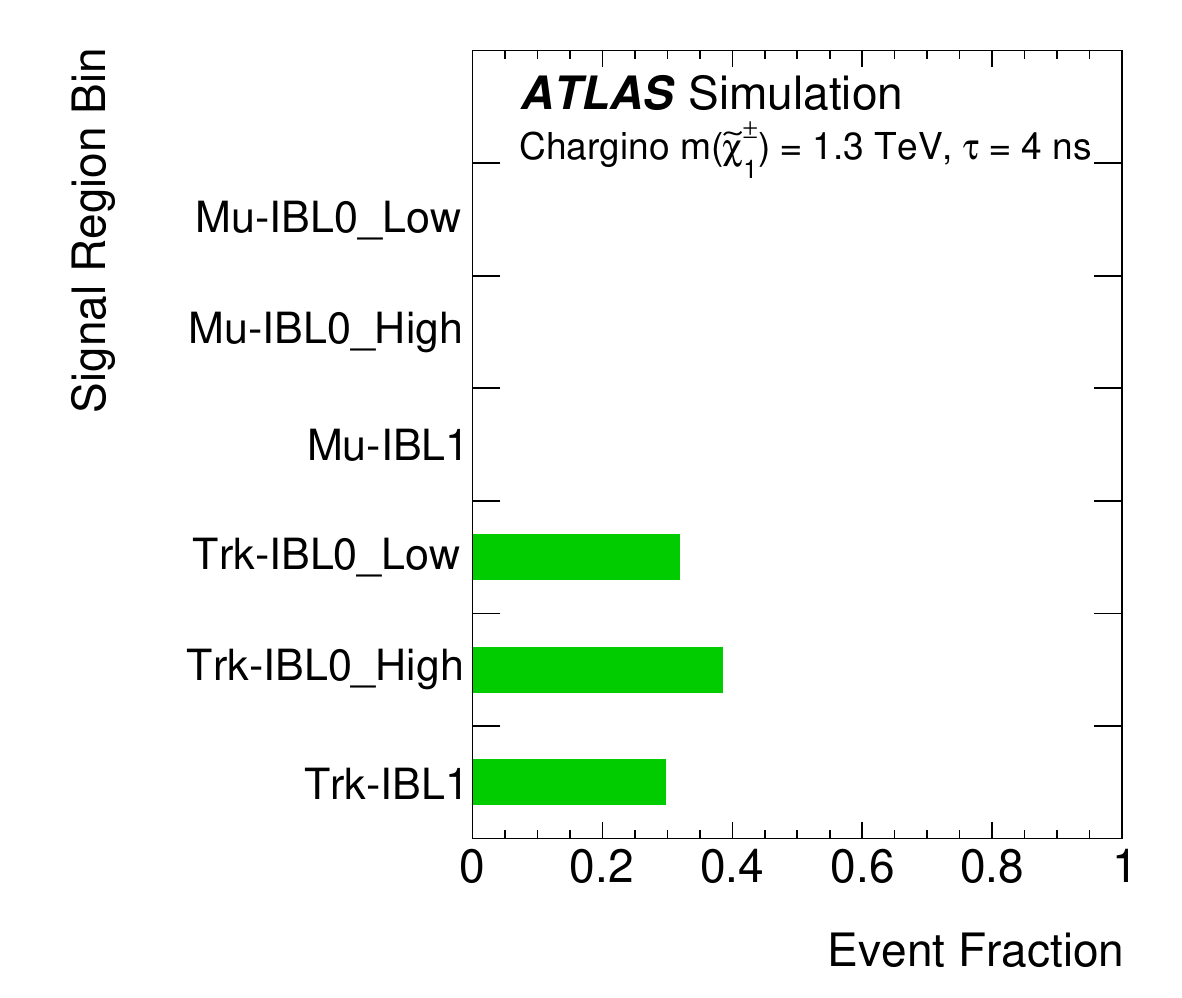}
}
\subfigure[$\tau=30~\mathrm{ns}$]{
\label{fig:chargino_bar_30}
\includegraphics[width=0.31\textwidth]{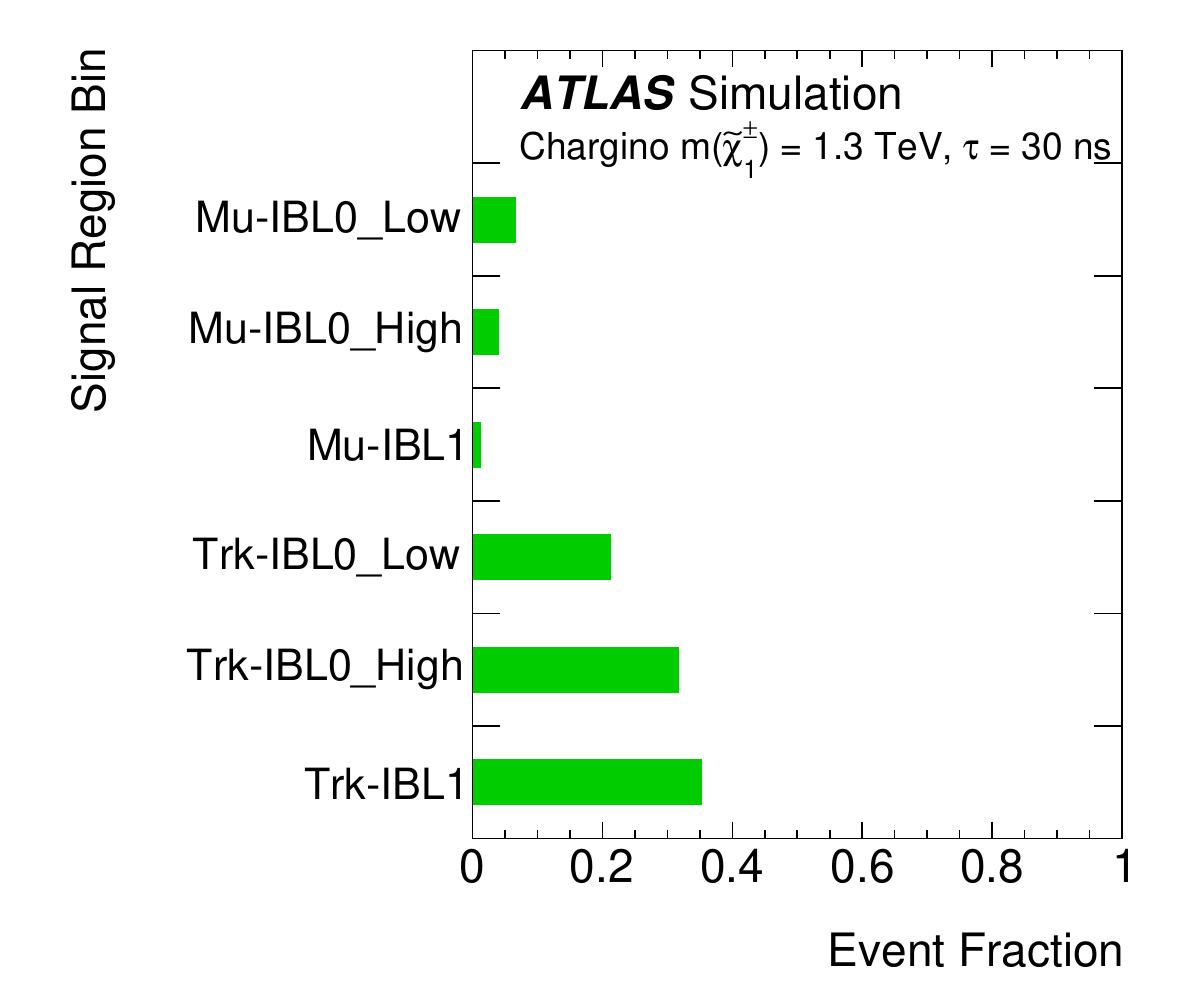}
}
\subfigure[stable]{
\label{fig:chargino_bar_3000}
\includegraphics[width=0.31\textwidth]{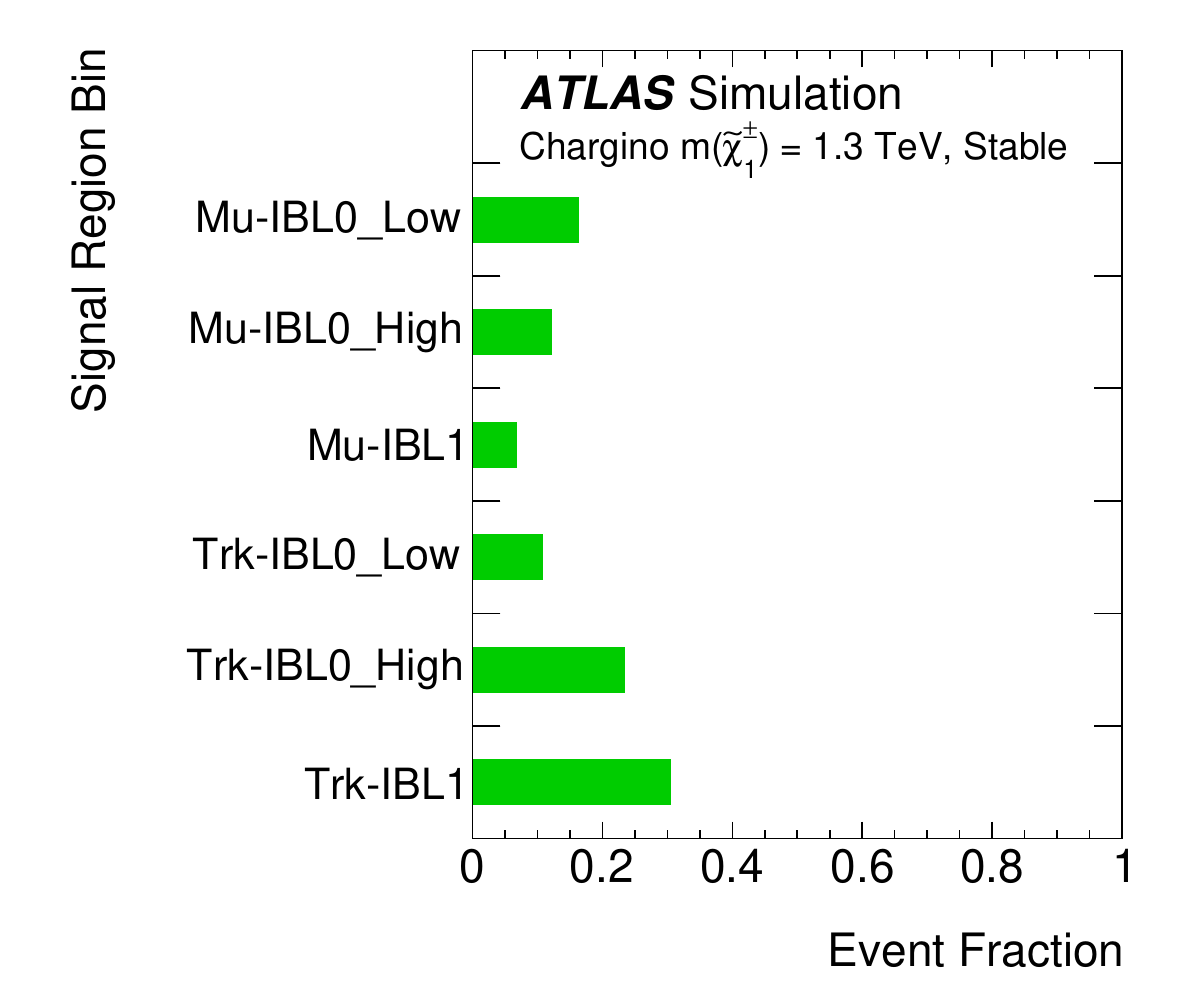}
}
\caption{Relative fraction of events in the signal-region bins for chargino scenarios with $m(\chinoonepm) = 1.3~\TeV$ and different lifetime values. The fractions of events in muon-identified bins increase with the lifetime. The total probability to pass the event selection is 7.3\% for $\tau=4~\mathrm{ns}$, 15.3\% for $\tau=30~\mathrm{ns}$ and 8.4\% if the chargino is stable.}
\label{fig:chargino_SRfraction}
\end{center}
 
\begin{center}
\subfigure[$\tau=3~\mathrm{ns}$]{
\label{stau_bar_3}
\includegraphics[width=0.31\textwidth]{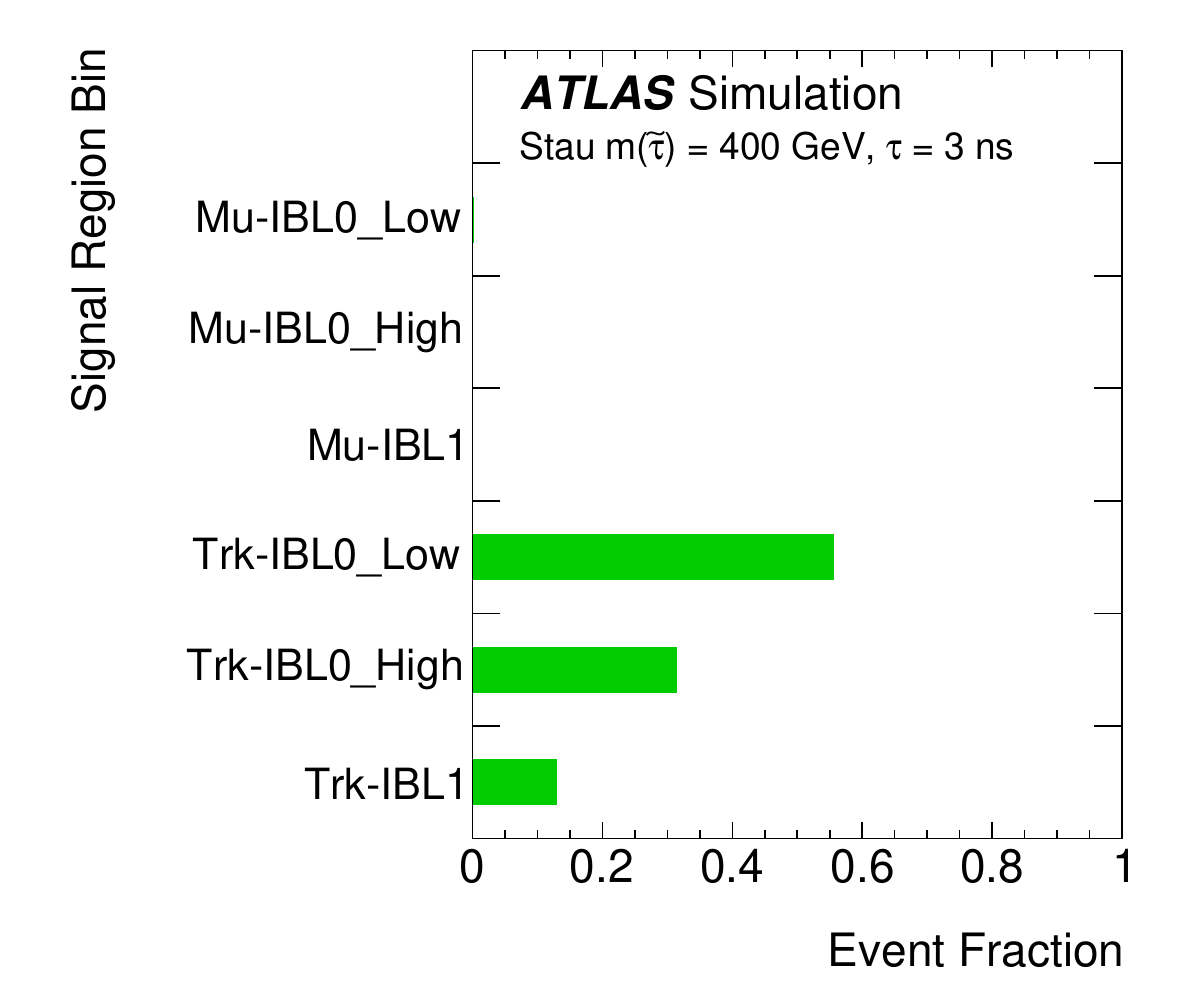}
}
\subfigure[$\tau=30~\mathrm{ns}$]{
\label{fig:stau_bar_30}
\includegraphics[width=0.31\textwidth]{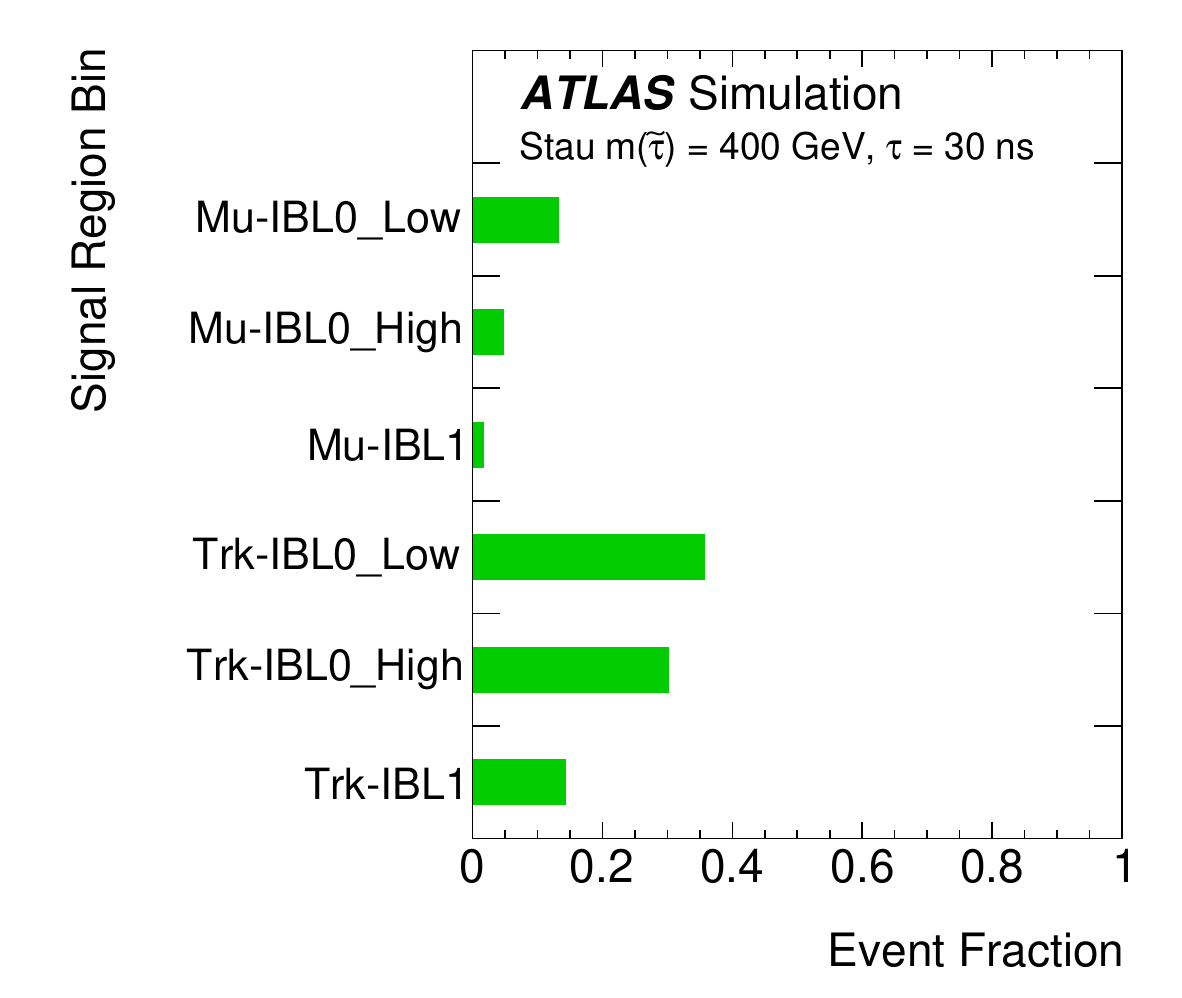}
}
\subfigure[stable]{
\label{fig:stau_bar_3000}
\includegraphics[width=0.31\textwidth]{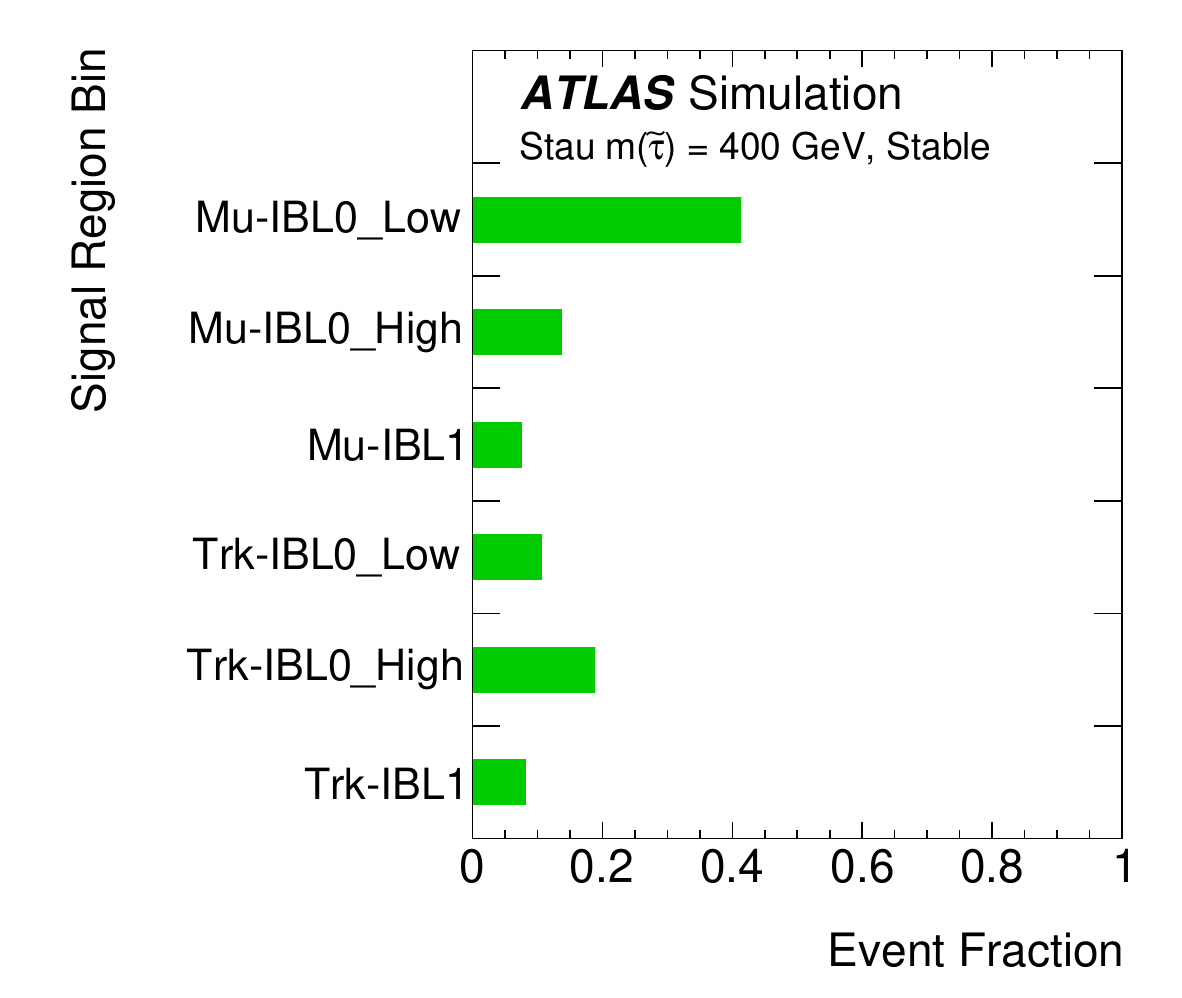}
}
\caption{Relative fraction of events in the signal-region bins for stau scenarios with $m(\stau) = 400~\GeV$ and different lifetime values. The fractions of events in muon-identified bins increase with the lifetime. The total probability to pass the event selection is 3.4\% for $\tau=3~\mathrm{ns}$, 4.0\% for $\tau=30~\mathrm{ns}$ and 1.4\% if the stau is stable.}
\label{fig:stau_SRfraction}
\end{center}
\end{figure}

A candidate track may be matched to a `combined muon' object when a track reconstructed in the MS is consistent with the candidate track when using `combined fitting' in the standard muon reconstruction algorithm~\cite{MUON-2018-03,PERF-2015-10}. Depending on muon identification by ID--MS track matching using the `medium' criteria~\cite{PERF-2015-10}, candidate tracks are categorised as either `muon tracks' (\texttt{SR-Mu}) or `tracks' (\texttt{SR-Trk}). This classification is useful because, as is described later, the \texttt{SR-Mu} category contains the majority of background tracks but only a limited fraction of the signal tracks, especially when the LLP's average decay length $\beta\gamma c\tau$ is insufficient to traverse the MS. Indeed, even for the stable lifetime case, a substantial fraction of the signal tracks remain in the \texttt{SR-Trk} category, as can be seen in Figures~\ref{fig:gluino_SRfraction}--\ref{fig:stau_SRfraction}. This is not only due to the intrinsic muon identification inefficiency, but also due to the fact that the standard ATLAS muon reconstruction algorithm assumes $\beta=1$ muons, and a substantial fraction of the signal tracks arrive at the MS much later than SM muons. For $R$-hadron signal samples, there is additional complexity because $R$-hadrons have strong interactions with the calorimeter material, and generally the probability of muon identification is expected to be even smaller.
For consistency, especially with the background estimation, the $\pt$ of the track always refers to that of ID tracks, even for muon-identified candidates.
 
The presence of an IBL overflow cluster on a track can provide a useful way to discriminate signal from background. The data sample can then be split into two independent subsets according to the presence (\texttt{IBL1}) or absence (\texttt{IBL0}) of an IBL overflow cluster on the candidate track. These subsets are treated independently and their sensitivity is combined.

\begin{table}[tbp]
\caption{Table of signal-region bins, showing their purpose (discovery or limit setting) and properties.}
\vspace{0.2cm}
\centering
\small
\renewcommand{\arraystretch}{1.3}
\begin{tabular}{l|ccccc}
\hline
\hline
SR name & Discovery & Limit setting & Track category & IBL overflow & $\dEdx~[\dEdxunit]$\\
\hline
\texttt{SR-Inclusive\_Low} & \checkmark & & \multirow{2}{*}{inclusive} &  \multirow{2}{*}{yes or no} & $[1.8,2.4]$\\
\texttt{SR-Inclusive\_High} & \checkmark & & & & $>2.4$\\
\hline
\texttt{SR-Trk-IBL0\_Low} & & \checkmark & \multirow{3}{*}{track} &  no & $[1.8,2.4]$\\
\texttt{SR-Trk-IBL0\_High} & & \checkmark & &  no & $>2.4$\\
\texttt{SR-Trk-IBL1} & & \checkmark & &  yes & $>1.8$\\
\hline
\texttt{SR-Mu-IBL0\_Low} & & \checkmark & \multirow{3}{*}{muon tracks} &  no & $[1.8,2.4]$\\
\texttt{SR-Mu-IBL0\_High} & & \checkmark & &  no & $>2.4$\\
\texttt{SR-Mu-IBL1} & & \checkmark & &  yes & $>1.8$\\
\hline
\hline
\end{tabular}
\label{tbl:SR_bins}
\end{table}
 
The analysis was therefore built to treat all the subsamples (IBL overflow yes/no, track identified as a muon yes/no) independently as well as all combined in a single sample, which is defined as the `\texttt{Inclusive}' sample. All samples are split into \texttt{Low} and \texttt{High} \dEdx intervals, with the exception of the \texttt{IBL1} events, where the tracks must satisfy the condition $\dEdx >1.8~\dEdxunit$. This choice is justified by the small number of \texttt{IBL1} tracks and by the better statistical significance expected in this sample.
 
In summary, the search is performed by splitting the data sample as illustrated in Table~\ref{tbl:SR_bins}. If multiple tracks in an event are found in a single signal-region bin after all signal selections listed in Table~\ref{tbl:signalSelections}, the track with the highest $\pt$ is selected. However, more than one track can be selected from the same event if they enter different signal-region bins. The probability of this occurrence is negligibly small for the background events, while for the signal models, depending on the mass and lifetime parameters, the probability is as large as 10\% when considering all events in the signal region. Figures~\ref{fig:gluino_SRfraction}--\ref{fig:stau_SRfraction} show how signal events are partitioned by this binning, depending on the LLP masses and lifetimes.

The two inclusive signal regions are less dependent on specific theory models and are therefore better suited to the search for an excess in the mass distribution. The six non-inclusive signal regions can be compared in detail to the LLP models considered in this analysis and are then used to derive limits directly related to these models.

\FloatBarrier
 
\subsection{Mass windows}
\label{sec:masswindows}
 
\begin{figure}[t]
\begin{center}
\subfigure[Short-lifetime regime]{
\label{fig:massWindows_short}
\includegraphics[width=0.47\textwidth]{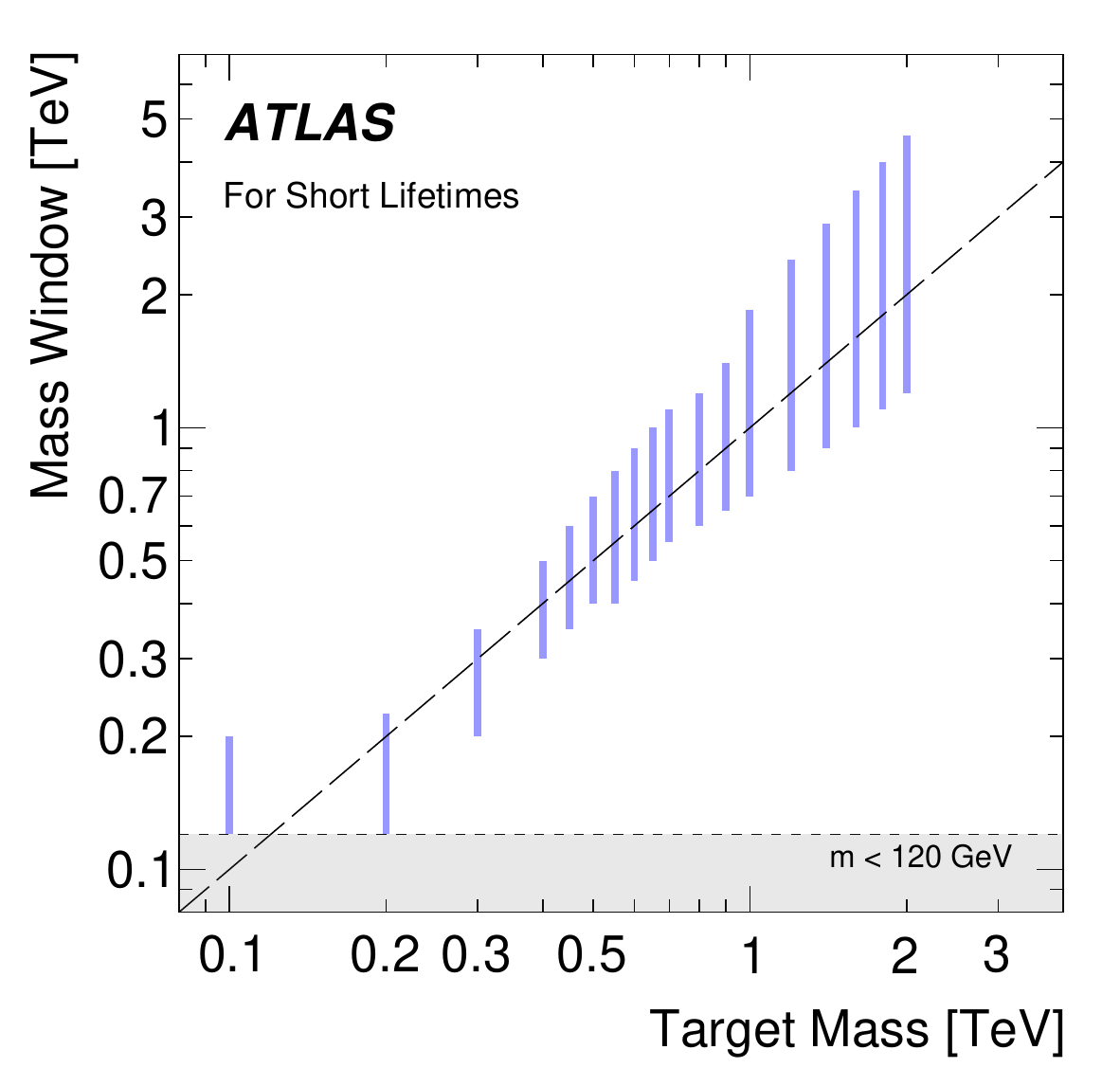}
}
\subfigure[Long-lifetime regime]{
\label{fig:massWindows_long}
\includegraphics[width=0.47\textwidth]{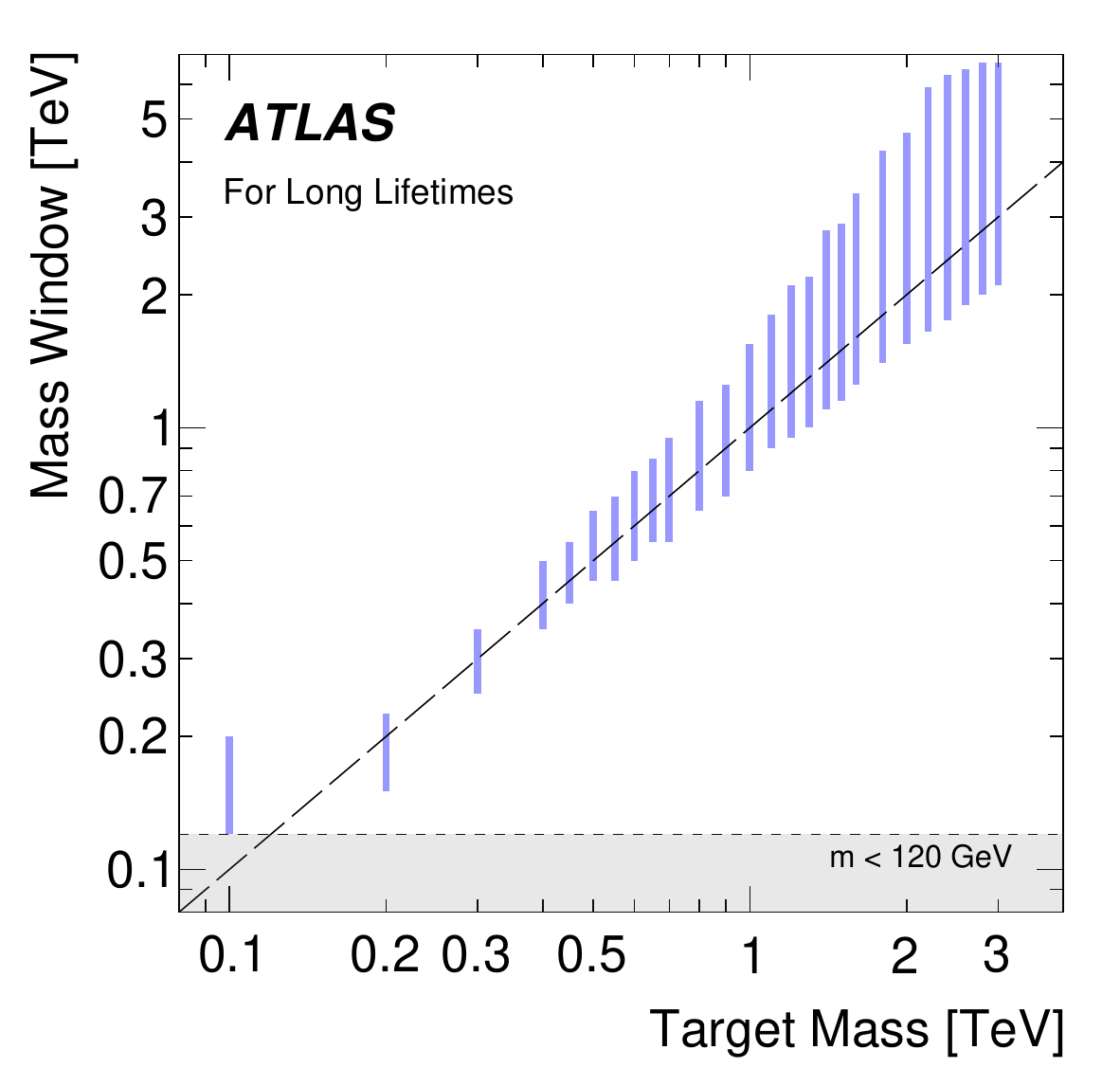}
}
\caption{The mass windows for various LLP masses for \subref{fig:massWindows_short} short lifetimes (i.e.\ $\le1$~ns) and \subref{fig:massWindows_long} long lifetimes. The range $m<120~\GeV$ is not considered for mass-window boundaries due to the kinematical event selection.}
\label{fig:massWindows}
\end{center}
\end{figure}

In the signal region, the expected background has a broad mass distribution which falls monotonically for $m\gtrsim 120~\GeV$ (see Section~\ref{sec:background}), while the signal forms a peak around its nominal mass. The signal mass resolution mainly reflects the momentum resolution of the tracks; the relative width of the core distribution is ${\cal O}(30\%)\times (m/\TeV)$, but it varies with the LLP's decay length and therefore its lifetime. Therefore, a mass window is defined for each target mass, and for both the `short' and `long' lifetime regimes, targeting $\tau \leq 1~\text{ns}$ and $\tau > 1~\text{ns}$, respectively. While the cross-sections span multiple orders of magnitude in the benchmarking models described in Section~\ref{sec:data_sim}, the mass-spectrum shapes of these models are similar for a common target mass, reflecting the common kinematic nature of the $\beta\gamma$ spectrum of pair-produced particles. Therefore, the mass window for a given LLP mass is defined by taking the average of the normalised mass distributions of these models and finding the mass range that captures approximately 70\% of the signal events in the signal region while excluding as much background as possible. Reflecting the steeply falling background distribution and the asymmetric long-tail mass distribution of the signal, especially at higher target masses, the window is asymmetric with respect to the target mass value and extends more to higher reconstructed masses. The lower boundary of the mass window is determined by maximising $1/\sqrt{B}$, where $B$ is the expected background yield. Values of the window boundaries are tested in steps of $10~\GeV~(m\le300~\GeV)$ or $50~\GeV~(m>300~\GeV)$. The range $m<120~\GeV$ is not used for the window definition owing to signal selection cut-offs. The window is common to all signal-region bins (see Table~\ref{tbl:SR_bins}) at a given target mass; minor acceptance differences for the various LLPs and the various lifetimes are taken care of in the sensitivity calculations.
The windows cover the full range of target masses from $100~\GeV$ to $3~\TeV$ as shown in Figure~\ref{fig:massWindows}. The `short'-lifetime regime has wider windows than the `long'-lifetime regime, reflecting poorer momentum resolution for shorter tracks. For consecutive target masses, the corresponding mass windows can largely overlap, and events common to such windows can be counted in each. The outcome of this search depends on the comparison of the data and background yields in each predefined mass window.
 
\section{Background estimation}
\label{sec:background}
 
\begin{figure}[tbp]
\begin{center}
\subfigure[Kinematic control regions]{
\label{fig:CRkin_dist_inclusive}
\includegraphics[width=0.47\textwidth]{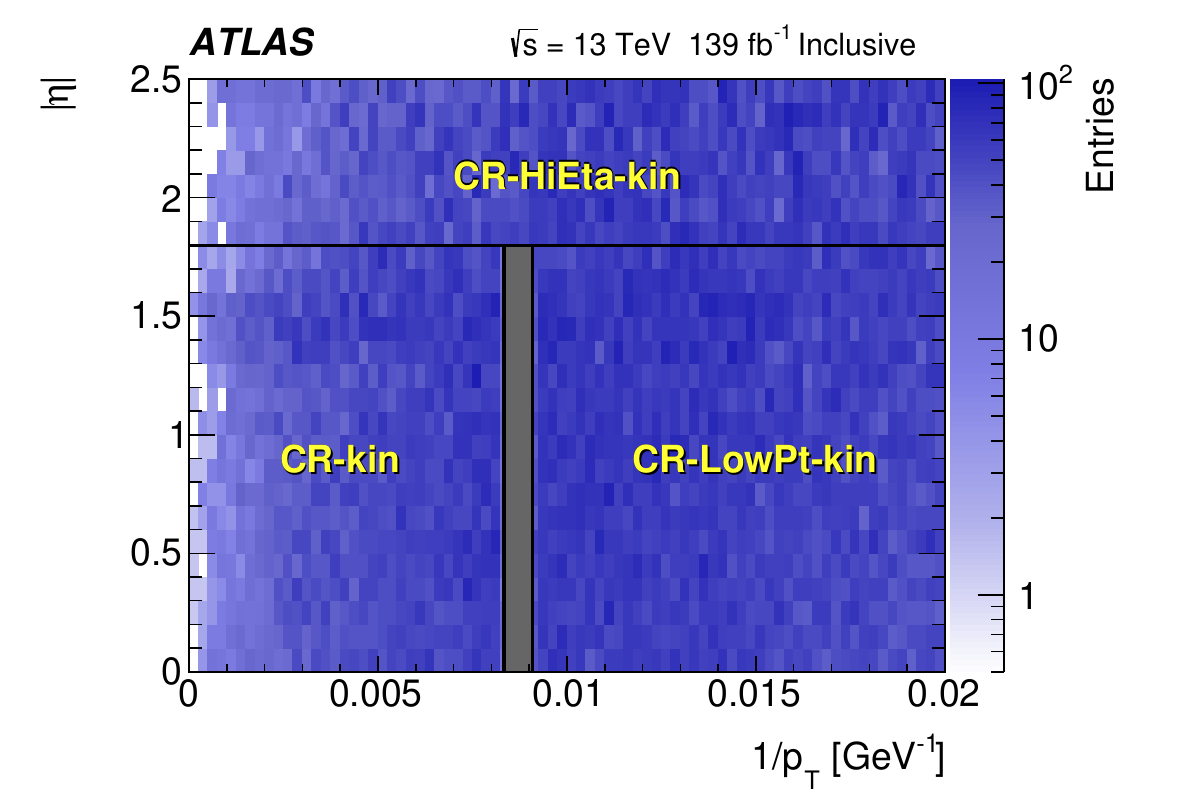}
}
\subfigure[$\dEdx$ control regions]{
\label{fig:CRdEdx_dist_inclusive}
\includegraphics[width=0.47\textwidth]{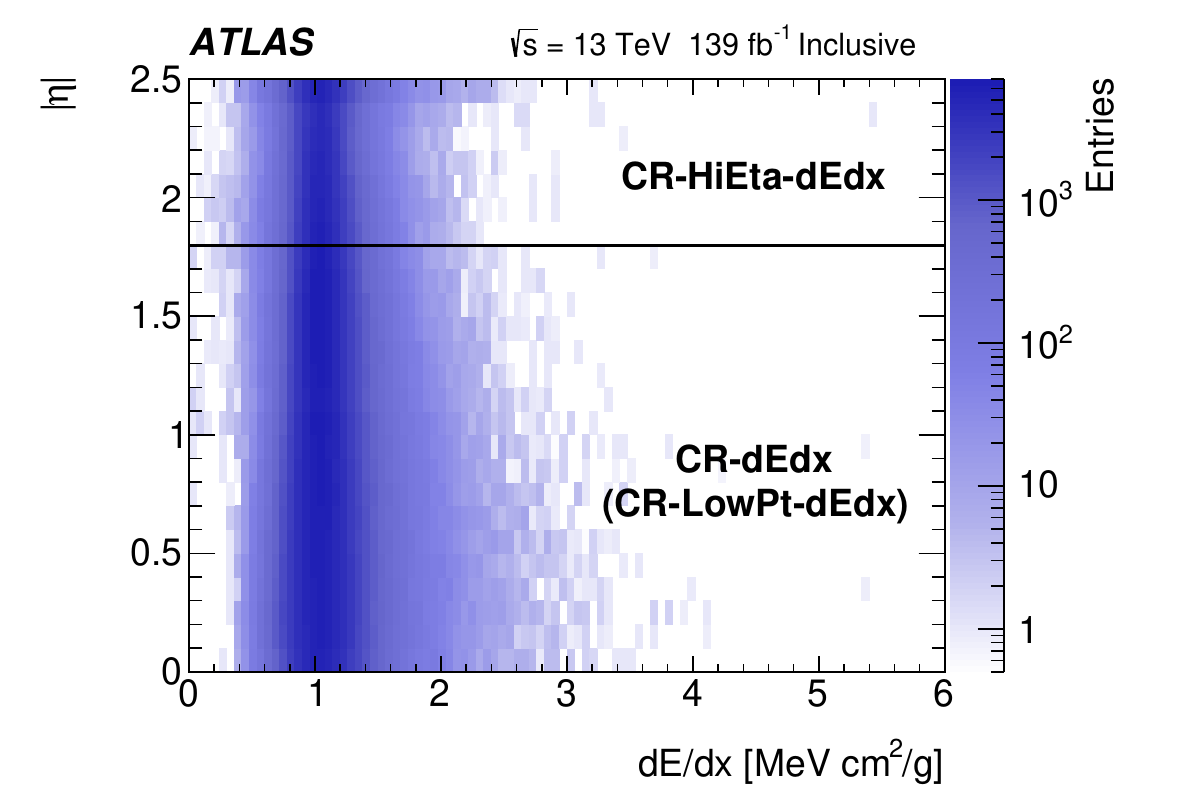}
}
\caption{Event distributions of \subref{fig:CRkin_dist_inclusive} kinematic control regions in the $1/\pt$--$|\eta|$ plane and \subref{fig:CRdEdx_dist_inclusive} $\dEdx$ control regions in the $\dEdx$--$|\eta|$ plane for the inclusive sample. }
\label{fig:CRdists_inclusive}
\end{center}
\end{figure}

The mass distribution of SM tracks in the signal region is estimated using a data-driven technique. Two control regions per signal region, \texttt{CR-kin} and \texttt{CR-dEdx}, are defined adjacent in phase space to the signal region (see Table~\ref{tbl:CRVR_definition}). \texttt{CR-kin} is defined by inverting the \dEdx requirement used in the signal region, and \texttt{CR-dEdx} is defined by inverting the \MET requirement used in the signal region and removing the $\dEdx$ requirement.\footnote{the \texttt{CR-dEdx} will include all the ionization fluctuations (e.g.~$\delta$-ray effects) which may remain after the calculation of the truncated mean. Thus these are taken into account in the background generation.} Here, for a narrow $\Delta\eta$ slice, \dEdx and $\pT$ are expected to be uncorrelated for minimum-ionising background tracks, and \dEdx and \MET are similarly expected to be uncorrelated; therefore, for each $|\eta|$ slice, the $1/\pt$ distribution\footnote{Instead of $\pt$, $1/\pt$ is used here in order to partition the samples with a sufficient number of tracks in all bins.} of tracks in \texttt{CR-kin} and the \dEdx distribution in \texttt{CR-dEdx} serve as the kinematic and \dEdx template distributions for the background mass prediction. The control-region distributions of tracks in the $1/\pt$--$|\eta|$ and $\dEdx$--$|\eta|$ planes are shown in Figure~\ref{fig:CRdists_inclusive}. In Figure~\ref{fig:CRkin_dist_inclusive} it is shown that the sample covers all kinematic space uniformly. Figure~\ref{fig:CRdEdx_dist_inclusive} indicates a slight dependence of the tails of the \dEdx distribution on $|\eta|$, which justifies slicing the background estimation in $|\eta|$. Because the mass of a track is defined by its momentum and $\beta\gamma$ (and therefore \dEdx), the background mass distribution can be constructed by sampling these two distributions.
 
\begin{table}[tbp]
\centering
\small
\caption{Definitions of the signal, control and validation regions.}
\vspace{0.2cm}
\begin{tabular}{l|llll}
\hline
\hline
\textbf{Region} & $\pt~[\GeV]$ & $|\eta|$ & $\met~\mathrm{[\GeV]}$ & $\dEdx~\mathrm{[\dEdxunit]}$\\
\hline
\texttt{SR}              & \multirow{3}{*}{$> 120$}    & \multirow{3}{*}{$<1.8$} & $>170$ & $> 1.8$\\
\texttt{CR-kin}          &                             &                         & $>170$ & $< 1.8$\\
\texttt{CR-dEdx}         &                             &                         & $<170$ & $> 0$\\
\hline
\texttt{VR-LowPt}        & \multirow{3}{*}{$[50,110]$} & \multirow{3}{*}{$<1.8$} & $>170$ & $> 1.8$\\
\texttt{CR-LowPt-kin}    &                             &                         & $>170$ & $< 1.8$\\
\texttt{CR-LowPt-dEdx}   &                             &                         & $<170$ & $> 0$\\
\hline
\texttt{VR-HiEta}        & \multirow{3}{*}{$> 50$}     & \multirow{3}{*}{$[1.8,2.5]$} & $>170$ & $> 1.6$\\
\texttt{CR-HiEta-kin}    &                             &                         & $>170$ & $< 1.6$\\
\texttt{CR-HiEta-dEdx}   &                             &                         & $<170$ & $> 0$\\
\hline
\hline
\end{tabular}
\label{tbl:CRVR_definition}
\end{table}

Although the \dEdx corrections described in Section~\ref{sec:correct_calib} equalise the MPV of the \dEdx distribution as a function of $|\eta|$, any $|\eta|$-dependent differences in the \dEdx tails remain after applying the corrections. As a result, the \dEdx template is sliced in bins of $|\eta|$.
The event rate in \texttt{CR-dEdx} changes over time, reflecting adjustments to the \MET trigger threshold during Run~2, while the event rate in \texttt{CR-kin} is stable since the $\met$ requirement of $170~\GeV$ is well above the trigger threshold. In order to compensate for this effect, events in the \dEdx template extracted from \texttt{CR-dEdx} are reweighted. The weight factor for the $i$-th run, $w_{i}$, is defined as $w_{i} \equiv R_{i_{0}}/R_{i}$ where the $R_{i}$ is the ratio of the numbers of events in \texttt{CR-LowPT-dEdx} and \texttt{CR-LowPt-kin} for a given run $i$ and $i_{0}$ is the reference run against which the events are calibrated. The weight factor differs from unity by up to $\pm2\%$.
 
The \texttt{CR-kin} region, used to predict the background in signal regions requiring tracks to have a hit in the IBL overflow, is statistically sparse. To mitigate this, the kinematic template in the \texttt{CR-kin} region, requiring tracks have no hits in the IBL overflow, is used instead, with a weight factor $n({\IBLOF=1})/n({\IBLOF=0})$ applied to each $|\eta|$ slice.
 
To form a `toy' background track, a pair of $1/\pt$ and $|\eta|$ values is sampled from the kinematic template. A \dEdx value is sampled from the corresponding $|\eta|$ bin of the \dEdx template. From these sampled values, the track mass, $m$, is calculated using the $\dEdx\mbox{--}\beta\gamma$ calibration. In total, 10 (40) million toy tracks are generated for the \texttt{Low}  (\texttt{High}) \dEdx range so that the number of toy samples does not limit the accuracy of predictions in any mass range. Because there is no restriction on the range of \dEdx in sampling from \texttt{CR-dEdx}, these toy samples predict the background distribution in both of the signal regions and \texttt{CR-kin} simultaneously. Finally, the toy samples are normalised to data in a sub-region of \texttt{CR-kin} which is expected to be depleted in signal, with $m < 160~\GeV$ and $\dEdx < 1.8~\dEdxunit$. After the normalisation, the mass distributions in the $m>160~\GeV$ range of \texttt{CR-kin} are in good agreement with those derived from the toy samples.
 
The background estimation procedure is validated in two sets of validation regions: one set (\texttt{VR-LowPt}) selects tracks with lower \pT and the other set (\texttt{VR-HiEta}) selects tracks with higher $|\eta|$ and has a looser \pT requirement than in the signal regions. The definitions of these regions and the corresponding control regions used for the background estimation are shown in Table~\ref{tbl:CRVR_definition}. The \dEdx requirement for the \texttt{VR-HiEta} region is loosened in order to probe a similar tail fraction of the narrower \dEdx distribution found at high $|\eta|$, and there is no subdivision of \texttt{VR-HiEta} into \texttt{Low} and \texttt{High} \dEdx ranges because of the limited sample size.
 
\begin{figure}[tbp]
\begin{center}
\subfigure[\texttt VR-LowPt-Inclusive\_High ]{
\label{fig:VR-LowPt-InclusiveHigh}
\includegraphics[width=0.47\textwidth]{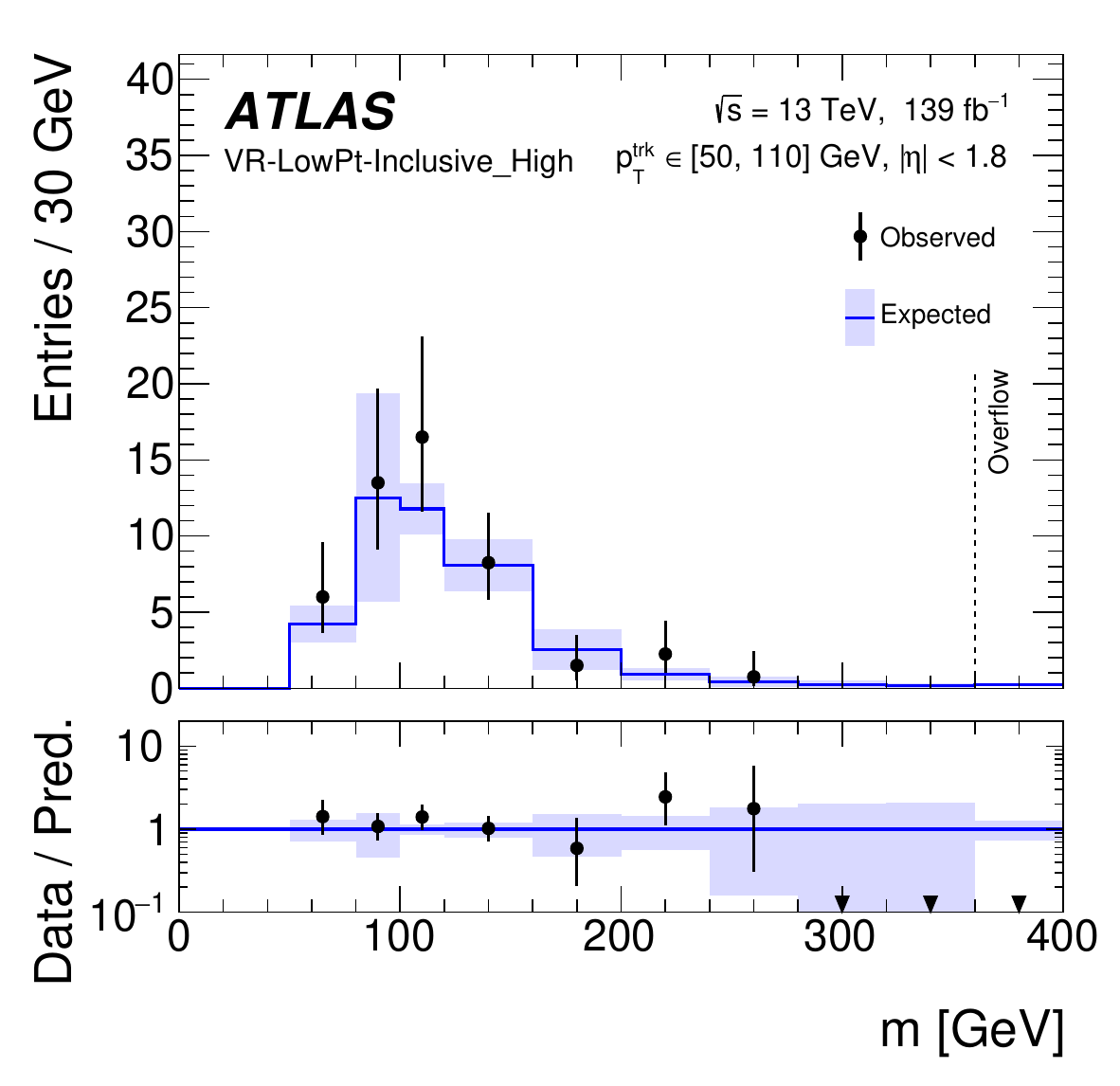}
}
\subfigure[\texttt VR-HiEta-Inclusive ]{
\label{fig:VR-HiEta-Inclusive}
\includegraphics[width=0.47\textwidth]{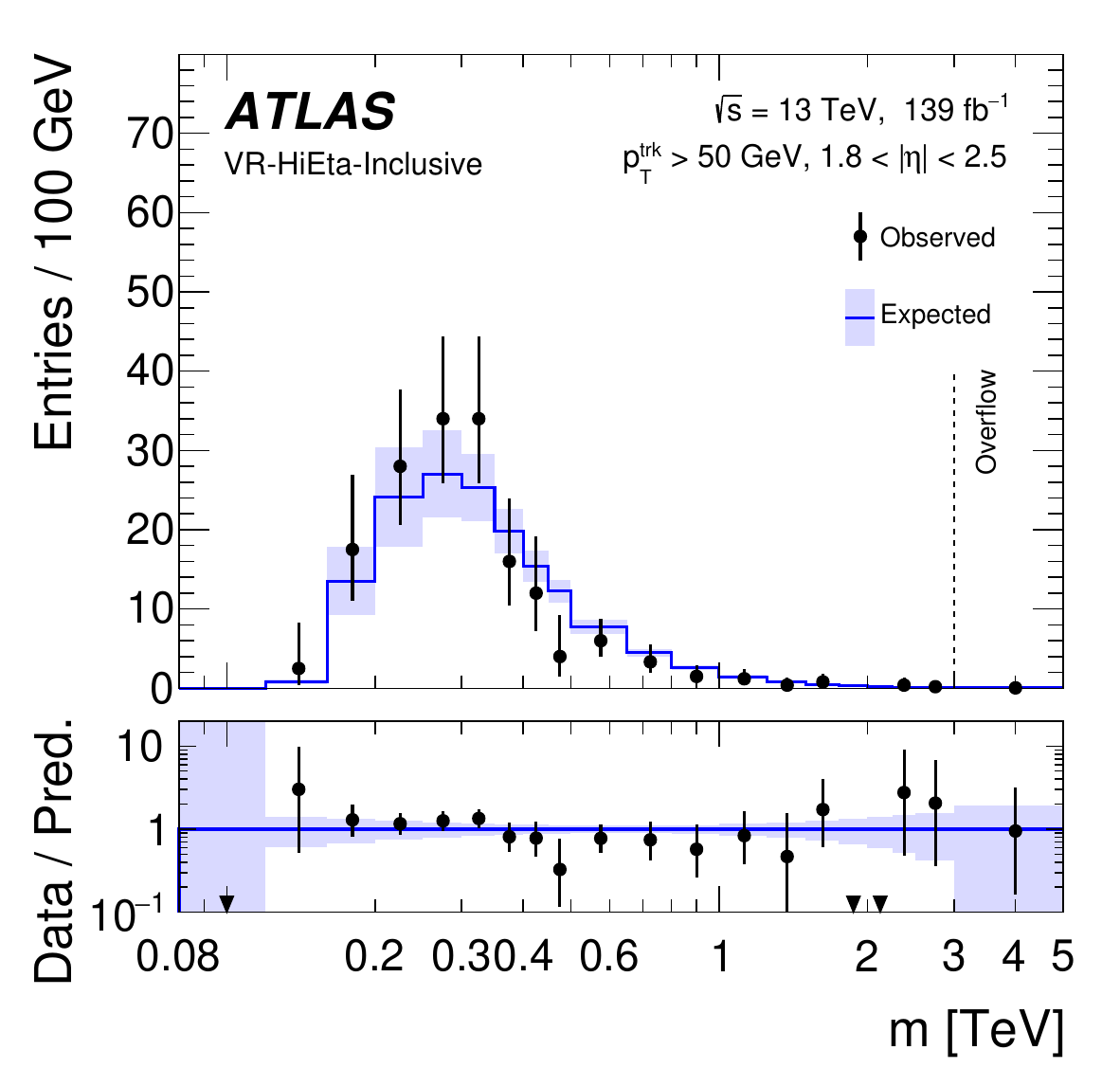}
}
\caption{Comparison of the observed and expected mass distributions in \subref{fig:VR-LowPt-InclusiveHigh} \texttt{VR-LowPt-Inclusive\_High} and \subref{fig:VR-HiEta-Inclusive} \texttt{VR-HiEta-Inclusive} validation-region bins. The band on the expected background estimation indicates the total uncertainty of the estimation. Downward triangle markers at the bottom of the panels indicate there are no events observed in the corresponding bin.}
\label{fig:VR-mass-distributions}
\end{center}
\end{figure}

The expected and observed mass distributions for two of the individual validation regions, \texttt{\seqsplit{VR-LowPt-Inclusive\_High}} (with $\dEdx > 2.4~\dEdxunit$) and \texttt{\seqsplit{VR-HiEta-Inclusive}} (with $\dEdx > 1.6~\dEdxunit$), are shown in Figure~\ref{fig:VR-mass-distributions}. The expected and observed distributions agree well across the lower mass range probed by \texttt{VR-LowPt-Inclusive\_High} and the higher mass range probed by \texttt{VR-HiEta-Inclusive}.
 
\begin{figure}[tbp]
\begin{center}
\subfigure[\texttt VR-LowPt ]{
\label{fig:VR-LowPt-ratio}
\includegraphics[width=0.47\textwidth]{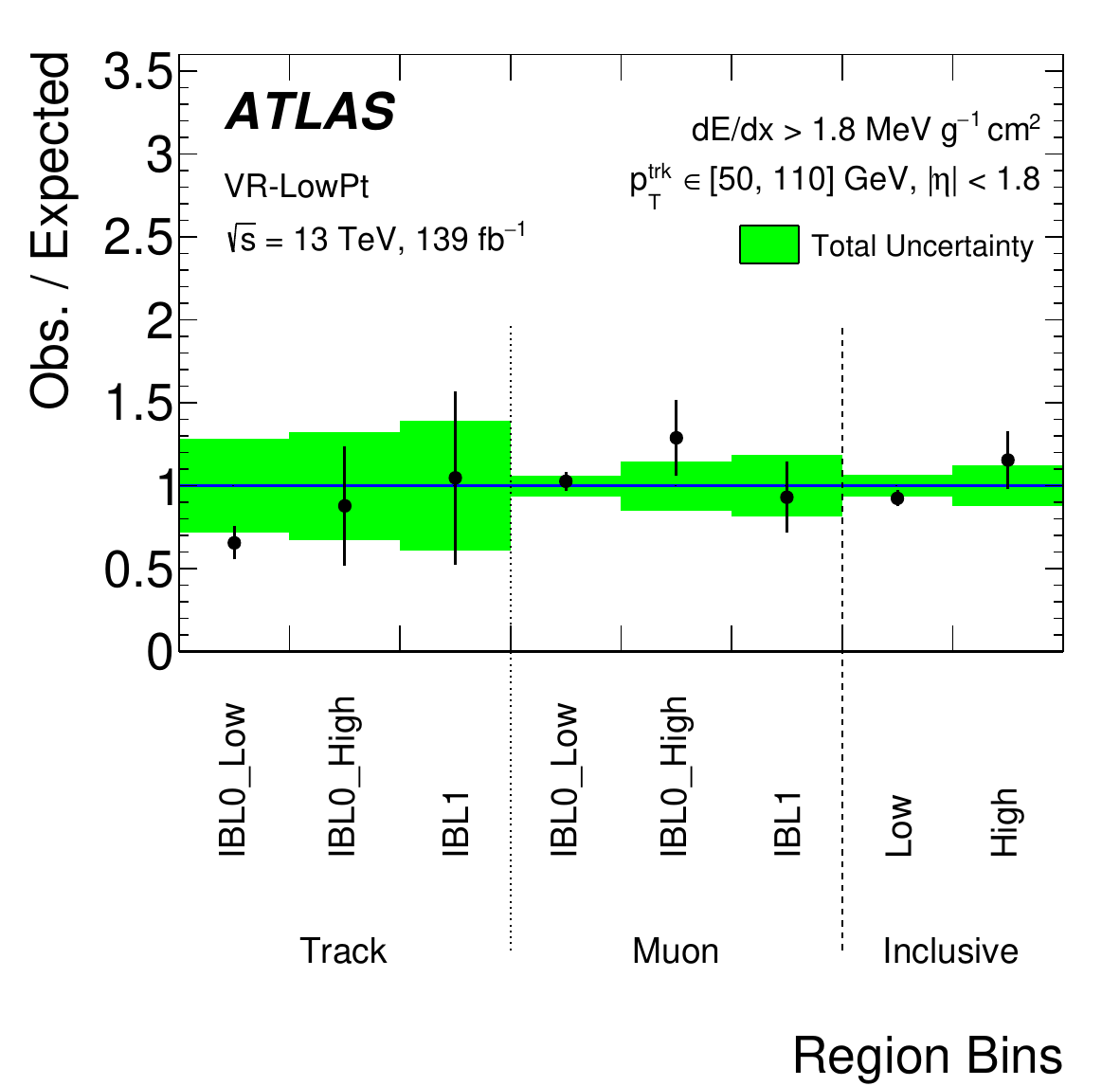}
}
\subfigure[\texttt VR-HiEta ]{
\label{fig:VR-HiEta}
\includegraphics[width=0.47\textwidth]{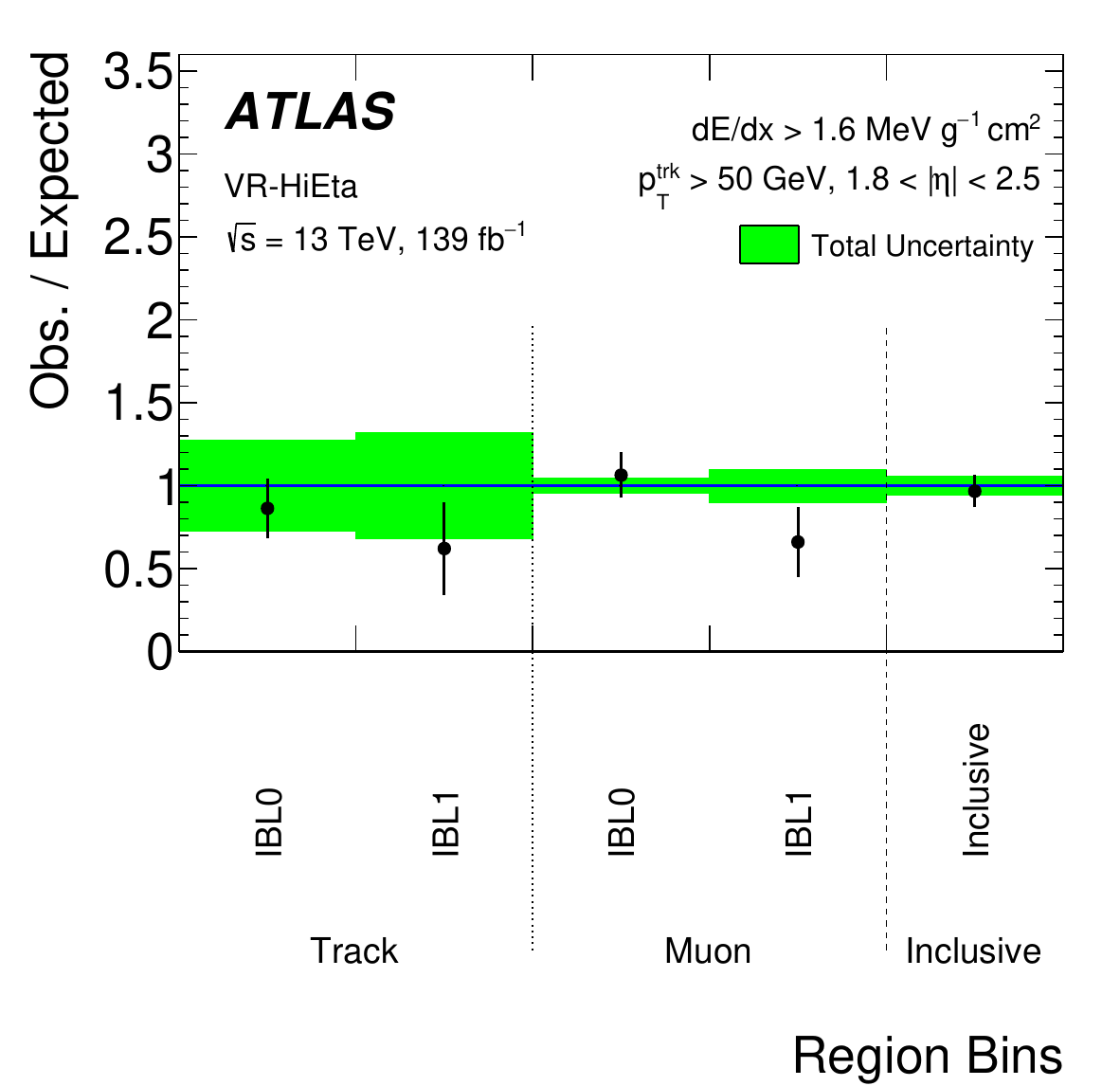}
}
\caption{Comparison of the observed and expected event yields in \subref{fig:VR-LowPt-ratio} \texttt{VR-LowPt} and \subref{fig:VR-HiEta} \texttt{VR-HiEta} validation-region bins.}
\label{fig:VR-yield-ratio}
\end{center}
\end{figure}

The expected and observed yields in all of the validation-region bins are shown in Table~\ref{tbl:vr_yields}, and the ratios of the observed to expected yields are shown in Figure~\ref{fig:VR-yield-ratio}. Good agreement is visible across all region bins except \texttt{VR-LowPt-Trk-IBL0\_Low}, where the observed yield is approximately $35\%$ lower than the prediction and unlikely to be a statistical fluctuation. In order to relieve this tension, an empirical systematic scale uncertainty is added. The details of this treatment are described in Section~\ref{sec:syst}. This additional uncertainty is already included in Figures~\ref{fig:VR-mass-distributions} and \ref{fig:VR-yield-ratio} and in Table~\ref{tbl:vr_yields}, combined with the rest of the systematic uncertainties.

\begin{table}[tbp]
\centering
\small
\caption{Expected and observed event yields in the validation-region bins.}
\vspace{0.2cm}
\begin{tabular}{l|ll|r@{~$\pm$~}lc}
\hline
\hline
\textbf{Region} & \textbf{Category} & \textbf{Bin} & \multicolumn{2}{l}{\textbf{Expected}} & \textbf{Observed}\\
\hline
\multirow{8}{*}{\texttt{VR-LowPt}}
& \multirow{3}{*}{\texttt{Trk}}
& \texttt{IBL0\_Low}       &   65.6 &  18.3 &  43 \\
&& \texttt{IBL0\_High}      &    6.8 &  2.2 &   ~~6 \\
&& \texttt{IBL1}            &    3.8 &  1.5 &   ~~4 \\ \cline{2-6}
& \multirow{3}{*}{\texttt{Mu}}
& \texttt{IBL0\_Low}       &  292   & 17   & 300~~ \\
&& \texttt{IBL0\_High}      &   24.8 &  3.6 &  32\\
&& \texttt{IBL1}            &   20.4 &  3.7 &  19\\ \cline{2-6}
& \multirow{2}{*}{\texttt{Inclusive}}
& \texttt{Low}             &  391   & 24   & 361~~ \\
&& \texttt{High}            &   37.2 &  4.4 &  43\\
\hline
\hline
\multirow{5}{*}{\texttt{VR-HiEta}}
& \multirow{2}{*}{\texttt{Trk}}
& \texttt{IBL0}       & 26.6 & 7.3 & 23  \\
&& \texttt{IBL1}       &  8.0 & 2.6 &  ~~5  \\ \cline{2-6}
& \multirow{2}{*}{\texttt{Mu}}
& \texttt{IBL0}       & 56.4 & 2.5 & 59  \\
&& \texttt{IBL1}       & 15.1 & 1.5 & 10  \\ \cline{2-6}
& \texttt{Inclusive} & ---
& 101   & 6   & 97  \\
\hline
\hline
\end{tabular}
\label{tbl:vr_yields}
\end{table}


\section{Systematic uncertainties}
\label{sec:syst}
 
\begin{figure}[htbp]
\begin{center}
\subfigure[\texttt SR-Inclusive\_Low, \text Long lifetime regime]{
\label{fig:syst-SR-IncLow}
\includegraphics[width=0.47\textwidth]{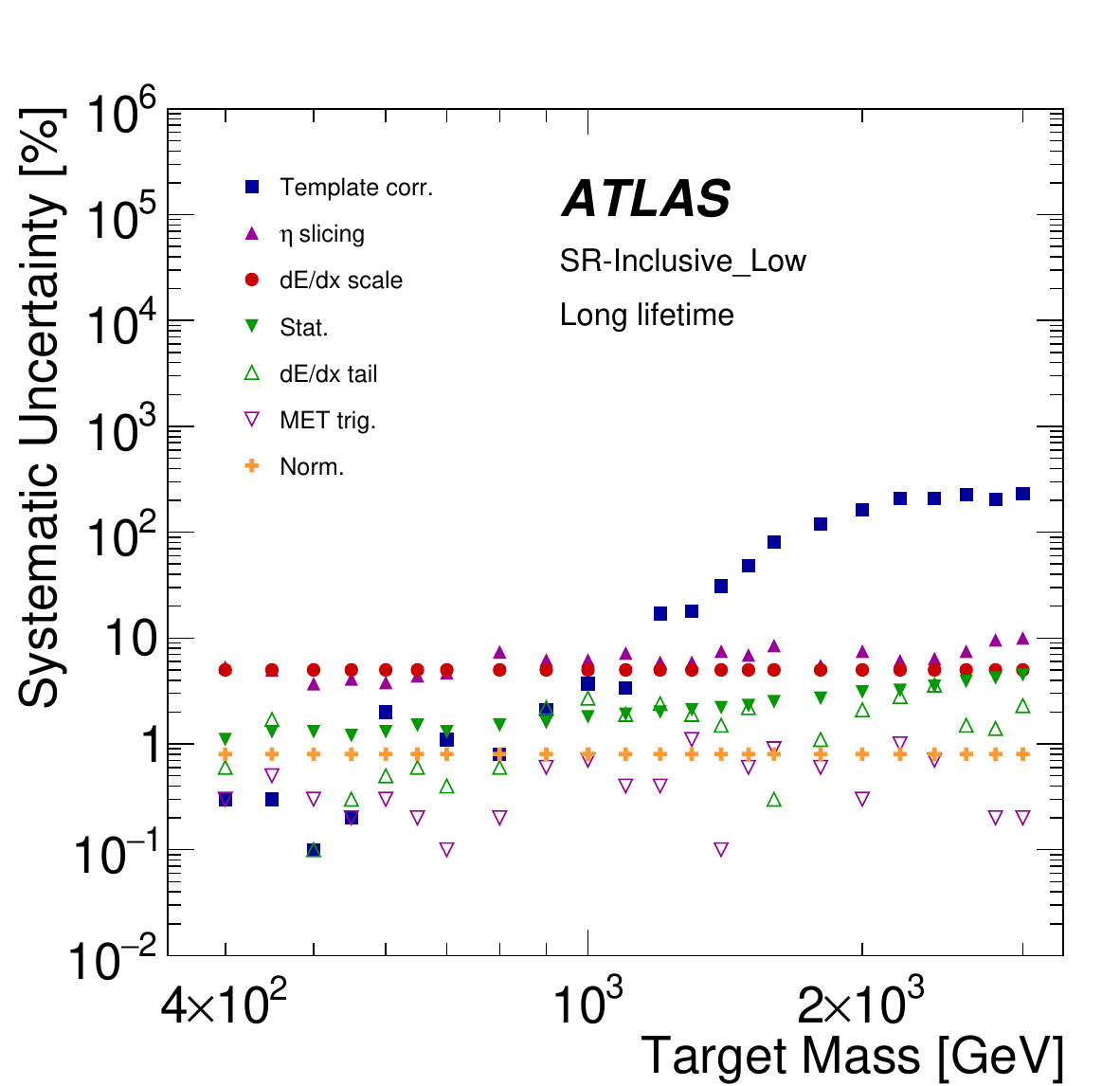}
}
\subfigure[\texttt SR-Inclusive\_High, \text Long lifetime regime]{
\label{fig:syst-SR-IncHigh}
\includegraphics[width=0.47\textwidth]{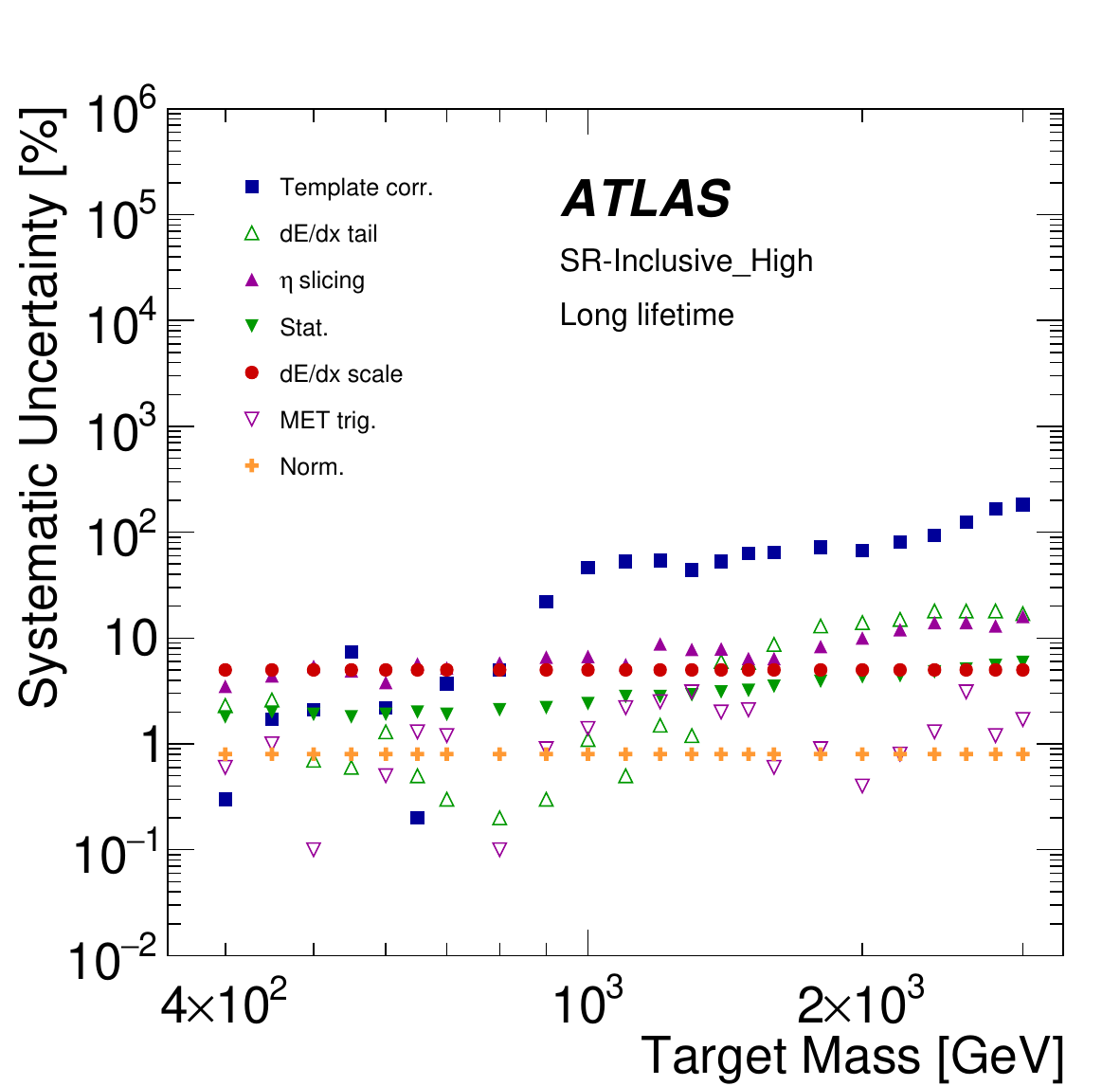}
}
\caption{Systematic uncertainties in the background estimate for each mass window targeting long lifetimes in the \subref{fig:syst-SR-IncLow} \texttt{SR-Inclusive\_Low} region and \subref{fig:syst-SR-IncHigh} \texttt{SR-Inclusive\_High} region. }
\label{fig:systs}
\end{center}
\end{figure}

The systematic uncertainties associated with the background estimate are evaluated for each mass window, as shown in Figure~\ref{fig:systs}. The leading uncertainty at high masses is the template correlation uncertainty, labelled `Template corr.', which estimates the effect of residual correlations between the template distributions used to generate the background. To test the assumption that the kinematic and \dEdx templates can be sampled separately to form a toy track, a pseudo signal region is defined in the \texttt{CR-dEdx} region (the one based on inverting the \MET requirement) by requiring $\dEdx>1.8~\dEdxunit$. The background estimation procedure is then executed, extracting both the $(\pT^{-1},|\eta|)$ pair and the \dEdx from the distributions in the \texttt{CR-dEdx} region.
The difference of the event counts in each mass window between the predicted and observed mass distributions in the pseudo signal regions is taken as a systematic uncertainty. The size of the uncertainty is evaluated for each signal-region bin separately. This uncertainty is the dominant uncertainty in the target mass above ${\sim}1~\TeV$.
 
In the background estimation procedure, the \dEdx templates are sliced in $|\eta|$ due to the remaining $|\eta|$-dependence of the \dEdx tails. To ensure that the background estimate is not heavily dependent on the choice of these bins, the background estimation is repeated with an alternative set of $|\eta|$ bins. The difference between the resulting mass distribution and the nominal one is taken as a systematic uncertainty, labelled `$\eta$ slicing'. The $\IBLOF=1$ region uses a reweighted version of the $\IBLOF=0$ region's kinematic template as explained in Section~\ref{sec:background}. An uncertainty is assigned to this method by generating an alternative background distribution using the original $\IBLOF=1$ region's kinematic template and comparing it with the background distribution generated by the reweighted kinematic template. This uncertainty only affects the \texttt{IBL1} SR.

Since the \MET trigger thresholds changed during Run~2, the \dEdx templates in the background estimation are reweighted to correct for any bias which may arise in the \texttt{CR-dEdx} region, which is populated by events with low \MET.
An uncertainty labelled `MET trig.', derived by comparing the predicted mass distributions with and without this reweighting, is applied to cover any deficiencies in the reweighting.
 
Uncertainties that account for the effect of statistical fluctuations in the control-region templates are also evaluated. The statistical uncertainties of the template histograms are derived by fluctuating each bin  in the template histograms according to a Poisson distribution with a mean equal to the number of entries in the bin. The root-mean-square deviation of alternative mass distributions generated from the randomly fluctuated templates is taken as the uncertainty, which is labelled `Stat.'. To quantify the effect of statistical fluctuations in the tail of the \dEdx template distributions and assign an uncertainty, labelled `\dEdx tail', each tail is fitted with a Crystal Ball distribution, which is then used instead of the data in the \dEdx template to generate an alternative mass distribution. A statistical uncertainty, labelled `Norm.', in the normalisation factor of the background prediction is also calculated and applied as a uniform uncertainty.
 
As described in the previous section, in validation-region bin \texttt{VR-LowPt-Trk-IBL0\_Low}, a deficit of 35\% relative to the prediction is observed in the range $m\lesssim 120~\GeV$, which is significantly larger than the statistical uncertainty. In the other bins, no significant excesses or deficits are found. This deficit can be attributed to the difference between the \dEdx tail distributions for $\dEdx\in[1.8,2.4]~\dEdxunit$ in the \texttt{CR-LowPt-dEdx} and \texttt{VR-LowPt} samples. An empirical scale uncertainty in the number of events, labelled `\dEdx scale',  is assigned based on the following considerations: the observed mismatch depends on the amount of \pileup, and for the \texttt{Mu} categories the \dEdx tail fraction is stable and the observed yield agrees with the prediction very well. The \dEdx response of well-isolated muon tracks is considered more robust against \pileup than that of other tracks. Therefore, different uncertainties are assigned to the \texttt{Trk} and \texttt{Mu} categories, and the uncertainty is correlated over all validation- and signal-region bins in the same category. With this global approach, a scale uncertainty of 27\% (3\%) is assigned to the \texttt{Trk} (\texttt{Mu}) category. For the \texttt{Inclusive} category, the combined scale uncertainty is found to be 5\%, and this is consistent with the fact that the majority of the background tracks in the \texttt{Inclusive} category are identified as muons. The derived `\dEdx scale' uncertainty is assumed to be uniform over mass and is dominant, uniquely for the \texttt{Trk} category, below ${\sim}1.0~\TeV$.
 
Additional systematic uncertainties associated with the modelling and predicted yield of the signals in simulation are also considered when setting limits on specific models. An uncertainty of 1.7\%  applied to the dataset's integrated luminosity is derived from $x\mbox{--}y$ beam-separation scans~\cite{ATLAS-CONF-2019-021}. Uncertainties in the QCD radiation modelling, which significantly impact the efficiency of triggering on signals with large initial-state radiation, are estimated using MC samples. These MC samples are generated identically to their corresponding signal MC samples but with variations in the factorisation, renormalisation, and merging scales, as well as in the parton showering tunes and radiation tunes. The differences between results from the nominal and alternative MC samples are then used as the systematic uncertainties.  The majority of these uncertainties are found to be of order of $1\%$. The only exceptions are parton shower and radiation tuning uncertainties that can grow to be $\sim 10\%$ for some MC samples.

For the MC signal samples, the \dEdx distribution (and therefore the mass distribution), as well as the probability of a track to have $\IBLOF=1$, is based on a template derived from a 2017 low-\pileup dataset. This dataset is also used to calibrate the $\dEdx\mbox{--}\beta\gamma$ relationship. Although the run-dependent corrections applied to the \dEdx measurement mitigate any time dependence of the most probable \dEdx measurement for a MIP-like particle, any remaining time dependence of the \dEdx distribution for particles with small $\beta\gamma$ in data is not accounted for in signal samples because the data-driven template is taken from a fixed point in time.
To quantify the potential impact of this, the $\dEdx\mbox{--}\beta\gamma$ relationship was calibrated again, this time using a similar low-\pileup dataset taken in 2018. The probability of a track to have $\IBLOF=1$ was also measured in this dataset. The difference between the two calibrations, quantified by comparing the resulting signal-region track masses, is negligible compared to the mass resolution and is less than ${\sim}3\%$. The probability of a track to have $\IBLOF=1$ was found to have increased significantly in 2018 due to a change in the IBL \ToT front-end configuration. An associated uncertainty is therefore calculated by comparing the probabilities of a track to have $\IBLOF=1$ in the 2017 and 2018 low-\pileup datasets and applied by using the $\beta\gamma$-dependent ratio of these probabilities to reweight the events.
 
Other uncertainties associated with the simulation modelling include those related to the \pileup distribution, track-level quantities, muon identification, the \MET trigger, and the offline \MET calculation. The largest of these are the modelling uncertainties associated with track momentum measurement errors due to detector misalignments, the \pileup distribution, and the offline \MET calculation.


\FloatBarrier

\begin{table}[p]
\centering
\caption{List of expected and observed numbers of events, $p_{0}$-value (capped at 0.5) and the corresponding local $Z$ significance, as well as the 95\% $\text{CL}_\text{s}$ upper limit on the expected and observed numbers of signal events ($S_{\mathrm{exp.}}^{95}$ and $S_{\mathrm{obs.}}^{95}$) in each mass window for \texttt{SR-Inclusive} bins in the `short' ($\tau \leq 1~\text{ns}$) and `long' ($\tau > 1~\text{ns}$) lifetime regimes.}
\vspace{0.2cm}
\scriptsize
\renewcommand{\arraystretch}{1.3}
\setlength{\tabcolsep}{1.2mm}
\begin{tabular}{rc|cccccc | cccccc }
\hline
\hline
\multirow{3}{8mm}{\textbf{Target mass [\GeV]}}
& \multirow{3}{10mm}{\textbf{Mass window [\GeV]}}
& \multicolumn{12}{c}{\textbf{Signal region bin}}\\ \cline{3-14}
&& \multicolumn{6}{c|}{\texttt{SR-Inclusive\_Low}} & \multicolumn{6}{c}{\texttt{SR-Inclusive\_High}}\\
&& \textbf{Exp.} & \textbf{Obs.} & \textbf{$p_{0}$} & \textbf{$Z_\mathrm{local}$}
& \textbf{$S_{\mathrm{exp.}}^{95}$} & \textbf{$S_{\mathrm{obs.}}^{95}$}
& \textbf{Exp.} & \textbf{Obs.} & \textbf{$p_{0}$} & \textbf{$Z_\mathrm{local}$}
& \textbf{$S_{\mathrm{exp.}}^{95}$} & \textbf{$S_{\mathrm{obs.}}^{95}$} \\
\hline
\hline
\multicolumn{2}{c}{}& \multicolumn{12}{c}{\textbf{Short lifetime}}\\
\hline
200               & [120, 225]        & 81 $\pm$ 4      & 76   & $5.00\times 10^{-1}$ & 0.0    & \,$21^{+8}_{-6}$        & 18~~~    & 5.6 $\pm$ 0.7   & 7    & $2.65\times 10^{-1}$ & 0.6     & $6.3^{+2.5}_{-1.7}$ & 7.8      \\
300               & [200, 350]        & 72 $\pm$ 4      & 72   & $4.72\times 10^{-1}$ & 0.1    & \,$20^{+8}_{-6}$        & 20~~~    & 9.2 $\pm$ 0.8   & 14~~ & $7.11\times 10^{-2}$ & 1.5     & $7.6^{+3.0}_{-2.1}$ & 12.5~~   \\
400               & [300, 500]        & 45.6 $\pm$ 3.3  & 43   & $5.00\times 10^{-1}$ & 0.0    & \,$16^{+6}_{-4}$        & 14~~~    & 5.8 $\pm$ 0.4   & 6    & $4.39\times 10^{-1}$ & 0.1     & $6.1^{+2.5}_{-1.8}$ & 6.5      \\
450               & [350, 600]        & 37.6 $\pm$ 2.7  & 44   & $1.72\times 10^{-1}$ & 0.9    & \,$15^{+6}_{-4}$        & 20~~~    & 5.1 $\pm$ 0.4   & 3    & $5.00\times 10^{-1}$ & 0.0     & $6.0^{+2.2}_{-1.6}$ & 4.6      \\
500               & [400, 700]        & 30.6 $\pm$ 2.2  & 42   & $3.41\times 10^{-2}$ & 1.8    & \,$13^{+5}_{-4}$        & 24~~~    & 4.3 $\pm$ 0.4   & 4    & $5.00\times 10^{-1}$ & 0.0     & $5.4^{+2.2}_{-1.3}$ & 5.2      \\
550               & [400, 800]        & 33.9 $\pm$ 2.5  & 45   & $4.74\times 10^{-2}$ & 1.7    & \,$14^{+5}_{-4}$        & 24~~~    & 4.8 $\pm$ 0.4   & 4    & $5.00\times 10^{-1}$ & 0.0     & $5.8^{+2.5}_{-1.8}$ & 5.4      \\
600               & [450, 900]        & 27.5 $\pm$ 1.9  & 35   & $9.48\times 10^{-2}$ & 1.3    & $12.1^{+5.3}_{-3.5}$  & 19.3     & 3.91 $\pm$ 0.31 & 2    & $5.00\times 10^{-1}$ & 0.0     & $5.5^{+2.2}_{-1.6}$ & 4.0      \\
650               & ~~[500, 1000]     & 22.5 $\pm$ 1.6  & 29   & $1.03\times 10^{-1}$ & 1.3    & $11.2^{+4.4}_{-2.8}$  & 17.2     & 3.22 $\pm$ 0.31 & 2    & $5.00\times 10^{-1}$ & 0.0     & $5.2^{+1.9}_{-1.6}$ & 4.4      \\
700               & ~~[550, 1100]     & 18.7 $\pm$ 1.4  & 23   & $1.71\times 10^{-1}$ & 0.9    & $10.3^{+4.0}_{-2.7}$  & 14.3     & 2.64 $\pm$ 0.31 & 2    & $5.00\times 10^{-1}$ & 0.0     & $4.7^{+1.9}_{-1.0}$ & 4.3      \\
800               & ~~[600, 1200]     & 15.6 $\pm$ 1.3  & 20   & $1.47\times 10^{-1}$ & 1.1    & ~~$9.5^{+3.8}_{-2.9}$ & 13.7     & 2.22 $\pm$ 0.24 & 3    & $2.86\times 10^{-1}$ & 0.6     & $4.5^{+1.8}_{-1.0}$ & 5.5      \\
900               & ~~[650, 1400]     & 13.8 $\pm$ 1.3  & 17   & $2.09\times 10^{-1}$ & 0.8    & ~~$9.1^{+3.5}_{-2.5}$ & 11.9     & 2.0 $\pm$ 0.3   & 4    & $9.74\times 10^{-2}$ & 1.3     & $4.3^{+1.6}_{-0.9}$ & 6.8      \\
1000              & ~~[700, 1850]     & 13.1 $\pm$ 1.3  & 17   & $1.54\times 10^{-1}$ & 1.0    & ~~$8.8^{+3.6}_{-2.3}$ & 12.7     & 1.9 $\pm$ 0.5   & 4    & $9.01\times 10^{-2}$ & 1.3     & $4.1^{+1.9}_{-0.7}$ & 7.0      \\
1200              & ~~[800, 2400]     & 11 $\pm$ 2      & 14   & $1.85\times 10^{-1}$ & 0.9    & ~~$8.6^{+3.3}_{-2.5}$ & 11.9     & 1.5 $\pm$ 0.7   & 6    & $9.10\times 10^{-3}$ & 2.4     & $4.0^{+1.6}_{-0.8}$ & 10.0~~   \\
1400              & ~~[900, 2900]     & ~~8.5 $\pm$ 2.1 & 11   & $2.37\times 10^{-1}$ & 0.7    & ~~$8.1^{+3.1}_{-2.6}$ & 10.5     & 1.1 $\pm$ 0.7   & 7    & $2.08\times 10^{-3}$ & 2.9     & $4.0^{+1.4}_{-0.7}$ & 11.5~~   \\
1600              & [1000, 3450]      & ~~6.9 $\pm$ 2.4 & ~~9  & $2.57\times 10^{-1}$ & 0.7    & ~~$7.8^{+3.0}_{-2.6}$ & 10.1     & 0.9 $\pm$ 0.5   & 7    & $6.03\times 10^{-4}$ & 3.2     & $3.6^{+1.5}_{-0.5}$ & 11.8~~   \\
1800              & [1100, 4000]      & ~~5.7 $\pm$ 2.6 & ~~8  & $2.35\times 10^{-1}$ & 0.7    & ~~$7.3^{+2.8}_{-2.3}$ & ~~9.9    & 0.8 $\pm$ 0.6   & 7    & $8.87\times 10^{-4}$ & 3.1     & $3.5^{+1.1}_{-0.2}$ & 11.9~~   \\
2000              & [1200, 4600]      & ~~5 $\pm$ 4     & ~~6  & $3.03\times 10^{-1}$ & 0.5    & ~~$7.3^{+3.0}_{-2.3}$ & ~~9.0    & 0.6 $\pm$ 0.5   & 5    & $4.92\times 10^{-3}$ & 2.6     & $3.1^{+1.1}_{-0.1}$ & 9.4      \\
\hline
\multicolumn{2}{c}{}& \multicolumn{12}{c}{\textbf{Long lifetime}}\\
\hline
100               & [120, 200]        & 68 $\pm$ 4             & 63   & $5.00\times 10^{-1}$ & 0.0     & \,$19^{+7}_{-5}$       & 16~~~    & 3.9 $\pm$ 0.6          & 5    & $2.81\times 10^{-1}$ & 0.6     & $5.4^{+2.1}_{-1.0}$ & 6.7      \\
200               & [150, 225]        & 63 $\pm$ 4             & 54   & $5.00\times 10^{-1}$ & 0.0     & \,$18^{+7}_{-4}$       & 13~~~    & 5.5 $\pm$ 0.6          & 7    & $2.61\times 10^{-1}$ & 0.6     & $6.1^{+2.6}_{-1.8}$ & 7.8      \\
300               & [250, 350]        & 40.9 $\pm$ 2.7         & 35   & $5.00\times 10^{-1}$ & 0.0     & \,$15^{+6}_{-4}$       & 11~~~    & 5.1 $\pm$ 0.5          & 7    & $2.01\times 10^{-1}$ & 0.8     & $5.9^{+2.4}_{-1.4}$ & 8.0      \\
400               & [350, 500]        & 29.2 $\pm$ 2.2         & 33   & $2.54\times 10^{-1}$ & 0.7     & $12.6^{+5.3}_{-3.2}$   & 16.0     & 3.83 $\pm$ 0.26        & 2    & $5.00\times 10^{-1}$ & 0.0     & $5.4^{+1.9}_{-1.3}$ & 4.2      \\
450               & [400, 550]        & 21.5 $\pm$ 1.6         & 30   & $5.03\times 10^{-2}$ & 1.6     & $11.0^{+4.2}_{-2.9}$   & 19.4     & 3.00 $\pm$ 0.23        & 2    & $5.00\times 10^{-1}$ & 0.0     & $5.1^{+1.7}_{-1.3}$ & 4.3      \\
500               & [450, 650]        & 19.4 $\pm$ 1.2         & 27   & $5.62\times 10^{-2}$ & 1.6     & $10.3^{+4.3}_{-2.6}$   & 17.4     & 2.73 $\pm$ 0.22        & 1    & $5.00\times 10^{-1}$ & 0.0     & $4.7^{+1.9}_{-0.9}$ & 3.9      \\
550               & [450, 700]        & 21.8 $\pm$ 1.5         & 29   & $7.73\times 10^{-2}$ & 1.4     & $11.0^{+4.2}_{-3.2}$   & 17.8     & 3.06 $\pm$ 0.32        & 2    & $5.00\times 10^{-1}$ & 0.0     & $5.0^{+2.1}_{-1.5}$ & 4.2      \\
600               & [500, 800]        & 18.4 $\pm$ 1.3         & 24   & $1.12\times 10^{-1}$ & 1.2     & \,$10^{+4}_{-3}$       & 15~~~    & 2.64 $\pm$ 0.19        & 2    & $5.00\times 10^{-1}$ & 0.0     & $4.4^{+2.2}_{-1.2}$ & 4.2      \\
650               & [550, 850]        & 15 $\pm$ 1             & 19   & $1.32\times 10^{-1}$ & 1.1     & ~~$9.1^{+3.7}_{-2.7}$  & 13.4     & 2.07 $\pm$ 0.17        & 2    & $5.00\times 10^{-1}$ & 0.0     & $4.5^{+1.5}_{-1.2}$ & 4.6      \\
700               & [550, 950]        & 16.6 $\pm$ 1.2         & 21   & $1.52\times 10^{-1}$ & 1.0     & ~~$9.7^{+3.8}_{-2.8}$  & 13.7     & 2.4 $\pm$ 0.2          & 2    & $5.00\times 10^{-1}$ & 0.0     & $4.5^{+2.0}_{-0.9}$ & 4.3      \\
800               & ~~[650, 1150]     & 12.0 $\pm$ 1.1         & 14   & $2.86\times 10^{-1}$ & 0.6     & ~~$8.4^{+3.5}_{-2.3}$  & 10.4     & 1.74 $\pm$ 0.16        & 3    & $1.79\times 10^{-1}$ & 0.9     & $4.1^{+1.8}_{-0.8}$ & 5.8      \\
900               & ~~[700, 1250]     & 10.4 $\pm$ 0.9         & 13   & $2.17\times 10^{-1}$ & 0.8     & ~~$8.1^{+3.0}_{-2.6}$  & 10.3     & 1.5 $\pm$ 0.4          & 3    & $1.35\times 10^{-1}$ & 1.1     & $3.9^{+1.8}_{-1.0}$ & 6.0      \\
1000              & ~~[800, 1550]     & ~~8.6 $\pm$ 0.8        & 11   & $2.16\times 10^{-1}$ & 0.8     & ~~$7.5^{+2.9}_{-2.5}$  & ~~9.6    & 1.2 $\pm$ 0.6          & 4    & $3.73\times 10^{-2}$ & 1.8     & $3.8^{+1.4}_{-0.8}$ & 7.5      \\
1100              & ~~[900, 1800]     & ~~7.1 $\pm$ 0.7        & 10   & $1.46\times 10^{-1}$ & 1.1     & ~~$7.0^{+2.5}_{-1.9}$  & ~~9.8    & 1.0 $\pm$ 0.5          & 4    & $2.13\times 10^{-2}$ & 2.0     & $3.7^{+1.2}_{-0.8}$ & 7.6      \\
1200              & ~~[950, 2100]     & ~~6.7 $\pm$ 1.3        & 10   & $1.38\times 10^{-1}$ & 1.1     & ~~$7.0^{+2.5}_{-2.3}$  & 10.2     & 0.9 $\pm$ 0.5          & 6    & $1.65\times 10^{-3}$ & 2.9     & $3.7^{+1.3}_{-0.6}$ & 10.4~~   \\
1300              & [1000, 2200]      & ~~6.1 $\pm$ 1.2        & ~~9  & $1.48\times 10^{-1}$ & 1.0     & ~~$6.5^{+2.9}_{-1.4}$  & ~~9.7    & 0.8 $\pm$ 0.4          & 6    & $5.47\times 10^{-4}$ & 3.3     & $3.5^{+1.2}_{-0.5}$ & 10.3~~   \\
1400              & [1100, 2800]      & ~~5.2 $\pm$ 1.7        & ~~8  & $1.76\times 10^{-1}$ & 0.9     & ~~$6.5^{+2.6}_{-2.0}$  & ~~9.6    & 0.7 $\pm$ 0.4          & 7    & $1.46\times 10^{-4}$ & 3.6     & $3.2^{+1.1}_{-0.1}$ & 11.9~~   \\
1500              & [1150, 2900]      & ~~4.9 $\pm$ 2.4        & ~~7  & $2.41\times 10^{-1}$ & 0.7     & ~~$6.6^{+2.8}_{-1.9}$  & ~~9.3    & 0.6 $\pm$ 0.4          & 6    & $6.09\times 10^{-4}$ & 3.2     & $3.2^{+1.2}_{-0.1}$ & 10.7~~   \\
1600              & [1250, 3400]      & ~~4.2 $\pm$ 3.4        & ~~5  & $3.24\times 10^{-1}$ & 0.5     & ~~$7.0^{+2.9}_{-2.2}$  & ~~8.4    & 0.54 $\pm$ 0.35        & 5    & $1.19\times 10^{-3}$ & 3.0     & $3.1^{+1.2}_{-0.1}$ & 9.5      \\
1800              & [1400, 4250]      & ~~$3^{+4}_{-3}$        & ~~4  & $2.74\times 10^{-1}$ & 0.6     & ~~$7.2^{+2.4}_{-1.5}$  & ~~8.3    & 0.44 $\pm$ 0.32        & 4    & $3.36\times 10^{-3}$ & 2.7     & $3.2^{+1.0}_{-0.1}$ & 8.1      \\
2000              & [1550, 4650]      & ~~$3^{+4}_{-3}$        & ~~3  & $3.14\times 10^{-1}$ & 0.5     & ~~$6.2^{+1.9}_{-2.1}$  & ~~6.9    & 0.36 $\pm$ 0.25        & 3    & $6.96\times 10^{-3}$ & 2.5     & $3.1^{+1.0}_{-0.1}$ & 6.8      \\
2200              & [1650, 5900]      & ~~$2^{+5}_{-2}$        & ~~4  & $2.18\times 10^{-1}$ & 0.8     & ~~$6.0^{+2.4}_{-2.2}$  & ~~8.2    & 0.33 $\pm$ 0.28        & 3    & $8.85\times 10^{-3}$ & 2.4     & $3.0^{+1.1}_{-0.1}$ & 6.8      \\
2400              & [1750, 6300]      & ~~$2^{+4}_{-2}$        & ~~3  & $3.17\times 10^{-1}$ & 0.5     & ~~$5.5^{+1.5}_{-1.5}$  & ~~6.7    & 0.29 $\pm$ 0.28        & 3    & $9.75\times 10^{-3}$ & 2.3     & $3.2^{+0.8}_{-0.0}$ & 6.9      \\
2600              & [1900, 6500]      & ~~$2^{+4}_{-2}$        & ~~1  & $5.00\times 10^{-1}$ & 0.0     & ~~$4.9^{+2.0}_{-1.6}$  & ~~4.0    & $0.25^{+0.31}_{-0.25}$~ & 3    & $9.71\times 10^{-3}$ & 2.3     & $3.1^{+0.7}_{-0.0}$ & 6.9      \\
2800              & [2000, 6700]      & \,~$1.5^{+3.1}_{-1.5}$ & ~~1  & $5.00\times 10^{-1}$ & 0.0     & ~~$4.5^{+2.1}_{-1.4}$  & ~~4.2    & $0.2^{+0.4}_{-0.2}$~    & 1    & $1.07\times 10^{-1}$ & 1.2     & $3.0^{+0.6}_{-0.0}$ & 4.0      \\
3000              & [2100, 6700]      & \,~$1.4^{+3.1}_{-1.4}$ & ~~1  & $5.00\times 10^{-1}$ & 0.0     & ~~$4.4^{+1.4}_{-1.1}$  & ~~4.3    & $0.2^{+0.4}_{-0.2}$~    & 1    & $9.43\times 10^{-2}$ & 1.3     & $2.9^{+0.4}_{-0.0}$ & 4.1      \\
\hline
\hline
\end{tabular}
\label{tbl:observed_p0_inclusive}
\end{table}

\begin{figure}[tbp]
\begin{center}
\subfigure[\texttt SR-Inclusive\_Low]{
\label{fig:SR_observed-inclusiveLow}
\includegraphics[width=0.6\textwidth]{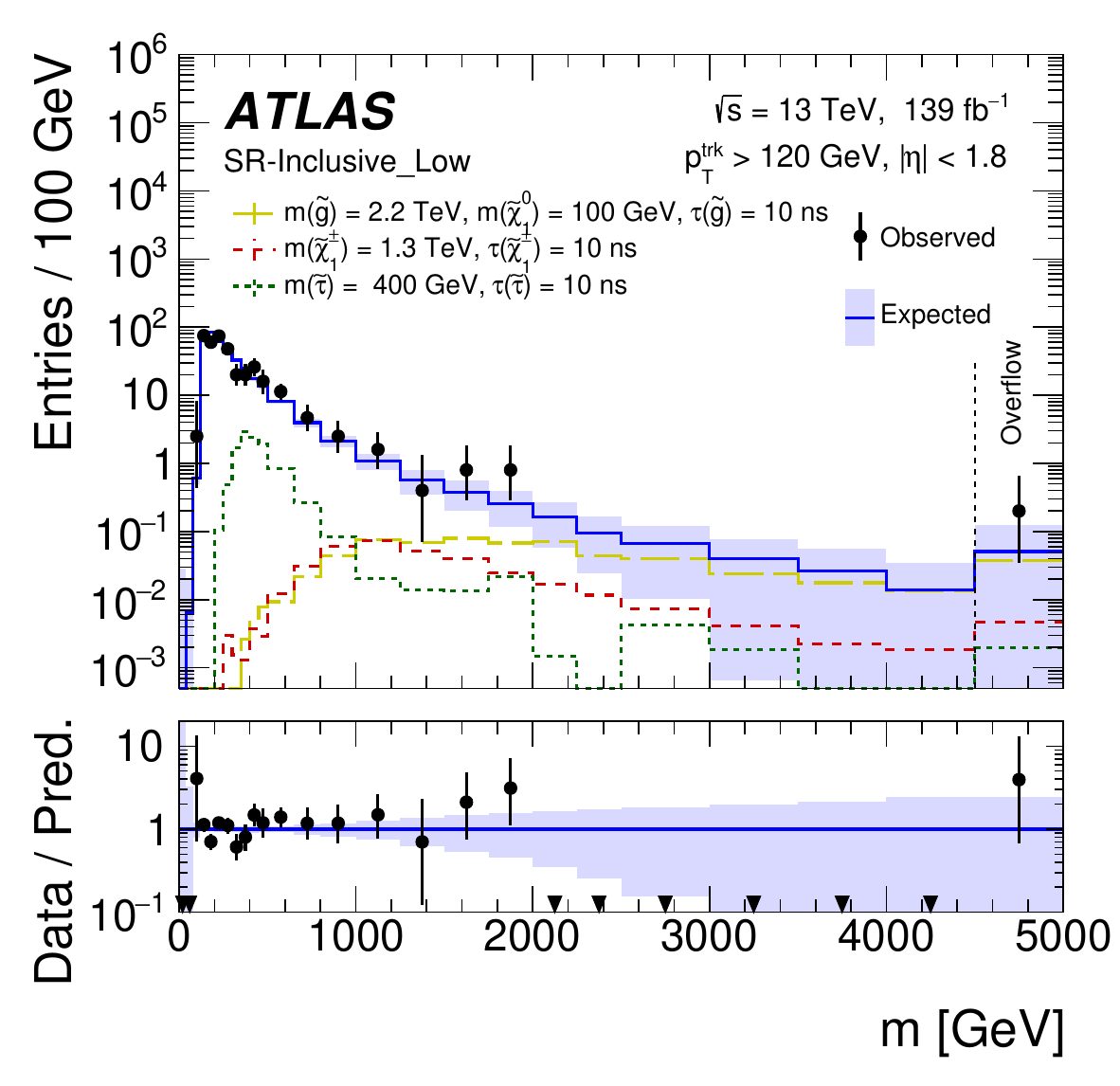}
}
\subfigure[\texttt SR-Inclusive\_High]{
\label{fig:SR_observed-inclusiveHigh}
\includegraphics[width=0.6\textwidth]{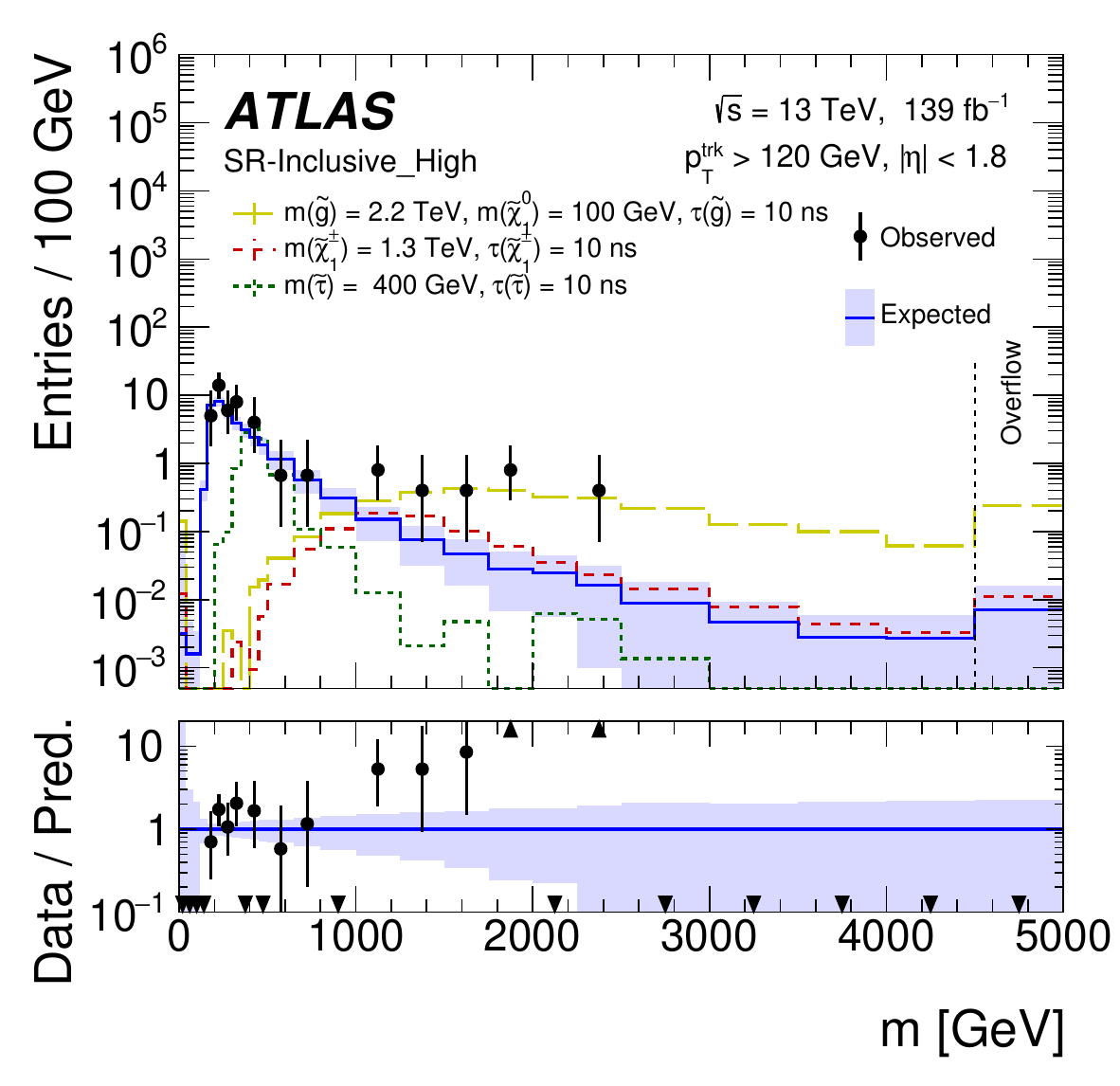}
}
\caption{The observed mass distribution in the \texttt{Inclusive} signal-region bins. The band on the expected background indicates the total uncertainty of the estimation. Several representative signal models are overlaid. Events outside the shown range are accumulated in the rightmost bin indicated as `Overflow'. Downward triangle markers at the bottom of the panels indicate that no events are observed in the corresponding mass bin, while the upward triangle markers at the top of the lower panel in (b) indicate that the observed data is beyond the range. The excess of events seen in the lower panel is further discussed in the text.}
\label{fig:SR_observed-inclusive}
\end{center}
\end{figure}

\begin{figure}[tp]
\begin{center}
\subfigure[\texttt SR-Inclusive\_Low]{
\includegraphics[width=0.38\textwidth]{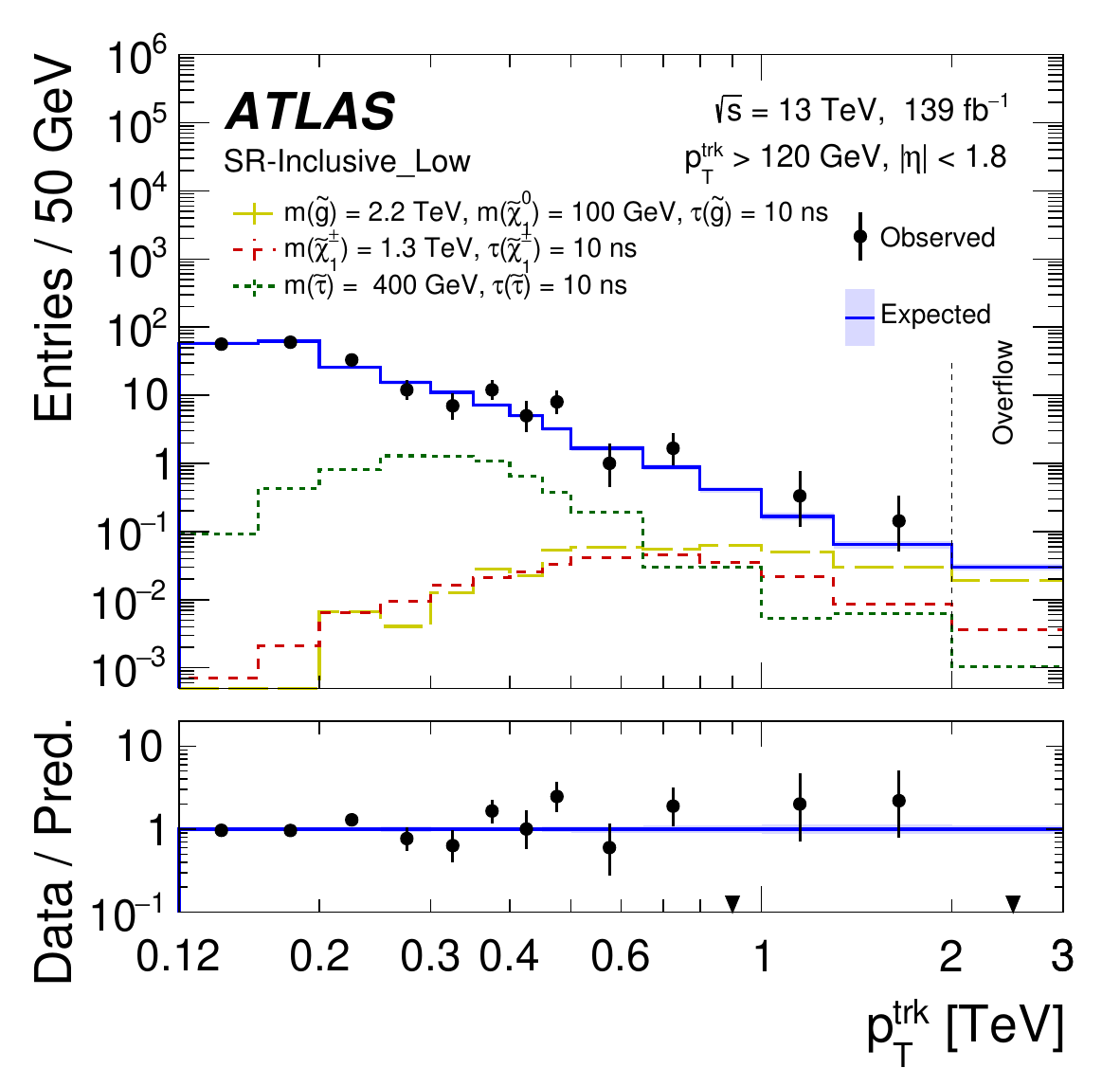}
}
\subfigure[\texttt SR-Inclusive\_High]{
\includegraphics[width=0.38\textwidth]{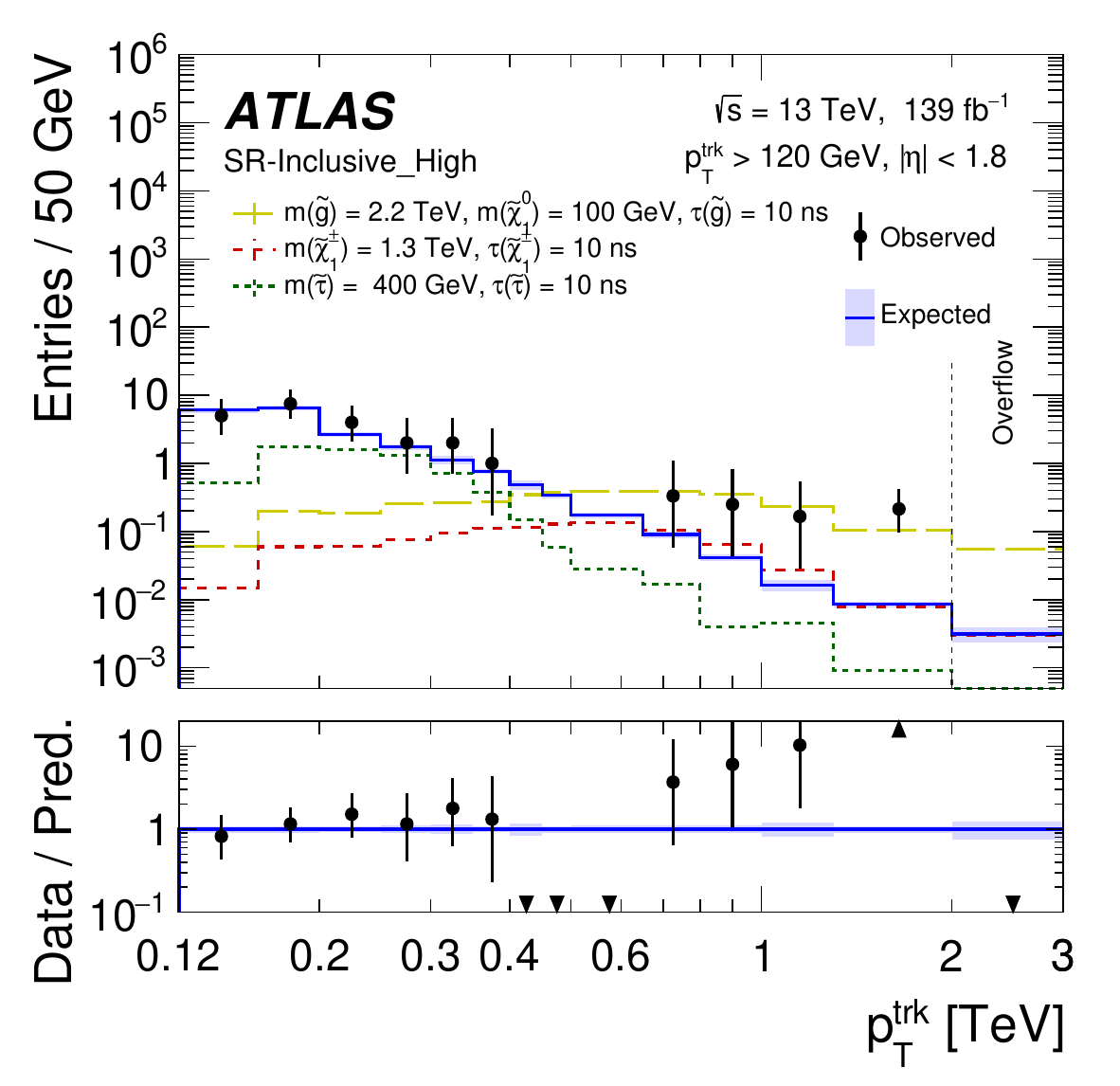}
}
\subfigure[\texttt SR-Inclusive\_Low]{
\includegraphics[width=0.38\textwidth]{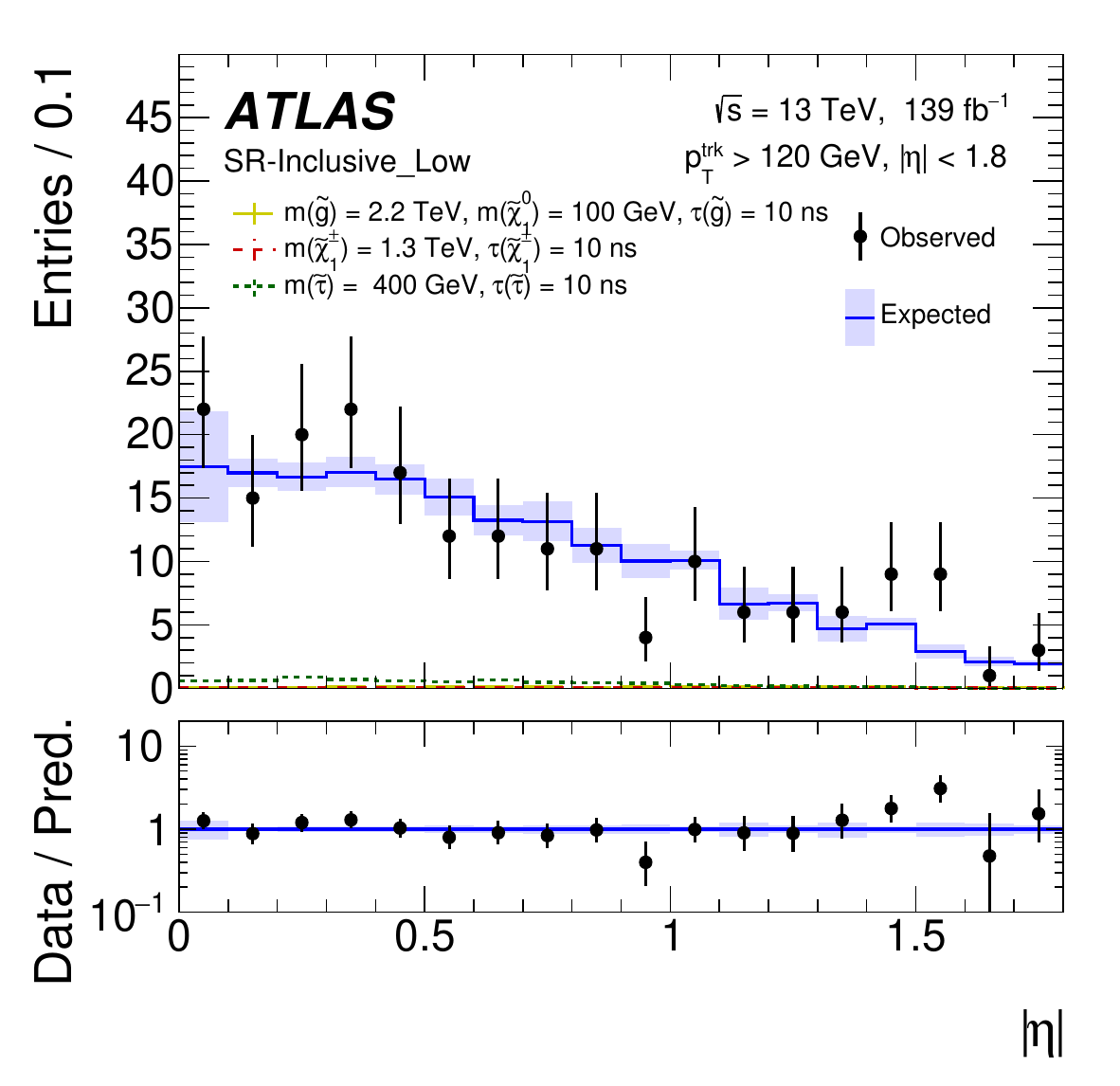}
}
\subfigure[\texttt SR-Inclusive\_High]{
\includegraphics[width=0.38\textwidth]{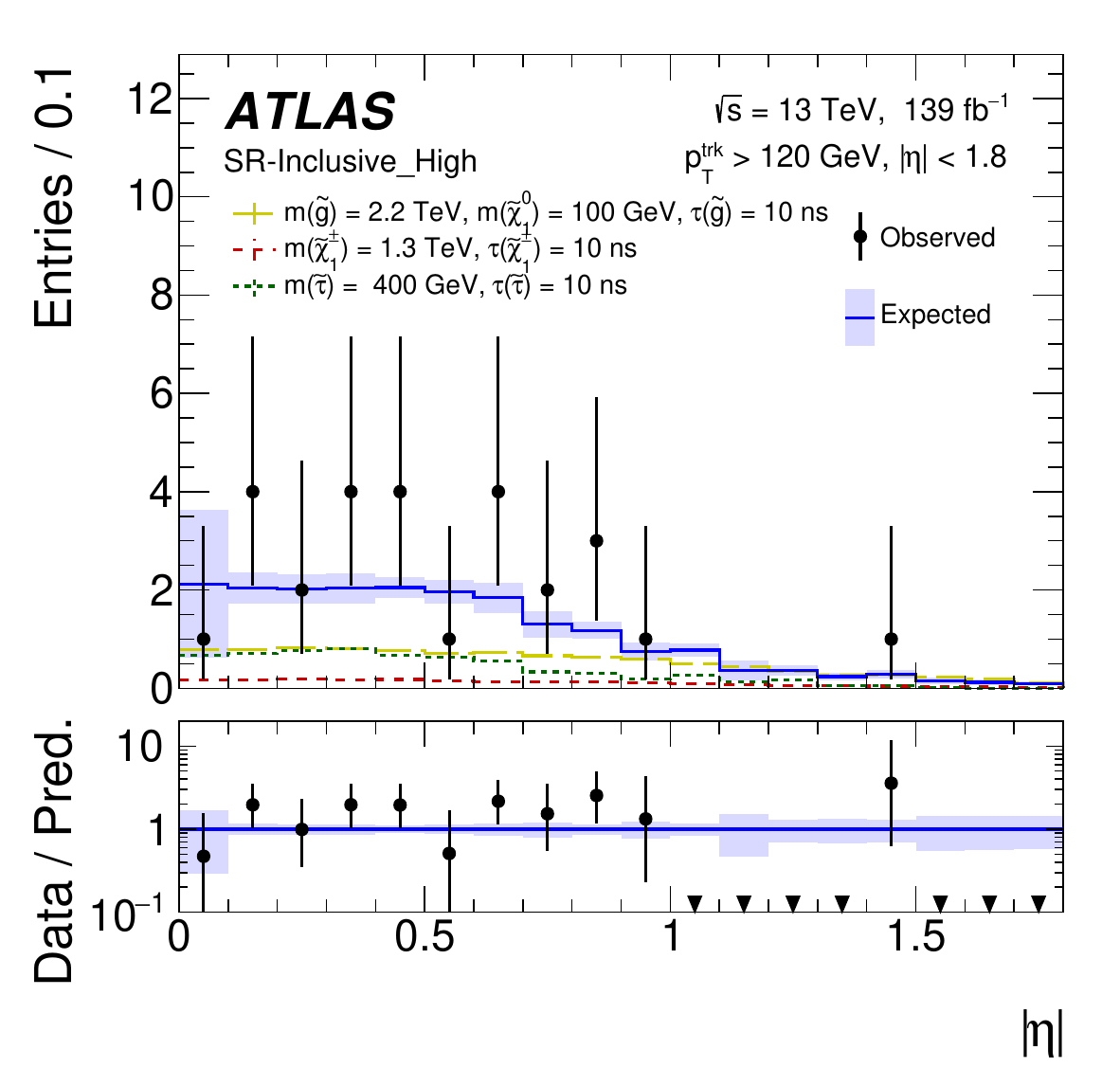}
}
\subfigure[\texttt SR-Inclusive\_Low]{
\includegraphics[width=0.38\textwidth]{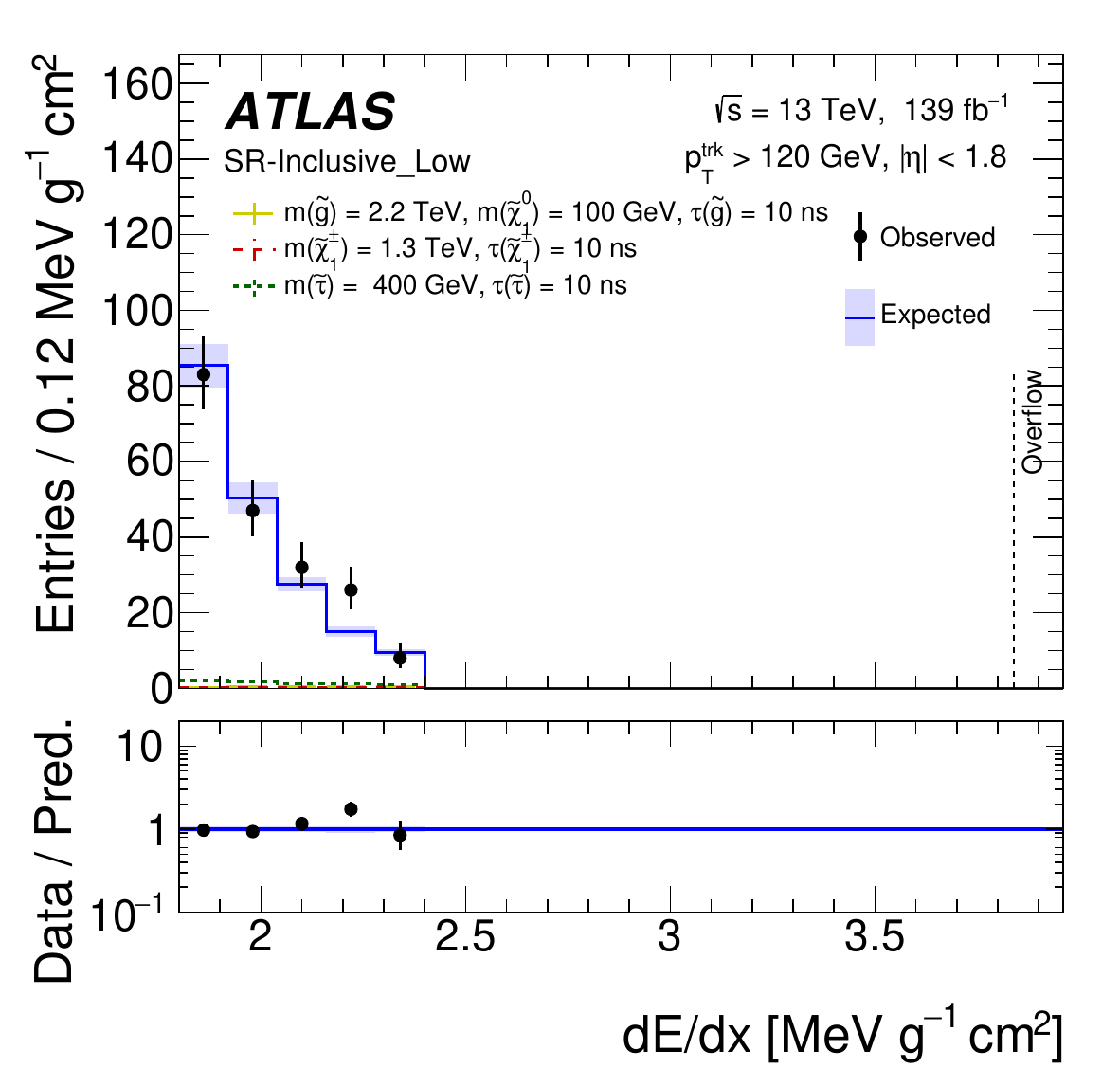}
}
\subfigure[\texttt SR-Inclusive\_High]{
\includegraphics[width=0.38\textwidth]{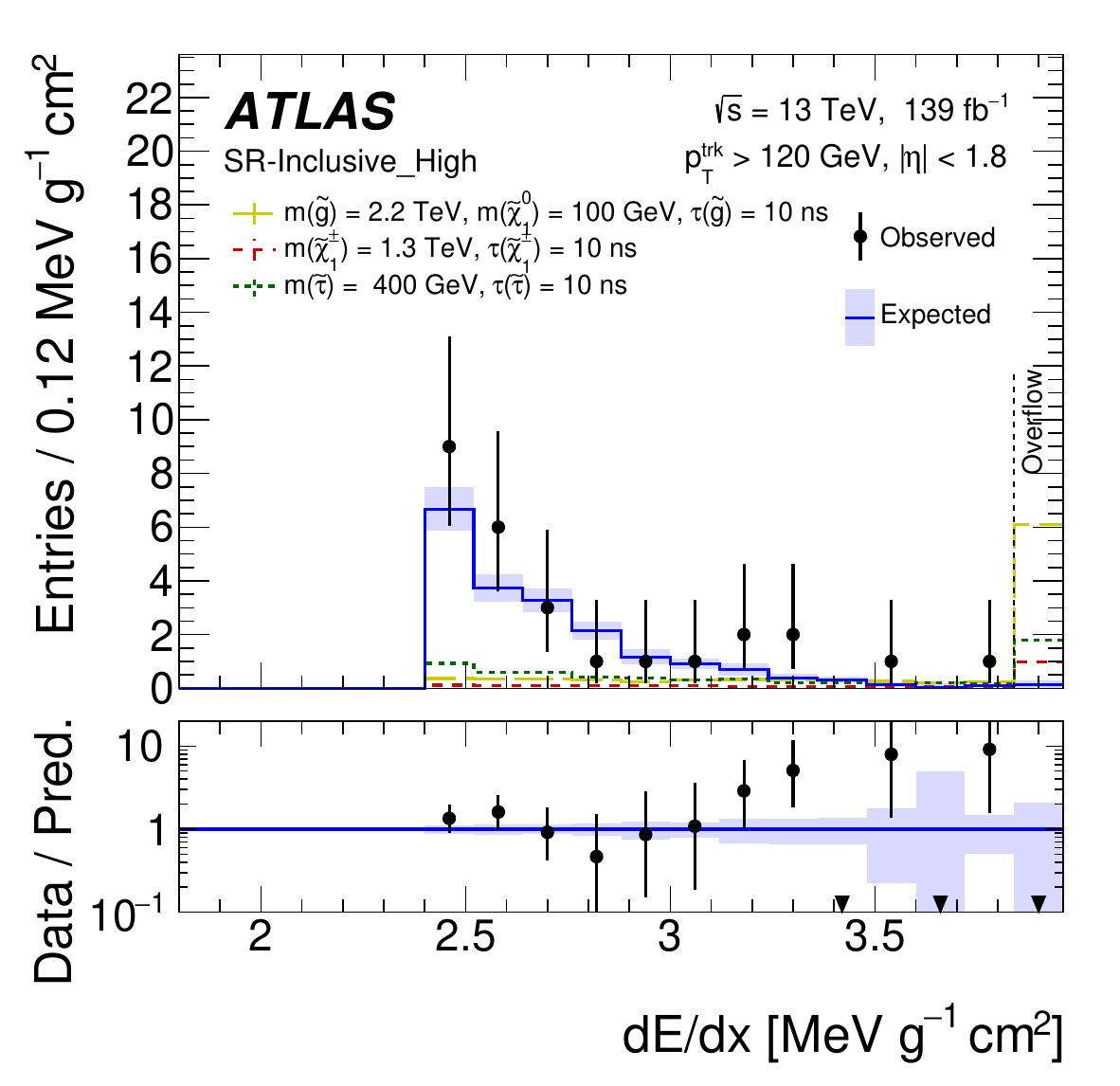}
}
\caption{The observed $\pt$, $|\eta|$ and $\dEdx$ distributions in the \texttt{Inclusive} signal-region bins. The band on the expected background indicates the total uncertainty of the estimation. Several representative signal models are overlaid. Events outside the shown range are accumulated in the rightmost bin indicated as `Overflow'. Downward triangle markers at the bottom of the panels indicate that no events are observed in the corresponding bin, while the upward triangle marker at the top of the lower panel in (b) indicates that the observed data is beyond the range.}
\label{fig:SR_observed-inclusive_others}
\end{center}
\end{figure}

\section{Results}
\label{sec:results}

In total, 223 events are observed in the signal region combining all exclusive bins. The observation in each mass window (defined for each hypothetic particle mass) for the \texttt{Inclusive} signal-region bins is presented in Table~\ref{tbl:observed_p0_inclusive}, and in total 196 (27) events are observed in the \texttt{SR-Inclusive\_Low} (\texttt{SR-Inclusive\_High}) bin over the full mass range (see Table~\ref{tbl:SR_bins}). The mass, $\pt$, $|\eta|$ and $\dEdx$ distributions in each of the \texttt{SR-Inclusive\_Low} and \texttt{SR-Inclusive\_High} bins are shown in Figures~\ref{fig:SR_observed-inclusive} and~\ref{fig:SR_observed-inclusive_others}, compared with the expected background yields and several representative signal points. Overall, the observed event distribution agrees well with the expected background distribution in these variables within the estimated uncertainties, except in the \texttt{SR-Inclusive\_High} bin in the high mass range.
 
The statistical analysis and likelihood construction were implemented in the \texttt{pyhf} software framework~\cite{pyhf}. For each mass window, the likelihood of the background-only hypothesis given the observed data was constructed from the background prediction and the associated systematic uncertainties. The effect of the systematic uncertainties is incorporated through nuisance parameters which are constrained to be Gaussian-distributed. Using a profile-likelihood-based test statistic~\cite{Cowan_2011}, independent $p_{0}$-values quantifying the level of agreement between the observed data and the background prediction were calculated for each of these windows. The lowest $p_{0}$-value of $1.46\times10^{-4}$ is associated with the \texttt{SR-Inclusive\_High} bin in the mass window of $[1100,2800]~\GeV$, corresponding to a target mass of $1.4~\TeV$ in the `long' lifetime regime. In this window, $0.7\pm 0.4$ events are expected and seven events are observed, and the $p_{0}$-value corresponds to a local $Z$ significance\footnote{By convention, the $Z$ significance of a $p_{0}$-value is defined as $Z\equiv\Phi^{-1}(1-p_{0})$ where $\Phi^{-1}$ is the inverse of the cumulative distribution for a unit Gaussian function.} of $3.6$.
The mild excess reported in Ref.~\cite{SUSY-2016-31} at around $600~\GeV$ is not confirmed in the dataset analysed here: the maximum local $Z$ significance observed in that region is $1.8$ for a target mass of $500~\GeV$ in the \texttt{SR-Inclusive\_Low} bin in the `short' lifetime regime.
 
After taking into account the look-elsewhere effect~\cite{Gross2010,ATL-PHYS-PUB-2011-011}, the corresponding global $Z$ significance for the $1.4~\TeV$ target mass in the \texttt{SR-Inclusive\_High} bin is evaluated to be approximately $3.3$ by using pseudo-experiments, including all `long' and `short' lifetime windows. The pseudo-experiments generate event distributions in the signal-region bins according to the background estimate, assuming the systematic uncertainty is correlated over all masses. In total, one million pseudo-experiments were generated per signal-region bin to evaluate the global significance. Reflecting the relatively broad mass resolution above $1~\TeV$, the look-elsewhere effect is found to be mild, consistent with the relatively small difference between the local and global $Z$ significances.

Events in the mass window with the lowest $p_{0}$-value were examined individually. Out of these seven events, four are in the \texttt{SR-Mu} category and have no IBL overflow, two of the remaining three events are in the \texttt{SR-Trk} category and have an IBL overflow, and the last event is in the \texttt{SR-Trk} category and has no IBL overflow. One of the three events in the \texttt{SR-Trk} category has a matched muon which does not satisfy the identification criterion applied in this analysis. Detailed pixel cluster information was extracted, and no obvious pathologies or instrumental data-quality problems were found. In particular, no signatures of multiple densely clustered particles, which could lead to poor track reconstruction, were found around the candidate tracks. Candidate tracks are well isolated both at the track level and at the calorimeter cluster level, as required by the signal selection. For each of the four \texttt{SR-Mu} category events and the one \texttt{SR-Trk} event that has a matched muon, the track momentum when reconstructed as a muon using both the inner detector and the muon spectrometer was compared with the nominal momentum obtained using only the inner detector. Allowing for the momentum resolution, the two reconstructed momentum values are compatible with each other for all five events.
 
The event topology of these excess events typically contains a counterbalancing jet, opposite in $\phi$ to the candidate signal track. Such a `back-to-back' topology is typical in the \texttt{CR-kin} region at high-\pt values above several hundred \GeV. One event lacks a counterbalancing jet, but a balancing amount of \MET is present. Overall, no obvious pathologies or instrumental issues were found.

The \dEdx values of the excess candidate tracks span $2.42\mbox{--}3.72~\dEdxunit$, and they map to particle speeds of $\beta=\text{0.62--0.52}$ according to the $\dEdx\mbox{--}\beta\gamma$ relationship in Eq.~(\ref{eq:mass_calib}), which assumes the given \dEdx value is the MPV. This range of $\beta$ should result in a significantly longer time-of-flight through the ATLAS detector than for $\beta=1$ SM particles, and this is used as a consistency test. As in the searches for stable charged particles~\cite{SUSY-2011-03}, two independent $\beta$ values were examined: one is the $\beta$ value measured by the muon spectrometer ($\beta_\mathrm{MS}$) when fitting muon tracks with $\beta$ as a free parameter; the other is the $\beta$ value obtained from calorimeter cell hits associated with the candidate track ($\beta_\mathrm{calo}$). In the latter case, $\beta$ is derived from the average ToF weighted by the timing resolution of the cells, which depends on the size of the energy deposit. The probability distributions of these two $\beta$ variables for the $\beta=1$ SM particles are modelled from the \texttt{CR-kin} dataset. Both $\beta$ probability distributions exhibit non-Gaussianity with approximately symmetric side-lobes. The FWHM of the peak divided by 2.35 for $\beta_\mathrm{MS}$ is 0.045, while it is 0.075 and 0.050 for $\beta_\mathrm{calo}$ in the \texttt{CR-kin-Mu} and \texttt{CR-kin-Trk} samples, respectively. The efficiency of obtaining a $\beta_\mathrm{MS}$ value from the \texttt{CR-kin-Mu} sample is 95\%, while that of obtaining a $\beta_\mathrm{calo}$ value is 85\% from  the \texttt{CR-kin-Mu} sample and 95\% for the \texttt{CR-kin-Trk} sample. Using MC signal samples, where particles have low $\beta$, it was found that the $\beta$ values from the ToF observables and the $\beta$ value deduced from the pixel \dEdx measurement agree within 6\%.
 
The $\beta$ values measured by ToF of the seven candidate tracks were all found to be compatible with $\beta=1$, with all the $\beta_\mathrm{MS}$ and $\beta_\mathrm{calo}$ values being well within the 95\% confidence interval of the distribution.
Therefore, the low particle speed suggested by the pixel \dEdx measurement for the seven candidate tracks in the excess was not confirmed by these ToF observables.
 
\begin{figure}[tp]
\begin{center}
\subfigure[\texttt SR-Trk\_IBL0\_Low]{
 
\includegraphics[width=0.38\textwidth]{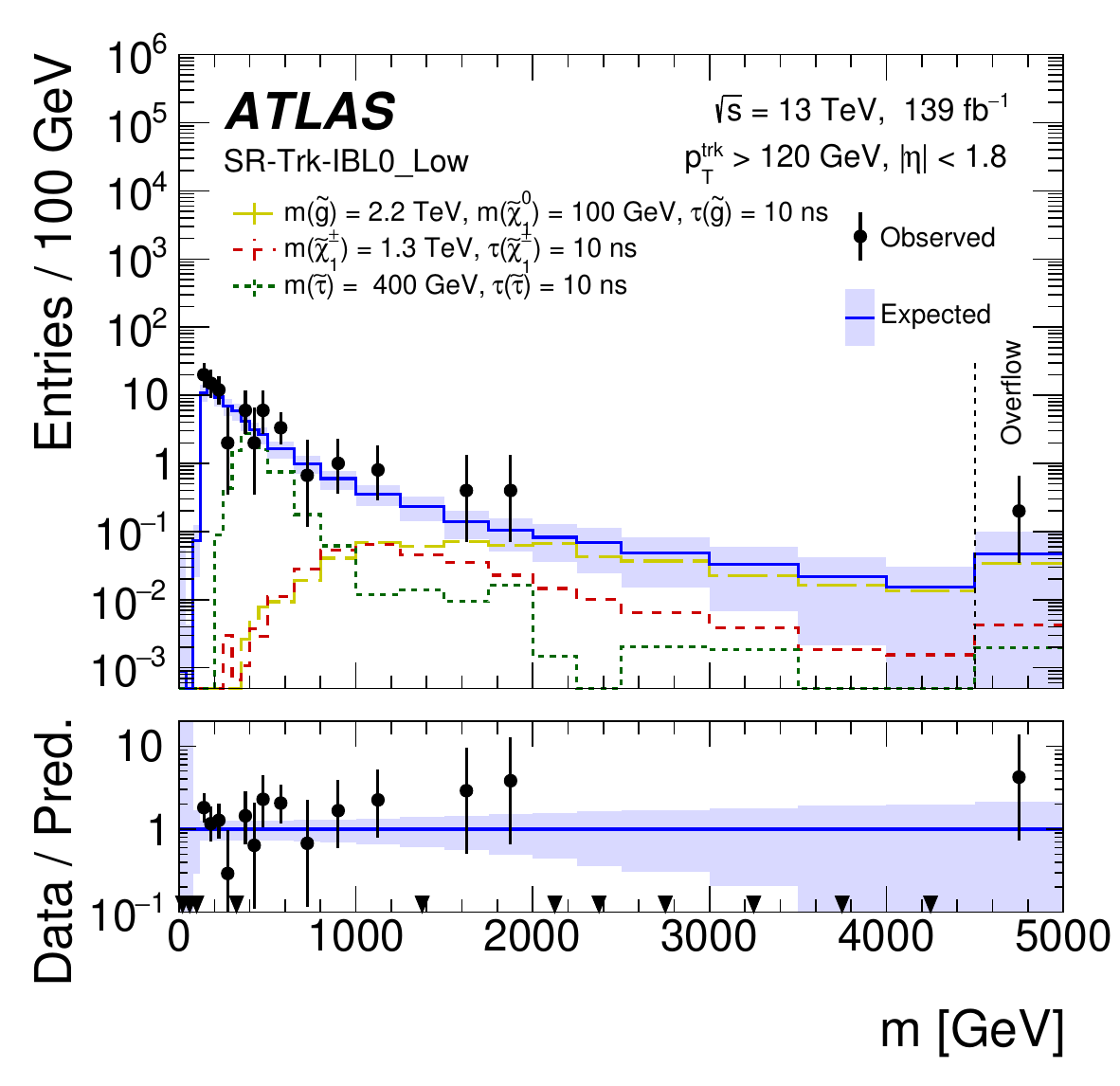}
}
\subfigure[\texttt SR-Mu\_IBL0\_Low]{
\includegraphics[width=0.38\textwidth]{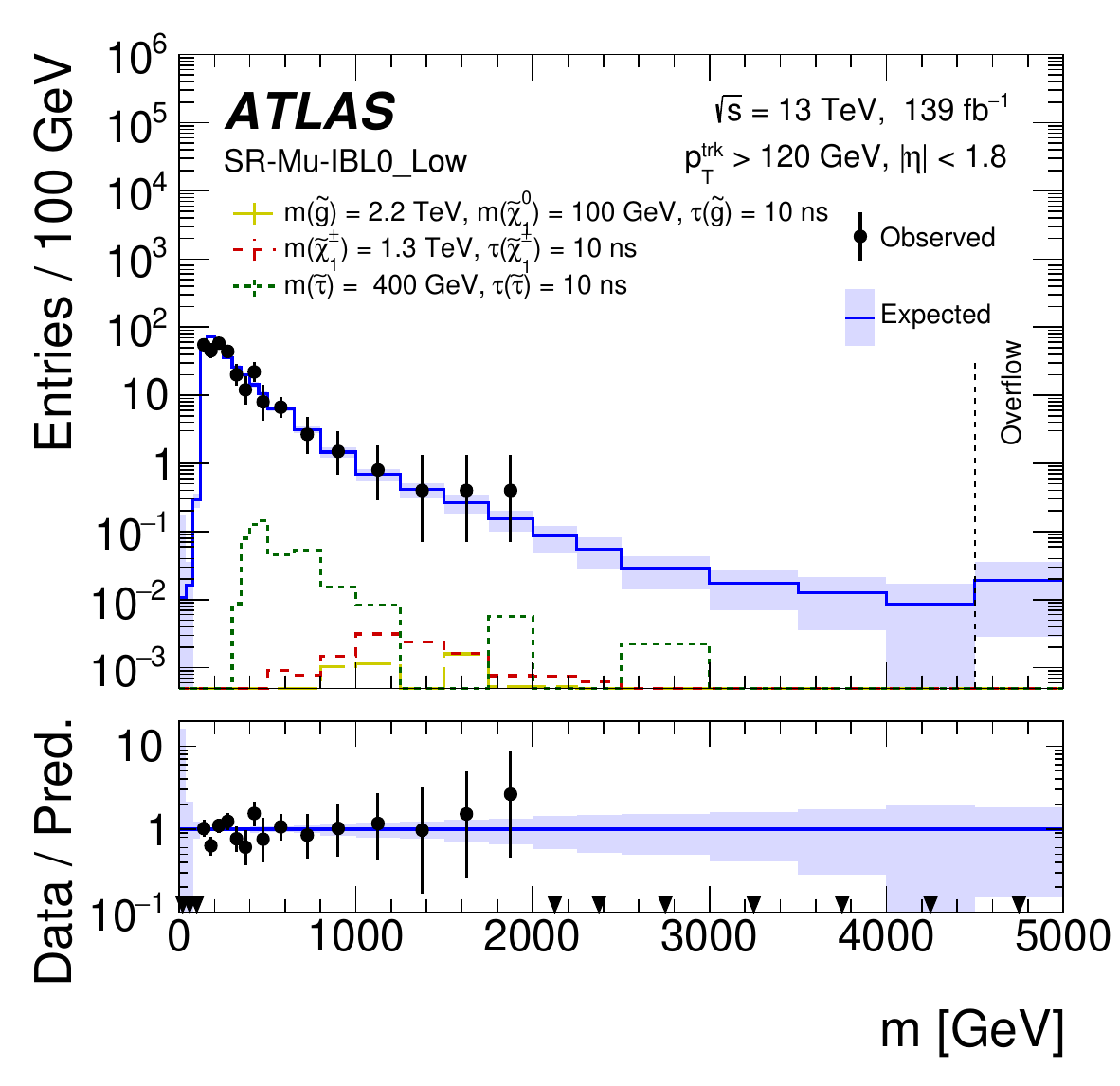}
}
\subfigure[\texttt SR-Trk\_IBL0\_High]{
\includegraphics[width=0.38\textwidth]{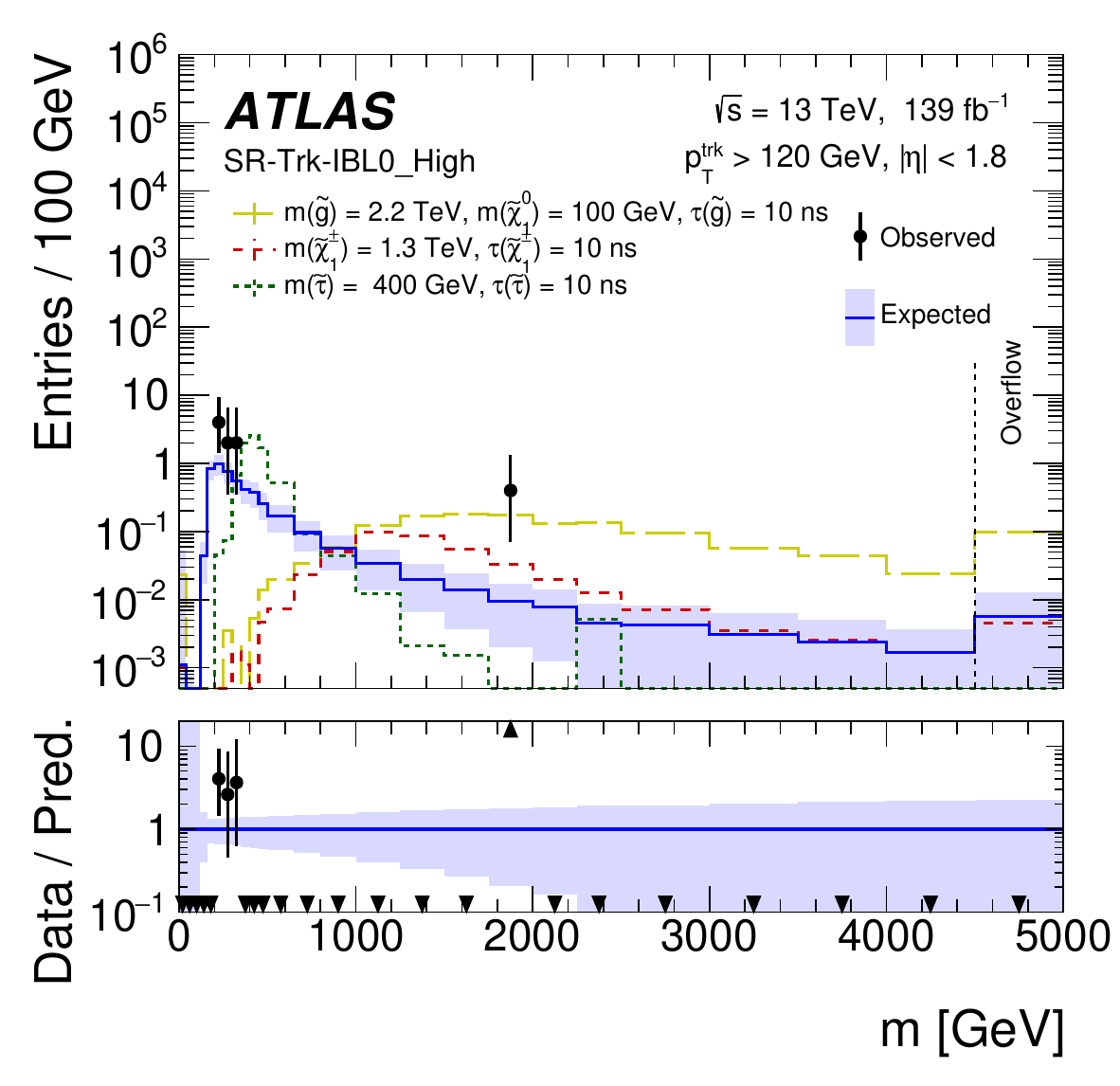}
}
\subfigure[\texttt SR-Mu\_IBL0\_High]{
\includegraphics[width=0.38\textwidth]{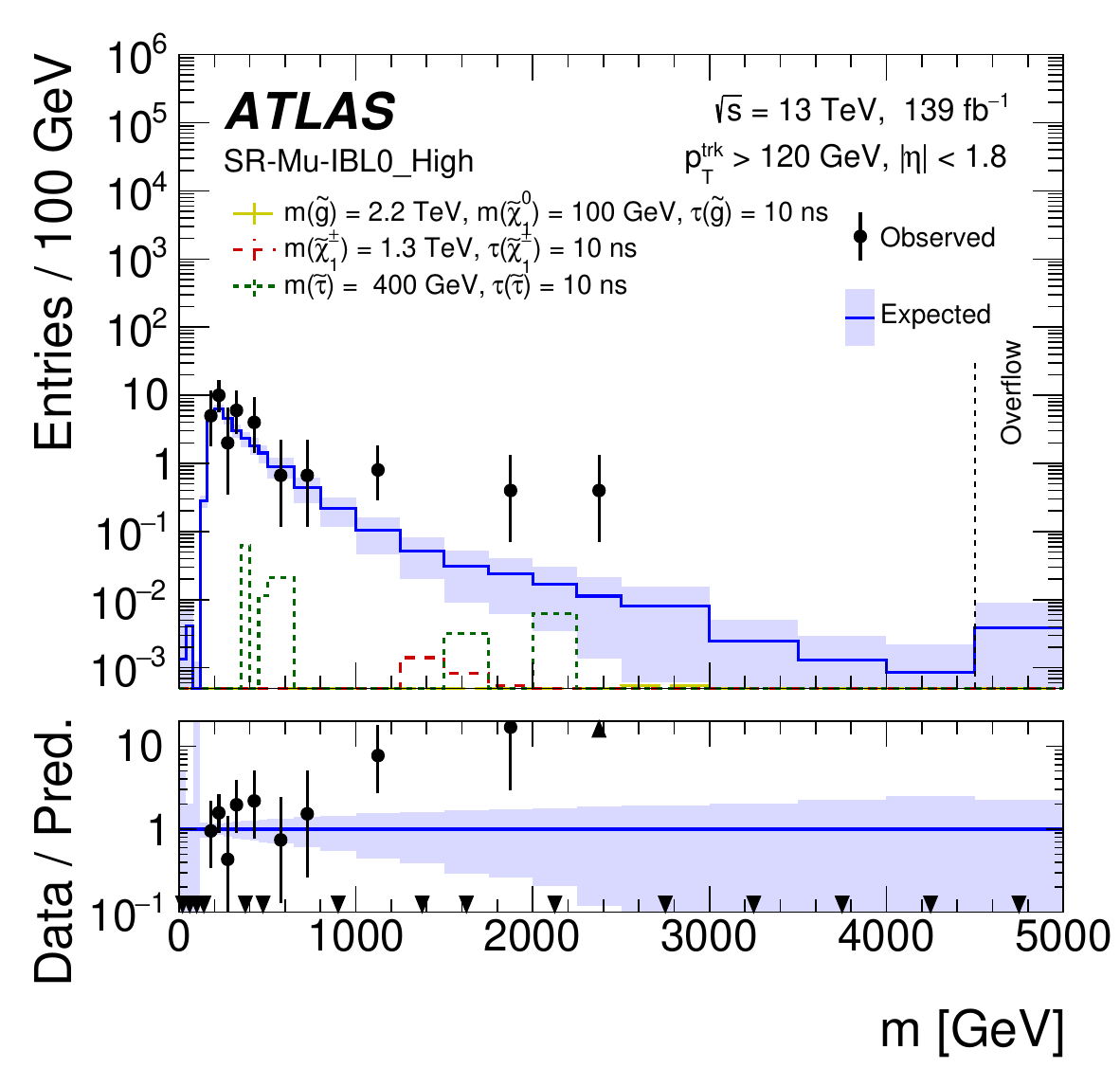}
}
\subfigure[\texttt SR-Trk\_IBL1]{
\includegraphics[width=0.38\textwidth]{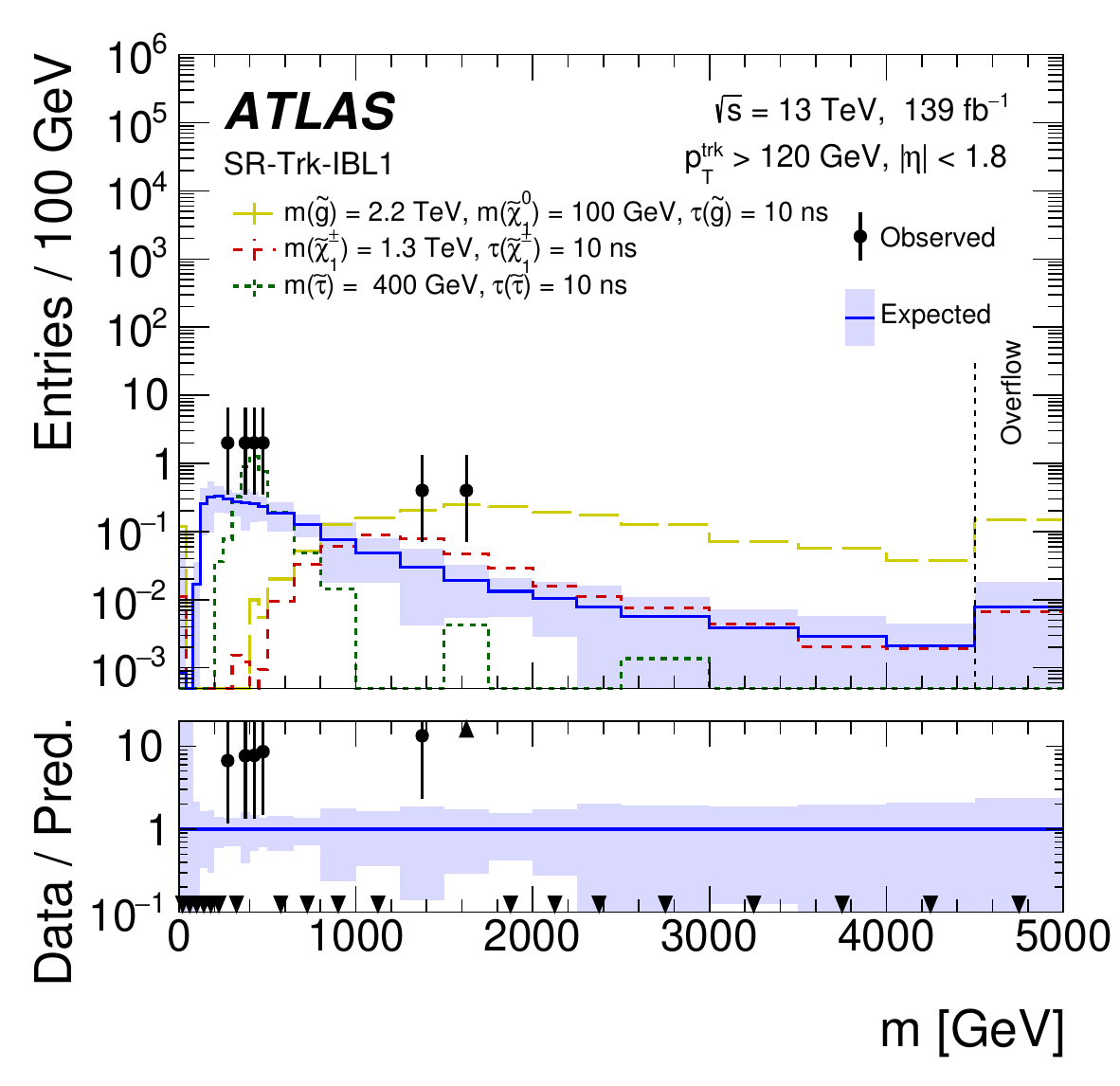}
}
\subfigure[\texttt SR-Mu\_IBL1]{
\includegraphics[width=0.38\textwidth]{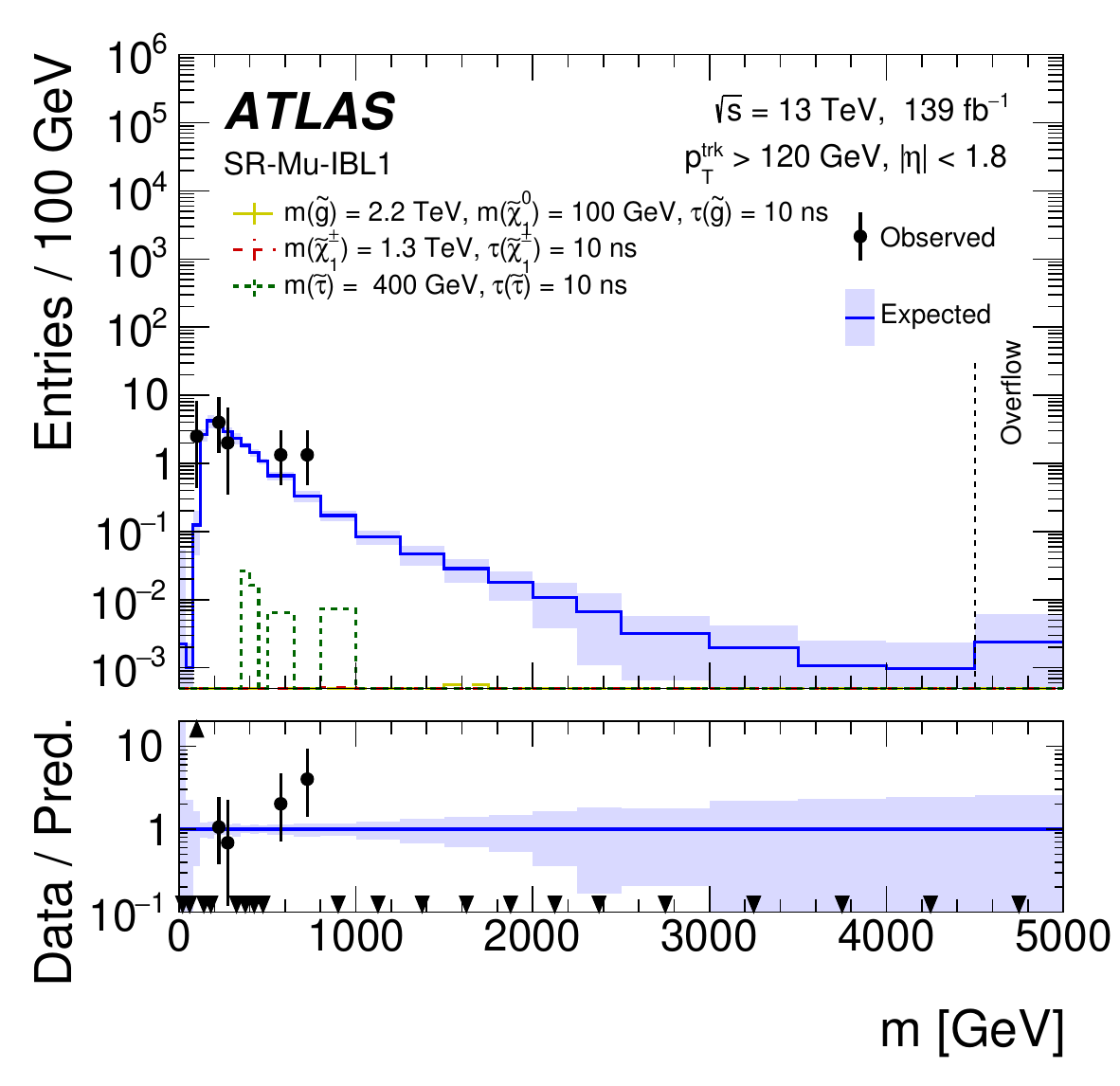}
}
\caption{The observed mass distribution in the \texttt{Trk} and \texttt{Mu} signal-region bins.  The band on the expected background indicates the total uncertainty of the estimation. Several representative signal models are overlaid. Events outside the shown range are accumulated in the rightmost bin indicated as `Overflow'. Downward triangle markers at the bottom of the panels indicate that no events are observed in the corresponding mass bin, while upward triangle markers in the lower panels indicate that the observed data is beyond the range.}
\label{fig:SR_observed}
\end{center}
\end{figure}

The results of this study are interpreted for the benchmark signal models considered, and the 95\% CL upper limit on the cross-section is extracted using the $\text{CL}_\text{s}$ prescription~\cite{Read:2002hq} for each signal mass and lifetime hypothesis, using a simultaneous fit of the six exclusive signal-region bins listed in Table~\ref{tbl:SR_bins}. These six signal-region bins are exclusive at the track level. The cross-section limit is derived by combining the six track-based regions as independent observations, with the exception of correlations through systematic uncertainties. According to simulation, the probability of having multiple tracks from the same event entering different bins is small and its impact is insignificant compared to the experimental uncertainty. The mass spectra of these exclusive bins are presented in Figure~\ref{fig:SR_observed}. When both the `short' and `long' lifetime windows are available, the window which provides the better expected limit is selected.
 
To more accurately probe the sensitivity of the analysis to LLP lifetimes other than those used in the generation of the signal samples, the same samples are reinterpreted for intermediate lifetime values by reweighting the LLP particle decay spectra. Intermediate lifetimes are modelled by reweighting the closest longer-lifetime sample to shorter lifetimes, except for
$\tau>30~\text{ns}$. The choice of target lifetimes for $\tau>30~\text{ns}$ is limited by the reduced size of the reweighted sample.

\begin{figure}[tbp]
\begin{center}
\subfigure[$R$-hadron, $m(\ninoone)=100~\GeV$]{
\label{fig:rhadron-limits-normal}
\includegraphics[width=0.47\textwidth]{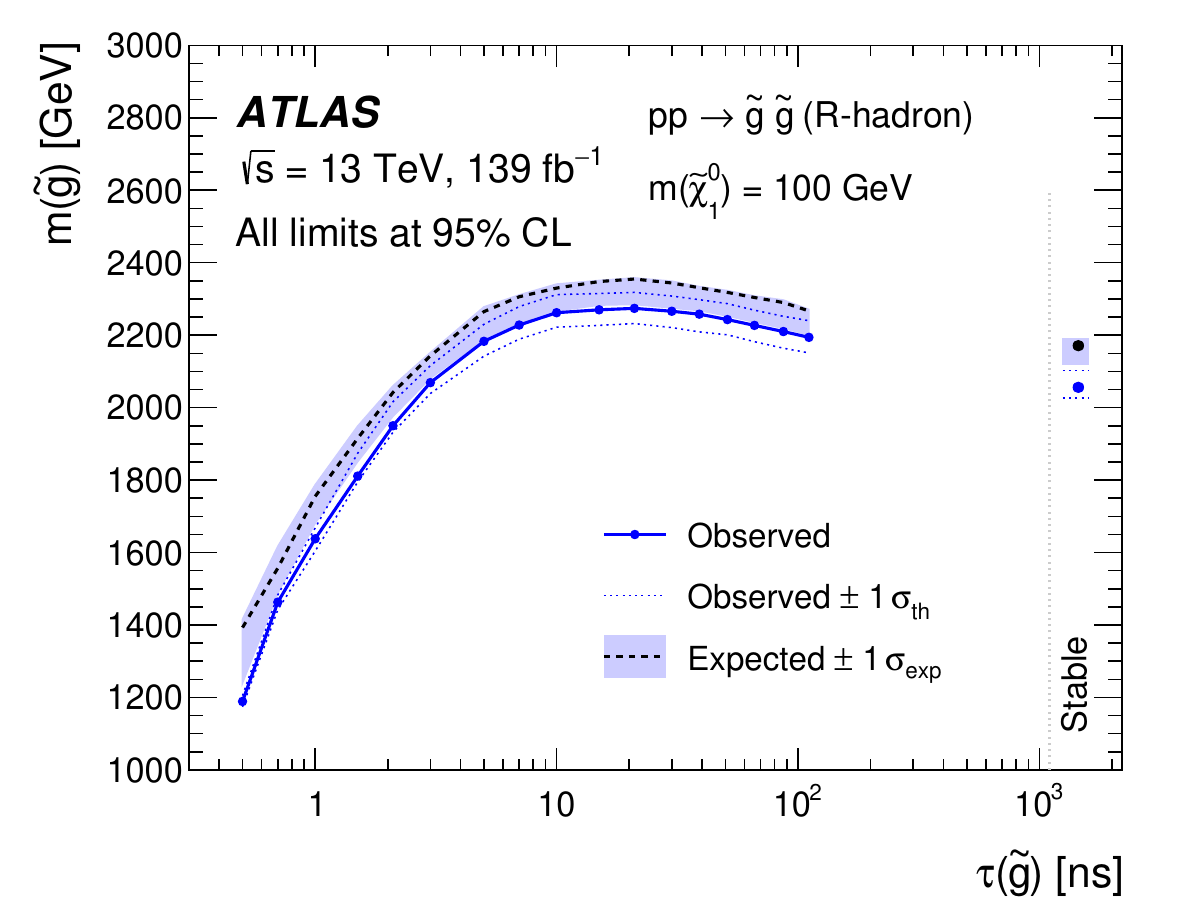}
}
\subfigure[$R$-hadron, $\Delta m(\gluino,\ninoone)=30~\GeV$]{
\label{fig:rhadron-limits-compressed}
\includegraphics[width=0.47\textwidth]{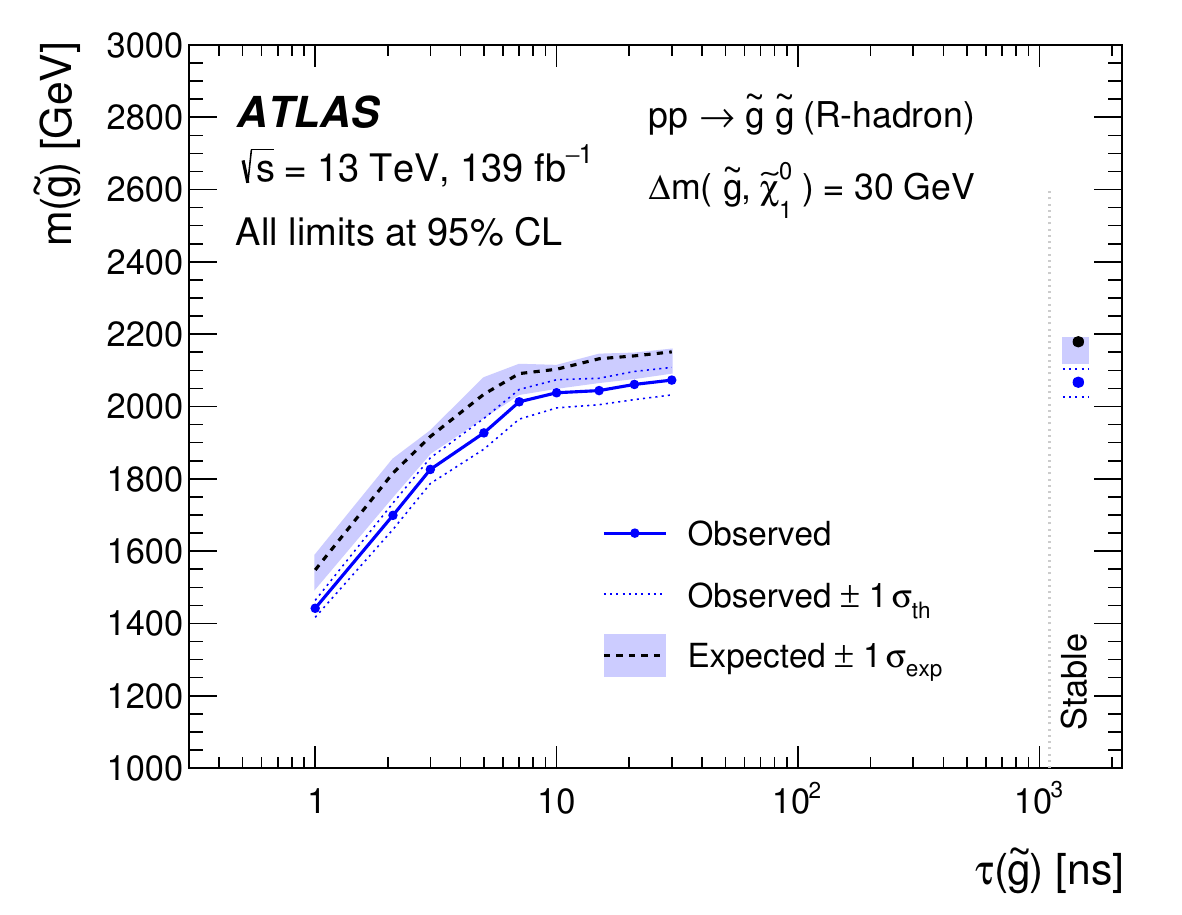}
}
\caption{Lower limits on the gluino mass, from gluino $R$-hadron pair production, as a function of gluino lifetime for two neutralino mass assumptions of \subref{fig:rhadron-limits-normal} $m(\ninoone)=100~\GeV$ and \subref{fig:rhadron-limits-compressed} $\Delta m(\gluino,\ninoone)=30~\GeV$. The upper 1~$\sigma_\text{exp}$ expected bound is very close to the expected limit for some lifetime values due to the expected background getting very close to 0 events.}
\label{fig:rhadron-limits}
\end{center}
\end{figure}

\begin{figure}[tbp]
\begin{center}
\subfigure[Chargino]{
\label{fig:chargino-limits}
\includegraphics[width=0.47\textwidth]{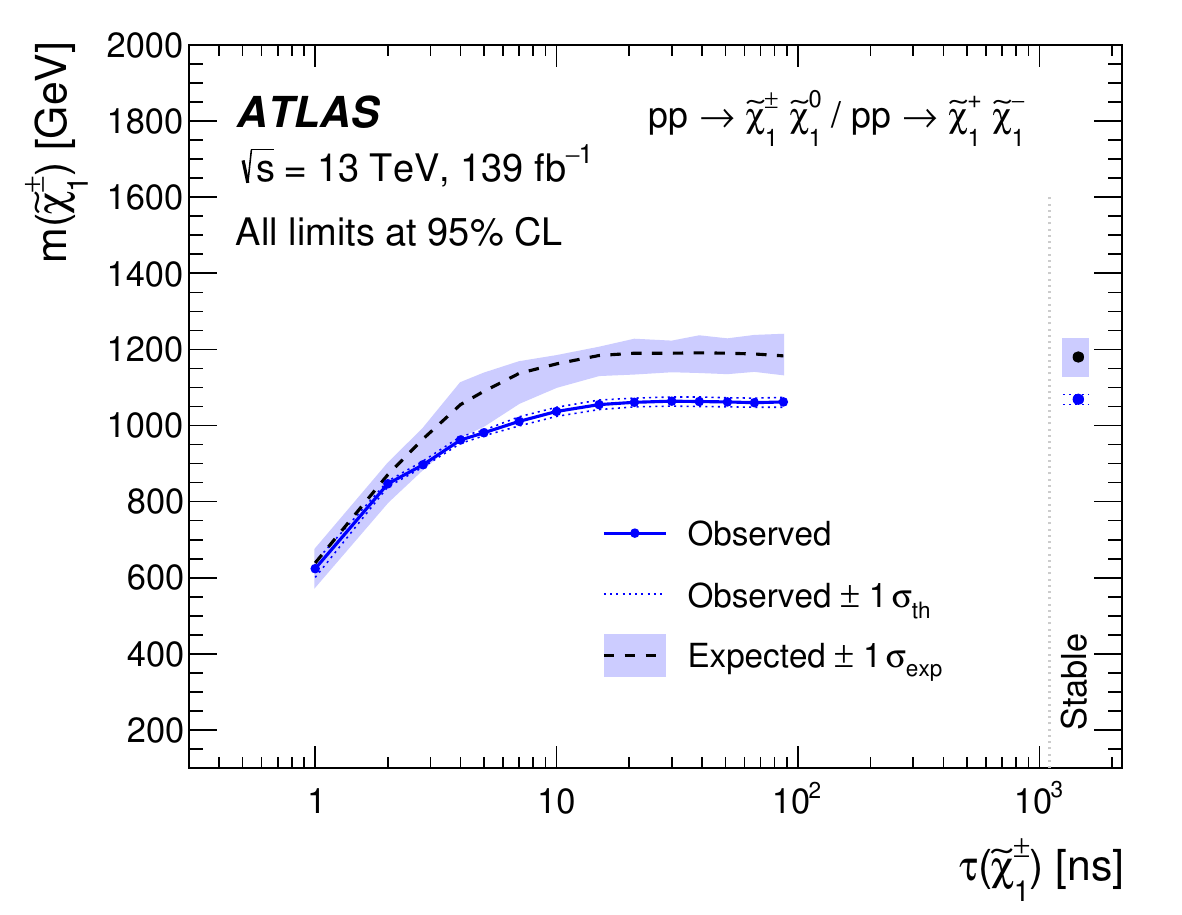}
}
\subfigure[Stau]{
\label{fig:stau-limits}
\includegraphics[width=0.47\textwidth]{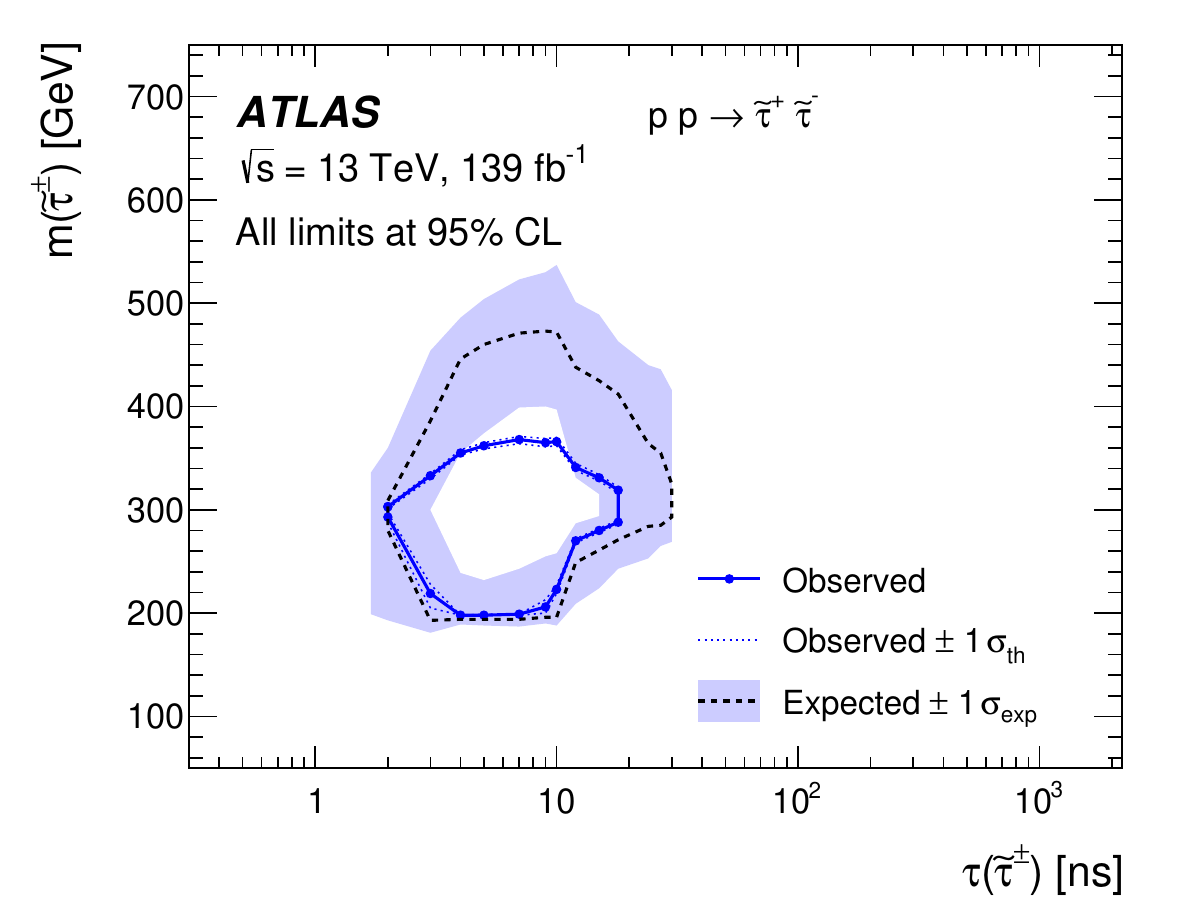}
}
\caption{\subref{fig:chargino-limits} Lower limits on the chargino mass as a function of lifetime, and \subref{fig:stau-limits} the contours around the excluded mass--lifetime region for stau pair production.}
\label{fig:electroweak-limits}
\end{center}
\end{figure}

Figure~\ref{fig:rhadron-limits} show the mass limits for gluino $R$-hadron pair production for both the $m(\ninoone) = 100~\GeV$ and $\Delta m(\gluino,\ninoone)=30~\GeV$ cases. The highest observed lower limit on the mass is $2.27~\TeV$ ($2.06~\TeV$) and is obtained at $\tau=20~\text{ns}$ ($\tau=30~\text{ns}$) for $m(\ninoone) = 100~\GeV$ $\left(\Delta m(\gluino,\ninoone)=30~\GeV\right)$, while the corresponding expected limit is $2.35~\TeV$ ($2.14~\TeV$). Similarly, the limits in the chargino and stau scenarios are shown in Figure~\ref{fig:chargino-limits} and Figure~\ref{fig:stau-limits}, respectively. The highest observed mass limit of $1.07~\TeV$ is obtained at $\tau=30~\text{ns}$ for the chargino production model, while the corresponding expected limit is $1.19~\TeV$. For the stau case, the mass range $220\mbox{--}360~\GeV$ is excluded for $\tau=10~\text{ns}$, while the corresponding expected exclusion is $200\mbox{--}470~\GeV$. Masses below the excluded mass range are not excluded by this search because of acceptance losses mainly due to the $\pt$ selection. These results provide the most stringent limits on these scenarios in the given lifetime range to date. Due to the observation of more events than expected at high masses, the observed limits are weaker than the expected limits for LLPs with mass $m\gtrsim 1~\TeV$.
 
\FloatBarrier

\FloatBarrier
\clearpage
\section{Conclusion}
\label{sec:conclusion}
 
A search is performed for heavy charged LLPs, with lifetimes sufficient ($\tau\gtrsim 1~\text{ns}$) to reconstruct inner-detector tracks, produced at the LHC in $139~\mathrm{fb^{-1}}$ of $pp$ collisions at $\sqrt{s}=13~\TeV$. The identification of LLPs is based on anomalously high specific ionisation measured by the \mbox{ATLAS} pixel detector for isolated high-momentum tracks in events with high \MET. The considerable increase in sensitivity compared to previous ATLAS searches 
is not only due to the higher integrated luminosity, but also to several significant improvements in the analysis strategy. The most noticeable are the use of a higher \dEdx threshold, the separate treatment of the tracks with an IBL overflow flag and the use of a data-driven \dEdx-response template instead of a simulated one, as well as a more optimised definition of sub-regions in the signal region for exclusion interpretations. Evaluation of systematic uncertainties was also improved through the adoption of a high-momentum validation region and the implementation of a pseudo signal region to test the background generation method.
 
Observed yields and distributions agree with the SM background expectations, with the exception of an accumulation of events in the high-\dEdx and high-mass range. The local (global) significance of this excess is $3.6\sigma~(3.3\sigma)$ in a sub-range of the signal region optimised for a target mass hypothesis of $1.4~\TeV$. The events in the excess region were examined in detail. Although no obvious pathologies were identified in the measurement of these events, the time-of-flight measurements in outer detector subsystems clearly indicate that none of the candidate tracks are from charged particles moving significantly slower than the speed of light.
 
Maximum sensitivity is reached for LLPs with lifetimes of around 10--30~$\text{ns}$. Masses smaller than $2.27~\TeV$ are excluded at the 95\% confidence level for gluino $R$-hadrons with a lifetime of 20~$\text{ns}$ and $m(\ninoone)=100~\GeV$. The mass limit for compressed-scenario $R$-hadrons, with $\Delta m(\gluino,\ninoone)=30~\GeV$ and a lifetime of 30~$\text{ns}$, is $2.06~\TeV$. Masses below $1.07~\TeV$ for charginos and in the range $220\mbox{--}360~\GeV$ for staus are excluded for lifetimes of 30~$\text{ns}$ and 10~$\text{ns}$, respectively. The limits in the mass--lifetime plane are the most stringent to date and provide further constraints on the $R$-hadron, chargino and stau production models considered.


\FloatBarrier
\section*{Acknowledgements}


We thank CERN for the very successful operation of the LHC, as well as the
support staff from our institutions without whom ATLAS could not be
operated efficiently.
 
We acknowledge the support of
ANPCyT, Argentina;
YerPhI, Armenia;
ARC, Australia;
BMWFW and FWF, Austria;
ANAS, Azerbaijan;
CNPq and FAPESP, Brazil;
NSERC, NRC and CFI, Canada;
CERN;
ANID, Chile;
CAS, MOST and NSFC, China;
Minciencias, Colombia;
MEYS CR, Czech Republic;
DNRF and DNSRC, Denmark;
IN2P3-CNRS and CEA-DRF/IRFU, France;
SRNSFG, Georgia;
BMBF, HGF and MPG, Germany;
GSRI, Greece;
RGC and Hong Kong SAR, China;
ISF and Benoziyo Center, Israel;
INFN, Italy;
MEXT and JSPS, Japan;
CNRST, Morocco;
NWO, Netherlands;
RCN, Norway;
MEiN, Poland;
FCT, Portugal;
MNE/IFA, Romania;
MESTD, Serbia;
MSSR, Slovakia;
ARRS and MIZ\v{S}, Slovenia;
DSI/NRF, South Africa;
MICINN, Spain;
SRC and Wallenberg Foundation, Sweden;
SERI, SNSF and Cantons of Bern and Geneva, Switzerland;
MOST, Taiwan;
TENMAK, T\"urkiye;
STFC, United Kingdom;
DOE and NSF, United States of America.
In addition, individual groups and members have received support from
BCKDF, CANARIE, Compute Canada and CRC, Canada;
PRIMUS 21/SCI/017 and UNCE SCI/013, Czech Republic;
COST, ERC, ERDF, Horizon 2020 and Marie Sk{\l}odowska-Curie Actions, European Union;
Investissements d'Avenir Labex, Investissements d'Avenir Idex and ANR, France;
DFG and AvH Foundation, Germany;
Herakleitos, Thales and Aristeia programmes co-financed by EU-ESF and the Greek NSRF, Greece;
BSF-NSF and MINERVA, Israel;
Norwegian Financial Mechanism 2014-2021, Norway;
NCN and NAWA, Poland;
La Caixa Banking Foundation, CERCA Programme Generalitat de Catalunya and PROMETEO and GenT Programmes Generalitat Valenciana, Spain;
G\"{o}ran Gustafssons Stiftelse, Sweden;
The Royal Society and Leverhulme Trust, United Kingdom.
 
The crucial computing support from all WLCG partners is acknowledged gratefully, in particular from CERN, the ATLAS Tier-1 facilities at TRIUMF (Canada), NDGF (Denmark, Norway, Sweden), CC-IN2P3 (France), KIT/GridKA (Germany), INFN-CNAF (Italy), NL-T1 (Netherlands), PIC (Spain), ASGC (Taiwan), RAL (UK) and BNL (USA), the Tier-2 facilities worldwide and large non-WLCG resource providers. Major contributors of computing resources are listed in Ref.~\cite{ATL-SOFT-PUB-2021-003}.


\printbibliography

\clearpage
 
\begin{flushleft}
\hypersetup{urlcolor=black}
{\Large The ATLAS Collaboration}

\bigskip

\AtlasOrcid[0000-0002-6665-4934]{G.~Aad}$^\textrm{\scriptsize 100}$,
\AtlasOrcid[0000-0002-5888-2734]{B.~Abbott}$^\textrm{\scriptsize 118}$,
\AtlasOrcid[0000-0002-7248-3203]{D.C.~Abbott}$^\textrm{\scriptsize 101}$,
\AtlasOrcid[0000-0002-2788-3822]{A.~Abed~Abud}$^\textrm{\scriptsize 35}$,
\AtlasOrcid[0000-0002-1002-1652]{K.~Abeling}$^\textrm{\scriptsize 54}$,
\AtlasOrcid[0000-0002-2987-4006]{D.K.~Abhayasinghe}$^\textrm{\scriptsize 93}$,
\AtlasOrcid[0000-0002-8496-9294]{S.H.~Abidi}$^\textrm{\scriptsize 28}$,
\AtlasOrcid[0000-0002-9987-2292]{A.~Aboulhorma}$^\textrm{\scriptsize 34e}$,
\AtlasOrcid[0000-0001-5329-6640]{H.~Abramowicz}$^\textrm{\scriptsize 149}$,
\AtlasOrcid[0000-0002-1599-2896]{H.~Abreu}$^\textrm{\scriptsize 148}$
% ...
\bigskip
\\

% ...
$^{28}$Physics Department, Brookhaven National Laboratory, Upton NY; United States of America.\\
% ...
$^{34}$$^{(a)}$Facult\'e des Sciences Ain Chock, R\'eseau Universitaire de Physique des Hautes Energies - Universit\'e Hassan II, Casablanca;$^{(b)}$Facult\'{e} des Sciences, Universit\'{e} Ibn-Tofail, K\'{e}nitra;$^{(c)}$Facult\'e des Sciences Semlalia, Universit\'e Cadi Ayyad, LPHEA-Marrakech;$^{(d)}$LPMR, Facult\'e des Sciences, Universit\'e Mohamed Premier, Oujda;$^{(e)}$Facult\'e des sciences, Universit\'e Mohammed V, Rabat;$^{(f)}$Institute of Applied Physics, Mohammed VI Polytechnic University, Ben Guerir; Morocco.\\
$^{35}$CERN, Geneva; Switzerland.\\
% ...
$^{54}$II. Physikalisches Institut, Georg-August-Universit\"{a}t G\"ottingen, G\"ottingen; Germany.\\
% ...
$^{93}$Department of Physics, Royal Holloway University of London, Egham; United Kingdom.\\
% ...
$^{100}$CPPM, Aix-Marseille Universit\'e, CNRS/IN2P3, Marseille; France.\\
$^{101}$Department of Physics, University of Massachusetts, Amherst MA; United States of America.\\
% ...
$^{118}$Homer L. Dodge Department of Physics and Astronomy, University of Oklahoma, Norman OK; United States of America.\\
% ...
$^{148}$Department of Physics, Technion, Israel Institute of Technology, Haifa; Israel.\\
$^{149}$Raymond and Beverly Sackler School of Physics and Astronomy, Tel Aviv University, Tel Aviv; Israel.\\
% ...
$^{a}$ Also Affiliated with an institute covered by a cooperation agreement with CERN.\\
$^{b}$ Also at Borough of Manhattan Community College, City University of New York, New York NY; United States of America.\\
$^{c}$ Also at Bruno Kessler Foundation, Trento; Italy.\\
$^{d}$ Also at Center for High Energy Physics, Peking University; China.\\
$^{e}$ Also at Centro Studi e Ricerche Enrico Fermi; Italy.\\
$^{f}$ Also at CERN, Geneva; Switzerland.\\
% ...

\end{flushleft}


\end{document}